\documentclass[
amsmath,amssymb,
aps,
prb,
twocolumn,
superscriptaddress
]{revtex4-2}

\usepackage{graphicx}
\usepackage{dcolumn}
\usepackage{bm}
\usepackage{hyperref}
\hypersetup{linktocpage,colorlinks=true,citecolor=blue}
\usepackage[final]{changes}
\newcommand{\stkout}[1]{\ifmmode\text{\sout{\ensuremath{#1}}}\else\sout{#1}\fi}
\setdeletedmarkup{\stkout{#1}}

\definecolor{newpink}{rgb}{1.0, 0.412, 0.71}

\def \be {\begin{equation}}
\def \ee {\end{equation}}
\def \bee{\begin{equation*}}
\def \eee{\end{equation*}}

\def \dd  {{\rm d}}

\def \bvec {\left( \begin{array}{c}}
	\def \evec {\end{array}\right)}

\def \l1{\overline{1}} 

\def \v{\rm{v}}

\newcommand{\nn}{\nonumber \\}
\newcommand{\np}{\; .}
\newcommand{\nc}{\; ,}
\newcommand{\nnc}{\; , \nonumber \\}
\newcommand{\pp}{\prime\prime}

\newcommand{\xfilll}[2][1ex]{%
	\dimen0=#2\advance\dimen0 by #1%
	\leaders\hrule height \dimen0 depth -#1\hfill%
}

\definecolor{deeppink}{rgb}{1.0, 0.08, 0.58}
\definecolor{brilliantrose}{rgb}{1.0, 0.33, 0.64}
\definecolor{darkorange}{rgb}{1.0, 0.55, 0.0}
\definecolor{darkpastelpurple}{rgb}{0.59, 0.44, 0.84}
\definecolor{red}{rgb}{1, 0.0, 0.0}

\newcommand{\norm}[1]{ 
	{ \left| #1 \right|} }

\begin{document}

\preprint{}

\title{A mechanism for quantum-critical  Planckian metal phase in high-temperature cuprate superconductors}

\author{Yung-Yeh Chang}
\affiliation{Institute of Physics, Academia Sinica, Taipei 11529, Taiwan}
\affiliation{Physics Division, National Center for Theoretical Sciences, Taipei 10617, Taiwan}
\affiliation{Department of Electrophysics, National Yang Ming Chiao Tung University, Hsinchu 30010, Taiwan}
\author{Khoe Van Nguyen}
\affiliation{Department of Electrophysics, National Yang Ming Chiao Tung University, Hsinchu 30010, Taiwan}
\author{Kim Remund}
\affiliation{Physics Division, National Center for Theoretical Sciences, Taipei 10617, Taiwan}
\affiliation{Department of Electrophysics, National Yang Ming Chiao Tung University, Hsinchu 30010, Taiwan}
\author{Chung-Hou Chung}%
\email{chung0523@nycu.edu.tw}
\affiliation{Physics Division, National Center for Theoretical Sciences, Taipei 10617, Taiwan}
\affiliation{Department of Electrophysics, National Yang Ming Chiao Tung University, Hsinchu 30010, Taiwan}
\affiliation{Center for Theoretical and  Computational Physics (CTCP), National Yang Ming Chiao Tung University, Hsinchu 30010, Taiwan}




\date{\today}

\begin{abstract}
The mysterious metallic phase showing perfect $T$-linear resistivity and a universal scattering rate $1/\tau = \alpha_P k_B T /\hbar$ with a universal prefactor $\alpha_P \sim 1$ and logarithmic-in-temperature singular specific heat coefficient, so-called ``Planckian metal phase" was observed in various overdoped high-$T_c$ cuprate superconductors over a finite range in doping. Here, we propose a microscopic mechanism for this exotic state based on quantum-critical bosonic charge Kondo fluctuations coupled to both spinon and a heavy conduction-electron Fermi surfaces within the heavy-fermion formulation of the slave-boson $t$-$J$ model. Using a controlled perturbative renormalization group (RG) analysis, we examine the competition between the pseudogap phase, characterized by Anderson's Resonating-Valence-Bond spin-liquid, and the Fermi-liquid state, characterized by the electron hoping (effective charge Kondo effect). We find a quantum-critical metallic phase with a universal Planckian $\hbar \omega/k_B T$ scaling in scattering rate near a localized-delocalized (pseudogap-to-Fermi liquid) charge  Kondo breakdown transition. Our results are in excellent agreement with the recent experimental observations on optical conductivity (without fine-tuning) in \textit{Nat. Commun.} \textbf{14}, 3033 (2023), universal doping-independent field-to-temperature scaling in magnetoresistance in \textit{Nature} \textbf{595}, 661 (2021), and the marginal Fermi-liquid spectral function observed in ARPES (\textit{Science} \textbf{366}, 1099 (2019)) as well as Hall coefficient in various overdoped cuprates in \textit{Nature} \textbf{595}, 661 (2021) and \textit{Annu. Rev. Condens. Matter Phys.} \textbf{10}, 409 (2019). Our mechanism offers a microscopic understanding of the quantum-critical Planckian metal phase observed in cuprates and its link to the pseudogap, \textit{d}-wave superconducting, and Fermi liquid phases.
\end{abstract}

\maketitle


\section{Introduction}

\begin{figure*}[ht]
    \centering
    \includegraphics[width=0.95\textwidth]{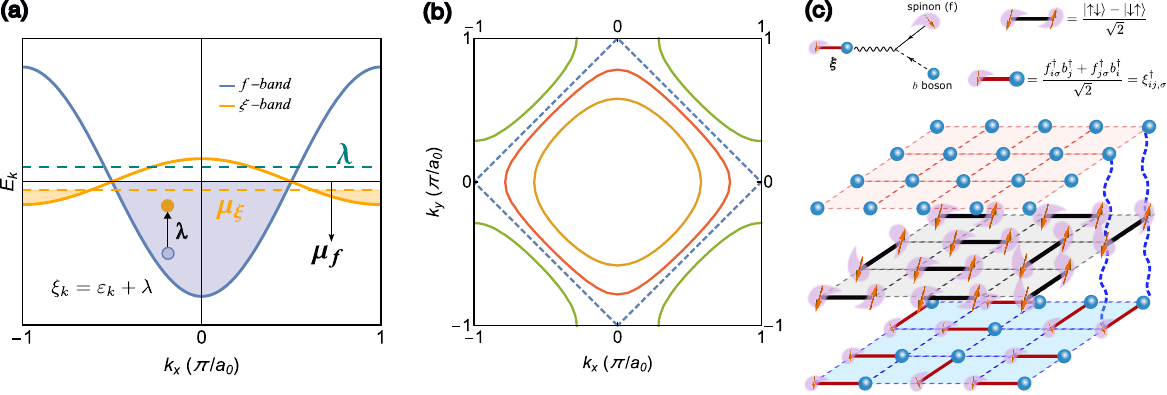}
    \caption{(a) Schematic plots for the dispersive $f$-spinon (blue curve) and the weakly dispersive $\xi$ (orange curve) bands (corresponding to hole doping $\delta$), associated with their chemical potentials, $\mu_f$ and $\mu_\xi$. The green dashed line denotes the energy level (Lagrange multiplier) $\lambda$ for the slave boson. The shaded areas represent filling of the $f$- and $\xi$-bands. $a_0$ here denotes the lattice constant. (b) Fermi surfaces of the $f$-spinon band (blue) and of the $\xi$ band with different levels of hole doping (orange for hole doping $\delta$, green for $1+\delta$, and red in between $\delta$ and $1+\delta$). (c) Main plot: Schematic plot of strange-metal state. The blue dashed curves represent the Kondo-like hopping term. Upper left: Feynman diagram for the interaction vertex of the Kondo-like hopping term. Upper right: the RVB spin-singlet bond in the $f$-spinon band and the spinon-holon bound state (the $\xi$ field).}
    \label{fig:schematics}
\end{figure*}

Over the recent three decades, metallic behavior that cannot be described within the Fermi liquid (FL) paradigm has commonly been observed in a wide variety of strongly correlated quantum materials. Yet, the emergence of such metals is poorly understood. This non-Fermi liquid (NFL) behavior often exists near a quantum phase transition, and shows ``strange metal (SM)" phenomena with (quasi-)linear-in-temperature decreasing resistivity and divergent logarithmic-in-temperature specific heat coefficient as $T \to 0$.

Of particularly intriguing class of SM states is the ``Planckian metal", observed in the normal state of unconventional superconductors, including cuprate superconductors \cite{Fournier-PCCO-PRL,Taillefer-planckian-2019}, iron pnictides and chalcogenides \cite{Hussey-pnictide0-RPP, Hussey-pnictide-Nature,Kasahara-PRB-pnictide,Jiang-JPCM-pnictide, FeSeS-PRR-2020}, organic \cite{Taillefer-Organic-PRB, Mackenzie-Science} and heavy-fermion compounds \cite{Gegenwart-YRS-PRL,Custers2003Nature, Custers2010prl, Mackenzie-Science}, and twisted bilayer graphene \cite{Cao-SM-TBG-PRL}. It shows perfect $T$-linear scattering rate, reaching ``Planckian dissipation limit" allowed by quantum mechanics, $1/\tau(T) = \alpha_P k_BT/\hbar$ with $\alpha_P \sim 1$. Since \textit{d}-wave superconductivity (dSC) emerges out of this exotic state of matter with decreasing temperatures, it was argued that revealing the mystery of the Planckian metal state is the key to understanding the mechanism for high-$T_c$ superconductivity in cuprates \cite{Zaanen-2004-nature,Hussey-Sci-Strangethanmetal}. Experimental observations in heavy-fermion superconductors suggest that $\alpha$ is a non-universal constant but depends on the strength of the Kondo correlations \cite{YYC-SM-115-NatComm}. By contrast, in overdoped cuprates the Planckian perfect $T$-linear scattering rate with the same $\alpha$ persists from very high temperatures ($T \sim 300$K) to very low temperatures ($T \to 0$)  over a wide range  of samples near optimal doping, indicating  that the Planckian state in overdoped cuprates might be a universal feature. Similar universal behavior has also been observed in magic angle twisted bi-layer graphene \cite{Cao-SM-TBG-PRL}.  
The Planckian behavior across the entire temperature range, in particular for the low temperature regime ($T\to 0$) in the  two mentioned systems is unlikely to be explained by phonons whose contributions are mainly at high temperatures. Meanwhile, perfect frequency-to-temperature ($\hbar \omega/k_B T$) scaling from optical conductivity measurement \cite{George-NatComm-SM} and field-to-temperature ($B/T$ ) scaling in magnetoresistance as well as other quantum critical-like properties extending over a range from the critical doping to the end of the dSC dome in various hole-doped \cite{hussey-incoherent-nature-2021} and electron-doped  \cite{green-e-doped-cuprate-scale-invariant,Greene-nature-e-doped-cuprate-spin-fluc} cuprates strongly suggest that a Planckian state is a quantum critical ``phase'' \cite{Hussey-Sci-Strangethanmetal}.  These observations lead to fundamental questions: What is the microscopic mechanism for this exotic phase of matter and its links to quantum criticality and the neighboring phases? The Planckain metal state lies in between the pseudogap and Fermi liquid phases with localized and itinerant characters of electrons, respectively. This points to an appealing scenario in which the Planckian metal phase may arise near a possible localized-delocalized quantum critical point due to competition between the pseudogap and Fermi liquid phases, and the \textit{d}-wave superconductivity is  reached by condensing this quantum critical metal.   

It is challenging to develop a controlled approach for this exotic state of matter. Recently, this state has been realized theoretically via controlled large-$N$ approach of the $t$-$J$ model with random hoping and exchange couplings, the ``SYK model" \cite{Patel-PRL-SM, Patel-2023-SYK-Sci,patel-pnas-overdamped-boson} and in the Hubbard model by Quantum Monte Carlo \cite{deveraux-SM-Hubbard-Sci}. Nevertheless, a microscopic understanding of this state in terms of critical spin and charge fluctuations within the well-established non-random Hubbard or \textit{t-J} model framework has yet to be developed. 

In this work, we address the microscopic mechanism for the Planckian metal phase in cuprates via different approaches from the previous ones \cite{Nagaosa-1990-PRL,Nagaosa-1992-PRB-cuprate,Nagaosa-2006-RMP, Varma-loopcurrent-RPP} based on recently developed heavy-fermion perspective of the two-dimensional slave-boson \textit{t}-\textit{J} model, known to offer qualitative understanding of cuprate phase diagram,  in the formulation of the Kondo-Heisenberg lattice model \cite{Punk-SBtJ-PRB}. Within our framework, the Planckian metal phase appears as a result of  local disordered bosonic charge fluctuations coupled to a fermionic spinon band and a heavy-electron band near a localized-deloclaized quantum critical point due to the competition between the  pseudogap and Fermi liquid phases. 
In this approach, the hoping of electrons is expressed in terms of effective Kondo hybridization between a composite fermionic spinon-holon bound state representing conduction band and a charge-neutral gapless fermionic spinon band with a Fermi surface, while as the Heisenberg exchange coupling is described by resonating-valence-bond (RVB)  spin-liquid with both hoping of fermionic spinons and singlet pairing between them. The mean-field theory of this approach captures qualitatively the pseudogap, Fermi liquid, and \textit{d}-wave superconducting phases. In particular, it advances the previous slave-boson \textit{t-J} approach by capturing aspects of coherent quasi-particle excitation observed at nodal Fermi pocket in the pseudogap phase. Meanwhile, the spin-liquid state here is further stablized by coupling to charge Kondo hybridization, similar to the Kondo-stablized spin-liquid mechanism in heavy-fermion systems \cite{Coleman-Andrei}. Here, by perturbative renormalization group analysis, we study the quantum phase transition of the model beyond mean-field and seek the possible emergence of Planckian metal state due to spin and charge fluctuations near criticality. Our approach is highly motivated by the striking similarity in strange metal phenomenology between cuprates \cite{Taillefer-annuphys-2019,hussey-incoherent-nature-2021} and heavy-fermion Kondo lattice systems \cite{Shishido-FS-jump,Friedemann-PNAS-2010}, in which the Fermi surface volume reconstructs over the entire strange metal region in both systems. This indicates a Kondo-breakdown-like physics observed in heavy-fermion systems may appear in cuprates where the SM state and Fermi surface reconstruction occur simultaneously near the Kondo breakdown QCP due to coupling of local charge fluctuations to the Fermi surface \cite{Chang-2018-SM,YYC-SSC-2019,YYC-SM-115-NatComm}. A related yet distinct heavy-fermion Kondo lattice approach was proposed in Ref. \cite{Sachdev-ancilla-PRR} to address the evolution from the pseudogap metal with small Fermi surfaces to the conventional Fermi liquid with a large Fermi surface. 
Via the controlled RG analysis, a stable quantum critical Planckian strange metal (SM) phase over a finite range in doping with universal $T$-linear scattering rate ($\alpha \sim O(1)$ being a universal constant independent of microscopic couplings) is realized near a localized-delocalized Kondo breakdown transition. Therein, the local bosonic charge (effective Kondo) fluctuations coupled to composite fermionic conduction band and gapless fermionic spinons. The universal quantum critical $\omega /T$-scaling is found in this phase in dynamical scattering rate, in excellent agreement without fine-tuning with the optical conductivity measurement \cite{George-NatComm-SM}, and the universal doping-independent field-to-temperature scaling in magnetoresistance over an extended doping range in Ref. \cite{hussey-incoherent-nature-2021} in the strange metal region of various overdoped cuprates. The marginal Fermi liquid single-electron spectral function and crossover in Fermi surface volume in this phase are both in good agreement with ARPES \cite{zxshen-sicience-incoherent-cuprate} and Hall measurements \cite{Taillefer-annuphys-2019, hussey-incoherent-nature-2021}, respectively. Our study indicates that this exotic phase is a quantum critical phase governed by the critical charge (effective Kondo) fluctuations at the localized-delocalized Kondo breakdown quantum critical point arising from the competition between the pseudogap (Cooper-pair formation) and Fermi liquid (electron hoping) phases. It provides an understanding of the strange metal state within the context of the global phase diagram of cuprate superconductors and reconciles the seemingly inconsistent scenarios of the strange metal state observed in cuprates.

\begin{figure}[t]
    \centering
    \includegraphics[width=0.42\textwidth]{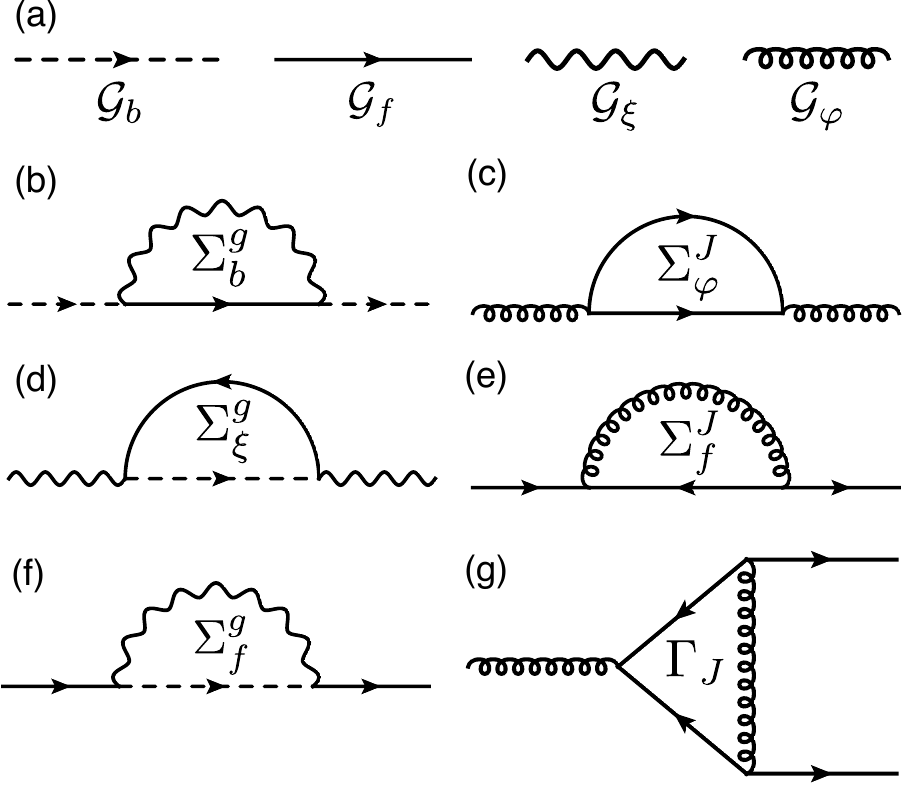}
    \caption{\textbf{The Feynman diagrams for the self-energy and vertex correction being used in the RG analysis.} (a) The graphical representation of the bare propagator of various fields/operators.  Feynman diagrams of the (b)-(f) self-energy and (g) vertex correction for $H_J$.}
    \label{fig:feyn-diag}
\end{figure}

\section{Results}
\subsection{Heavy-fermion formulation of the slave-boson $t$-$J$ model}
We start from the Hamiltonian of  the slave-boson representation of $t$-$J$ model on a 2D lattice \cite{Punk-SBtJ-PRB}, describing as $ H=H_{t}+H_{J}$ with $H_{t} =-t\sum_{\langle i,j \rangle , \sigma} c_{i\sigma}^{\dagger} c_{j\sigma}-\mu \sum_{i\sigma} c_{i\sigma}^{\dagger} c_{i\sigma}$ and $	H_J  = J_{H} \sum_{\langle i,j\rangle} \bm{S_{i}} \cdot\bm{S_{j}}.$ Here, $(t,\, \mu, \, J_H)$ denotes the (hopping strength, chemical potential, Heisenberg coupling), and $\langle i,j \rangle$ the nearest-neighboring sites, and the local spin operator $ \bm{S}_{i}=\frac{1}{2} \sum_{\sigma\sigma^\prime} c_{i\sigma}^{\dagger} \bm{\sigma}_{\sigma\sigma^\prime} c_{i\sigma^\prime} $. 
Under the slave-boson representation $c_{i\sigma}^\dagger \to  f^\dagger_{i\sigma} b_i$ with $f_{i\sigma} \, (b_i)$ being fermionic charged-neutral spinon (bosonic spinless charged holon), we further factorize the $H_t$ and $H_J$ terms via Hubbard–Stratonovich transformation following Ref. \cite{Punk-SBtJ-PRB}, i.e., $H_{t} \to t\sum_{\langle i,j\rangle,\sigma} \left[\left(f_{i\sigma}^{\dagger}b_{j}^{\dagger}+f_{j\sigma}^{\dagger}b_{i}^{\dagger}\right)\xi_{ij,\sigma}+H.c.\right]$ and $ H_{J} \to \sum_{\langle i,j\rangle\sigma}\left(-\chi_{ij}f_{i\sigma}^{\dagger}f_{j\sigma}+ \Delta_{ij}\tilde{\sigma}f_{i\sigma}^{\dagger}f_{j,-\sigma}^{\dagger}  +H.c.\right), ~ \tilde{\sigma} \equiv \text{sgn}(\sigma)$.

\begin{figure*}[ht]
    \centering
    \includegraphics[width=0.98\textwidth]{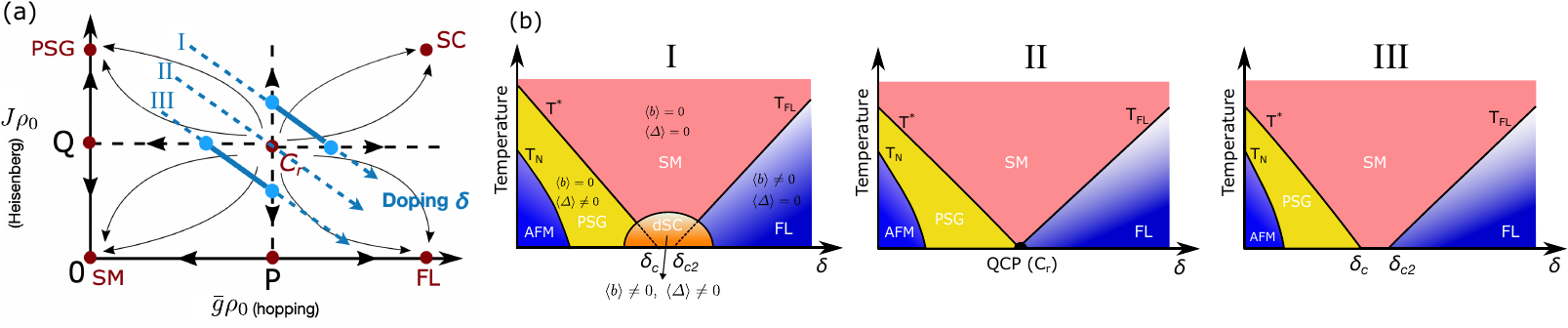}
    \caption{(a) Plot of the RG flow for the Heisenberg and hopping terms. Paths I, II, III in (a) indicate the three types of phase diagram that can be accessed from the RG equations. (b): three possible distinct types of phase diagrams (I, II, III) corresponding to paths (I, II, III) from the RG flow diagram in (a). $T_N$, $T^*$, and $T_{FL}$ denote the temperature crossovers for the anti-ferromagnetic (AFM) order, pseudogap, and  FL states, respectively. In (b)-I, $\delta_c$ and $\delta_{c2}$ represent the critical points associated respectively with PSG-SM and SM-FL transitions once superconductivity is suppressed.}
    \label{fig:RG-flow-PD}
\end{figure*}

Note that we decompose the Heisenberg interaction into both the particle-hole and particle-particle forms. Here, $\xi_{ij,\sigma}$, $\chi_{ij}$, and $\Delta_{ij}$ represent the auxiliary Hubbard-Stratonovich fields living between sites $i$ and $j$, as a result, their degrees of freedom for the bond fields are twice as many as that for $b$ and $f$. The $\chi_{ij}$ and $\Delta_{ij}$ fields represent the spinon hopping and pairing bond fields, respectively, whose condensate play the role of the effective hopping and  pre-formed singlet \textit{d}-wave Cooper pairing of spinons in the RVB spin liquid. The  model shows a U(1) gauge symmetry: $f_{i\sigma} \to f_{i\sigma} e^{i \theta_i}$, $b_i \to b_i e^{i \theta_i}$, $\chi_{ij} \to \chi_{ij} e^{i (\theta_i - \theta_j)}$ and $\xi_{ij,\sigma} \to \xi_{ij,\sigma} e^{i (\theta_i+\theta_j)}$. A finite and uniform mean-field $\chi=\langle \chi_{ij} \rangle$ is assumed. The $\xi_{ij,\sigma}$ field, previously proposed in Refs. \cite{Moon-Sachdev-PRB,Punk-PNAS-dimer,Punk-SBtJ-PRB},  describes a gauge charge-2$e$ spinon-chargeon bound state; it further advances the original slave-boson approach by capturing the nodal Fermi pocket (arc) quasi-particle excitations in the pesudogap phase of cuprates. The hoping term $H_t$, after the decomposition, behaves as an effective Kondo coupling in the charge sector. If the slave boson $b$ gets Bose condensed, the $\xi$ fermion becomes a physical conduction electron. Terms containing slave boson occupation number $n_b^i$ are neglected in the Hubbard-Stratonovich transformation for $H_t$ and $H_J$. The constraints $b_{i}^{\dagger} b_{i} + \sum_{\sigma} f_{i\sigma}^{\dagger} f_{i\sigma} = 1$  is imposed \cite{Coleman-SB,Kotliar-PRL-dwave,Punk-SBtJ-PRB}. Four distinct mean-field phases are realized in this model, depending on whether or not the two boson fields $b$ and $\Delta$ get Bose-condensed [see Fig. \ref{fig:RG-flow-PD}(b) and Ref. \cite{Punk-SBtJ-PRB}]: the pseudogap phase, known as $Z_2$ fractionalized Fermi liquid or $Z_2$ FL$^\star$, (Landau Fermi liquid, FL) is reached when $\Delta \neq 0, \, \langle b \rangle =0 \, (\Delta =0,  \langle b\rangle \neq 0)$; while the U(1) FL$^*$ (\textit{d}-wave superconducting) phase arises when $\langle b \rangle = \Delta =0 \, (\langle b \rangle \neq 0, \, \Delta \neq 0)$.  The $Z_2$ FL$^\star$ and U(1) FL$^\star$ phases realized here are examples of the previously proposed fractionalized Fermi liquid states in Ref. \cite{Senthil-prl-fractionalized-FL} \added{in the context of heavy-fermion Kondo lattice systems} with a small Fermi surface volume and charge-neutral fractionalized spinon excitations which carry a gauge charge characterizing the topological order. \added{The spinons in these two fractionalized spin-liquid phases are deconfined due to the existence of a spinon Fermi surface, and are thus stable against U(1) gauge field fluctuations \cite{SSLee-PRB-2008-stablility,Hermele-stability-spinliquid}.} Meanwhile, \added{these two fractionalized phases are energetically more stable} compared to that in the earlier approach of slave-boson \textit{t-J} model \cite{Kotliar-PRL-dwave}  by the charge Kondo hybridization term, similar to the Kondo-stablized spin-liquid mechanism in heavy-fermion systems \cite{Coleman-Andrei, sm-phase-PNAS}.

We shall explore the phase diagram in U(1) FL$^*$ phase beyond mean-field by including fluctuations of both $t$- and $J$-terms. We will not include U(1) gauge fluctuations in our RG analysis since the spinons are deconfined and stable against them due to the presence of a spinon Fermi surface \cite{SSLee-PRB-2008-stablility}.  The leading  effective action in units of the half-bandwidth of the \textit{f}-spinon band $D\approx 4\chi \approx 4J_H$ beyond the mean-field level reads ($\hbar = k_B = 1$) \cite{Punk-SBtJ-PRB}
     \begin{align}
         S=&-\sum_{k\sigma}f_{k\sigma}^{\dagger}\mathcal{G}^{-1}_f (k)f_{k\sigma}-\sum_{k}b_{k}^{\dagger}\mathcal{G}^{-1}_b (k)b_{k}  \nn 
         &+ \sum_{k\sigma a}\xi_{k\sigma}^{a\,\dagger} \mathcal{G}^{-1}_\xi(k) \xi_{k\sigma}^{a} + \sum_{k a}\varphi_{k}^{a\,\dagger}\mathcal{G}^{-1}_\varphi(k) \varphi_{k}^{a} \nn
         & +\frac{g}{\sqrt{\beta N_{s}}}\sum_{kp\sigma a}\left(f_{k\sigma}^{\dagger}b_{p}^{\dagger}\xi_{k+p,\sigma}^{a}+H.c.\right)  \nn
         & +\frac{J}{\sqrt{\beta N_{s}}} \sum_{kpa}\left[f_{k\uparrow}^{\dagger}f_{p \downarrow}^{\dagger}\varphi_{k+p}^{a}+H.c.\right] 
    \label{eq:action-S}
     \end{align}
with $k = (\bm{k},\omega)$ and $p = (\bm{p},\nu)$, $\varphi$ being the fluctuating field for the \textit{d}-wave RVB pairing order parameter $\Delta$. \added{In the phases with $\langle b\rangle=0$, the bare hoping parameter $t$ is strongly suppressed by the disordered bosons, leading to an effective hoping $g \equiv 2t \sqrt{\delta}$. Here, $ J = 2J_H$ is the effective exchange}, $\mathcal{G}_f (k)= (i\omega-\varepsilon_{\bm{k}})^{-1}$, $\mathcal{G}_\xi (k) = \zeta^{-1} \left( i\omega-\xi_{\bm{k}} \right)^{-1}$, $\mathcal{G}_b = (i\omega - \lambda)^{-1}$, and $\mathcal{G}_\varphi = (J/2)^{-1}$ denote the bare Green's functions (see Supplementary Notes 2). The momenta relevant for the bosonic $\varphi$ field are near the antinode. \added{Our perturbative expansion in bare couplings $g$, $J$ is controlled since $g/D, \, J/D <1$ (for an  estimated $J_H/t \sim 0.3$).}
To study the charge dynamics and transport properties, we go beyond static mean-field level of $\xi$ field by generating its dynamics and dispersion via second-order hoping process at fixed $g=g^*_0$ such that $\zeta^{-1} \equiv ( g^2 \rho_0/D)^{-1}$ appears as a prefactor in $\mathcal{G}_\xi$, with $\rho_0 = 1/D$ being the constant density of states at Fermi level for $f$ spinon band.  
The $f$-spinon band is approximated by a linear-in-momentum dispersion, namely $\varepsilon_{\bm{k}}=h_{\bm{k}}-\mu_{f} \approx v |\bm{k}|$ with  $h_{\bm{k}}\equiv-2\chi\left(\cos k_{x}+\cos k_{y}\right)$ and $\mu_{f}=\mu-\lambda$ being the effective chemical potential for $f$ spinon ($\mu$ is the chemical potential for the original conduction electron, see Supplementary Notes 1). Here, the spinon band is fixed at half-filling, $\mu_f=0$.  The band structure for the $\xi$ fermion shows a hole-like dispersion, $\xi_{\bm{k}}=-\zeta\varepsilon_{\bm{k}}-\mu_\xi$. The Lagrange multiplier $\lambda >0$ is introduced to enforce the local constraint for slave boson. The slave boson ($b$ field) is effectively treated as a local boson with a flat band of energy $\lambda$ and with a negligible dispersive band (see Supplementary Notes 2). In the pseudogap and U(1) FL$^\star$ phases, the chemical potential of the $\xi$ band $\mu_\xi$ fixes hole doping $\delta$ for the system such that  $N_s^{-1} \sum_{ij,\sigma}  \langle \xi^\dagger_{ij, \sigma}\xi_{ij, \sigma} \rangle = \delta$. Note that Eq. (\ref{eq:action-S}) is also applicable when \textit{d}-wave preformed Cooper pair order, relevant for temperatures slightly above the superconducting transition temperature of cuprates, is replaced by the pair density wave state, considered as a hallmark of pseudogap phase at higher temperatures \cite{Berg_Kivelson_pdw-njp,patrick-prb-pdw}. 

\subsection{Renormalization group analysis}


The perturbative RG analysis is applied to the effective action of the modified slave-boson model Eq. (\ref{eq:action-S}) by considering the one-loop diagrams shown in Fig. \ref{fig:feyn-diag}. \added{It is convenient to define the dimensionless bare coupling constants $g \to g/D$, $J \to J /D$, and absorb $1/\zeta$ in $\mathcal{G}_\xi$ into $g$ by a rescaling, $g \to \bar{g} = g/\sqrt{\zeta}, \, \xi \to \sqrt{\zeta} \xi $. Our RG analysis is perturbatively controlled as $\bar{g}<1$ 
(the bare value $g<g_0^*$ is set), and $J<1$.}
Our RG approach can be further controlled by the $\epsilon$-expansion technique with a small parameter $\epsilon=d-z$ within the  convergence radius $|\epsilon|\leq 1$. We set the dynamical exponent $z=1$ due to the linearized dispersion of $f$, and spatial dimension $d=2$ here. The RG scaling equations of the running renormalized dimensionless couplings, $\bar{g}(\ell) \rho_0$ and $J(\ell) \rho_0$ read (see Supplementary Notes 3)
\begin{align}
     \frac{d(\bar{g}(\ell)\rho_{0})}{d\ell}	 & =-\left(\frac{\epsilon}{2}\right)(\bar{g}(\ell)\rho_{0}) + \left(\bar{g}(\ell)\rho_{0}\right)^{3}, \nn
     \frac{d(J(\ell)\rho_{0})}{d\ell}	& =-\left(\frac{\epsilon}{2}\right)(J(\ell)\rho_{0})+\left(J(\ell)\rho_{0}\right)\left(\bar{g}(\ell)\rho_{0}\right)^{2}\nn
     &\qquad   +\frac{3}{2}\left(J(\ell)\rho_{0}\right)^{3},
 \label{eq:beta-fun-shift}
\end{align}
where scaling parameter $\ell\equiv-\ln D>0$. The RG flow equations of Eqs. (\ref{eq:beta-fun-shift}) is shown in Fig. \ref{fig:RG-flow-PD}(a) where the critical fixed point occurs at $C_r = \left(J^*\rho_{0},\, \bar{g}^*\rho_{0} \right) = (\sqrt{\epsilon/2},\sqrt{\epsilon/3})$. \added{For simplicity, the U(1) gauge-field fluctuations are not included in our RG analysis; nevertheless, its effect on transport and thermodynamic properties are addressed in Discussions (see also Supplementary Notes 5).} 


\textit{RG flow and phase diagram.} As shown in Fig. \ref{fig:RG-flow-PD}(a), our one-loop RG flow  diagram supports a quantum critical point (QCP) at $C_r$, separating the four phases: the pseudogap PSG (FL) phase is reached when the Heisenberg coupling $J$ becomes relevant (irrelevant), while the hopping (effective Kondo) term $\bar{g}$ becomes irrelevant (relevant); the \textit{d}-wave superconducting state (U(1) FL$^*$, SM) phase arises for both $\bar{g}$ and $J$ being relevant (irrelevant), respectively. \added{Near $C_r$, the correlation length $\eta (\ell)$ shows a power-law divergence as:  $\eta (\ell)\sim \ell^{-\nu}$ with the exponent $\nu = 1$, indicating linear-in-doping crossover scales $T^*, \, T_{FL} \propto \eta^{-z}  \propto |\delta-\delta_c|$, corresponding to the PSG$\to$SM and SM $\to$ FL crossovers, respectively \cite{Sato-YBCO-nature}.} We shall see below that the systems with initial couplings flowing to the mean-field U(1) FL$^*$ fixed point show strange metal features with universal quantum critical Planckian scaling in electron scattering rate; therefore beyond mean-field under RG, it corresponds to the strange metal fixed point.


\begin{figure}[h]
    \centering
    \includegraphics[width=0.48\textwidth]{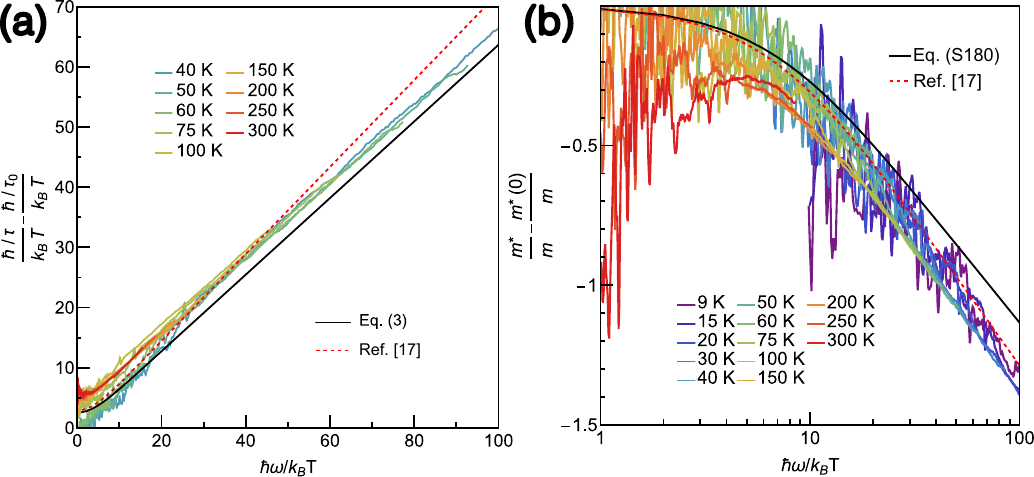}
    \caption{\textbf{$\omega/T$ scaling of scattering rate and the effective mass enhancement}. Theoretical fitting (black curve) for both (a) scattering rate $\hbar/\tau$ and (b) the effective mass enhancement  from our theory exhibits a universal scaling behavior $m^*/m - 1=g(x)$ as a function of $x\equiv \hbar\omega/k_B T$. Theoretical result of $m^*/m - 1$ (red dashed line) obtained from Eq. (S180) in Supplementary Notes 6B is added in (b) for comparison. The experimental data shown here is reproduced from Ref. \cite{George-NatComm-SM}.}
    \label{fig:htau-scaling}
\end{figure}

\subsection{Physical observables near the QCP}
\subsubsection{Conductivity and scattering rate}
The total conductivity $\sigma_{tot} = \rho^{-1}_{tot}$ can be computed via the rule by Ioffe and Larkin \cite{larkin-rule-PRB}:  $\sigma_{tot} = \sigma_\xi \sigma_f/ (4\sigma_\xi + \sigma_f)$. Because the slave boson is localized and has negligible dispersion, it does not contribute to the conductivity.
Since the effective mass of the $\xi$ band is much heavier than that for the $f$ spinons, $m_\xi/m_f \sim 1/\zeta \gg 1$,  resistivity is dominated by the $\xi$-band, leading to  $\rho_{tot} = 1/\sigma_{tot} \approx 1/\sigma_\xi= \rho_\xi$.    

In the theory of local (momentum independent) self-energy, the vertex corrections are negligible, and the optical conductivity can be obtained directly from the self-energy, $\sigma(\omega) = \left( i\Phi(0)/\omega\right)\int^\infty_{-\infty} d\varepsilon \left[n_F (\varepsilon)-\right. \left. n_F (\varepsilon+\hbar \omega)\right] / [ \hbar \omega + \Sigma_c^* (\varepsilon)-\Sigma_c(\varepsilon+\hbar \omega)]$, where $\Phi (\varepsilon)$ is the transport function (see Ref. \cite{George-NatComm-SM} for details). Therefore, electron transport time is equivalent to the relaxation time since $\sigma_\text{tot} \sim  \sigma_\xi$. The total scattering rate is hence dominated by the contribution from the $\xi$ field: $1/\tau_\text{tot} \sim  1/\tau_\xi$. In the SM state with $\langle b \rangle=\Delta=0$, the gauge-invariant (physical) electron operator can be constructed from $\xi_{ij}$ and $\varphi_{ij}$ as $c= \xi \sqrt{2} \varphi_{ij}^* \sqrt{J}$. The dynamical scattering (relaxation) rate at $T = 0$ is calculated via the imaginary part of the electron self-energy: $\tau(\omega) =-\hbar /2\Sigma_c^{\prime\prime}(\omega)$ and $\Sigma_c^{\prime\prime}(\omega)=(1/2) \Sigma^{\prime\prime}_{\xi}(\omega)$,  where $\Sigma^{\prime\prime}_{\xi}(\omega)$ at $T=0$ is obtained via  second-order RG renormalized perturbation close to the QCP with bare couplings being replaced by the renormalized ones i.e. $\bar{g}, \, \bar{g}^*_0\to \bar{g}(\ell)$, showing a universal local MFL self-energy insensitive to couplings \cite{Chowdhury-2023-prb}, including a constant and a linear-in-$|\omega|$ term, i.e.  $\Sigma_{\xi}^{\prime\prime}(\omega)= \alpha-\varsigma\left|\omega\right|$ with $\alpha \approx \frac{5}{2}\bar{g}^{2}\rho_{0} = \frac{5}{2} D$ and $ \varsigma=\frac{2}{\pi}$. 
This local self-energy $\Sigma_\xi$ comes as a consequence of the local nature of the slave boson in our approach.
The constant $\alpha$ is generated by the fluctuating non-condensed local slave-bosons through self-energy of the $\xi$ field, reminiscent of electrons in metals being scattered by disordered random impurities.

\subsubsection{Quantum critical phase with universal Planckian scattering rate }
Close to the QCP in the scaling regime where conformal symmetry is present, the scattering rate at finite temperatures can be derived from the scattering rate $T = 0$,  $1/\tau(\omega,T = 0)$, by a conformal transformation \cite{georges-multichannel-kondo-PRB,George-NatComm-SM,Georges-2021-PRR-seeback}.  Strikingly, following path III of Fig. \ref{fig:feyn-diag}(a) over the  region where both $\bar{g}(\ell)$ and $J(\ell)$ are irrelevant, we discover a quantum critical Planckian strange metal ``phase" with universal quantum-critical   $\omega/T$-scaling in the scattering rate that is independent of coupling constant (see Supplementary Notes 5):
\begin{align}
    \frac{1}{\tau(x)}-\frac{1}{\tau_0} = \frac{4}{\pi} k_B T  f(x/2), ~ f(x/2)=\frac{x}{2}\coth \left(\frac{x}{4}\right)  
    \label{eq:scatt-rate-scaling}
\end{align}
with $x\equiv \hbar \omega/k_B T$ and $\tau_0 \equiv \tau (\omega=0,T=0)$. Surprisingly, as shown in Fig. \ref{fig:htau-scaling}(a), the universal Planckian scattering rate of Eq. (\ref{eq:scatt-rate-scaling}) is in excellent agreement with the recent optical conductivity measurement in Ref. \cite{George-NatComm-SM} without fine-tuning. In the high-frequency, low-temperature limit $x \gg 1$, the scattering rate divided by $k_B T$ shows a universal scaling behavior, $\frac{\tau^{-1}(T}{k_B T} \approx (2/\pi)  x$. Conversely, in the DC-limit ($x \to 0$), the scattering rate manifests the Planckian scattering rate, revealing a universal feature that is insensitive to microscopic coupling constants: $1/\tau \approx \alpha_P k_B T/\hbar$ with $\alpha_P \approx 8/\pi$, in good agreement with DC-scattering rate estimated in various overdoped cuprates \cite{George-NatComm-SM,Taillefer-planckian-2019}. We find this universal feature of scattering rate that is insensitive to coupling constants in our theory is originated from the cancellation of  the same RG renormalized running coupling constant $g^2 (\ell)$ in $\Sigma_\xi$ and in the denominator of $\mathcal{G}_\xi$ (arising from the second-order hoping process that generates dynamics and energy dispersion of the $\xi$-band). Due to this cancellation, all initial (bare) values of $\bar{g}$ and $J$ which flow to the U(1) FL$^*$ fixed point at $\bar{g}=J=0$ form an extended ``phase" by showing the same Planckian strange-metal behavior [see Eq. (\ref{eq:scatt-rate-scaling})].  Our finding is in excellent agreement with  the doping-independent universal $B/T$ scaling in magnetoresistance observed over the extended strange metal phase of Tl2201  \cite{hussey-incoherent-nature-2021} (see Supplementary Notes 6). 



Meanwhile, the electron mass renormalization  $m^*_e - m^*_e(0 )$ is well captured by  a universal scaling form: $m^*/m -1 = g(x)$ with the scaling function \textit{g(x)} obtaining from the real part of the dynamical self-energy $\Sigma_{\xi}^\prime$ via Kronig-Kramers relation [see Fig. \ref{fig:htau-scaling}(b) and Supplementary Notes 7]. 

\subsubsection{Single-particle spectral function}
\begin{figure*}[ht]
    \centering
    \includegraphics[width=0.9\textwidth]{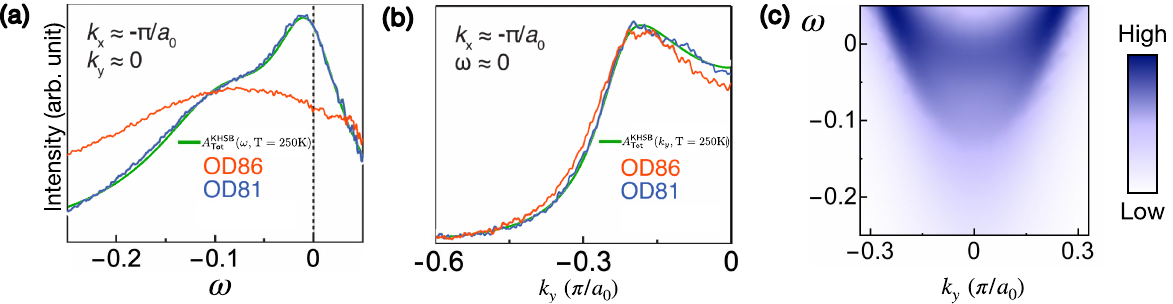}
    \caption{\textbf{Spectral intensity for the SM phase}: (a) Spectral weight as a function of frequency $\omega$ at the anti-nodal point, and (b) spectral weight as a function of momentum $k_y$ at a zero frequency. Experimental data OD86 (red) and OD81 (blue) are reproduced from Ref. \cite{zxshen-sicience-incoherent-cuprate}.  Greens curves in (a) and (b) corresponds to our theoretical fitting to OD81, with temperature being fixed at $T = 250$K. (c) Density plot for the spectral weight as a function of $\omega$ and $k_y$ (see supplementary Notes 7 for details.) }
    \label{fig:spectral}
\end{figure*}

Our theory offers a mechanism for the phenomenological MFL ansatz \cite{varma-mfl-prl}: The complete electron self-energy $\Sigma_\xi$, is reminiscent of the MFL behavior with a distinction that $\Sigma_\xi$ here is insensitive to coupling constant. The resulting single-particle spectral function $A(\omega, \bm{k}, T) \equiv -\pi^{-1}G^R_c (\omega, \bm{k}, T) = -\pi^{-1} \text{Re} \left[ \omega -\varepsilon_{\bm{k}}-\Sigma_c\right]^{-1} $ thus shows an excellent fit to the recent ARPES measurement for overdoped cuprates in the SM region \cite{zxshen-sicience-incoherent-cuprate} (see Fig. \ref{fig:spectral} and Supplementary Notes 7).

\subsubsection{Strange metal state: Signature of quantum critical point vs. quantum critical phase.}
Our RG results offer further insights to the issue on the origin of the Planckian strange metal state.  The RG flow in Fig. \ref{fig:RG-flow-PD}(a) in general may lead to three distinct temperature-doping ($T\,,\delta$) phase diagrams by following the three paths I, II, and III depending on initial values of $t$ and $J$ with increasing doping, see Figs. \ref{fig:RG-flow-PD}(b)-I, \ref{fig:RG-flow-PD}(b)-II, \ref{fig:RG-flow-PD}(b)-III. Experimentally, the three paths can be tuned by magnetic fields: When the external magnetic field is weak, the couplings lie on path I in which a finite range of couplings on this path flow to the superconducting phase. This corresponds to Fig. \ref{fig:RG-flow-PD}(b)-I. At a critical magnetic field, couplings follow path II and pass through the QCP at $C_r$; consequently, the strange metal state at finite $T$ is controlled by a single QCP. The most interesting case, however, is path III, occurring at a larger field which completely suppresses superconductivity. For path III, there is  a finite range of doping where all initial values of $t$ and $J$  [blue solid line in path III of Fig. \ref{fig:RG-flow-PD}(a)] flow to the SM fixed point. In this case,  the Planckian strange metal behaviors at finite T persist over a finite doping range as $T \to 0$, corresponding to the quantum critical strange metal ``phase". Experimental signatures of both quantum critical point \cite{michon-nature-2019-QC-cuprate,Taillefer-annuphys-2019} and quantum critical phase \cite{hussey-incoherent-nature-2021} were reported in different cuprates. These seemingly incoherent results can be coherently unified in our generic RG phase diagram: the former may follow a path close to path II, while the latter follows path III.

\subsubsection{Thermodynamic property: divergence of specific heat coefficient at the QCP}
We further computed the specific heat coefficient in the SM phase near the QCP showing logarithmic-in-$T$ MFL behavior: $\gamma \equiv C_V/T \sim -[A+ B(J)] \ln T$ with $A$ being a constant and $B(J)$ the power-law divergent prefactor arising from the antiferromagnetically short-ranged spin fluctuations (the $J$ term) being quantum critical: $B(J) \sim |J-J_c|^{-p}$ with $p = \nu/2 = 1/2$, in good agreement with the power-law divergent $\gamma$ coefficient $\gamma(T\to0) \sim |\delta-\delta_c|^{-m}$ with $m \sim 0.5$ observed in Eu-LSCO and Nd-LSCO near pseudogap critical point \cite{michon-nature-2019-QC-cuprate,Taillefer-annuphys-2019}. We find the dominating logarithmic-in-$T$ term $-B(J) \ln T$ is originated from the van Hove singularity of the half-filled $f$-spinon band on a 2D square lattice [see the spinon Fermi surface in Fig. \ref{fig:schematics}(b) and Supplementary Notes 4], while the sub-leading constant term $-A \ln T$ is contributed from the self-energy $\Sigma_{c,f}$. 

\subsubsection{Fermi surface reconstruction across the SM region}
Our heavy-fermion two-band formulation of the slave-boson \textit{t-J} model  offers a natural explanation on the enlargement of Fermi surface (FS) volume from a smaller FS with hole density $\delta$ in pseudogap phase to a larger one with hole density $1+\delta$ in the FL phase $(\langle b \rangle =\sqrt{\delta}, ~ \xi_\sigma \sim  \sqrt{\delta} f_\sigma, ~ \sum_\sigma f^\dagger_{i\sigma} f_{i\sigma} = 1-\delta)$ across the strange metal region [Fig. \ref{fig:schematics}(b)] \cite{hussey-incoherent-nature-2021,taillefer-NatLett-changeofcarrier-YBCO-2016,taillefer-PRB-FS-transformation-LSCO}. 

In the pseudogap and SM phases with non-condensed bosons, the  $\xi$-band shows a small hole Fermi surface (hole doping $\delta$). When bosons get condensed across the FL phase boundary, the Fermi surface volume undergoes a sudden jump from $\delta$ to $1+\delta$, in parallel to that occurred at Kondo-breakdown transition in heavy-fermion systems  \cite{Friedemann-PNAS-2010,Shishido-FS-jump}. This is due to the participation of the half-filled spinon band to the Fermi surface by Kondo hybridization between the $f$- and the $\xi$-bands; therefore the hole Fermi surface volume of a normal conduction electron band with a hole density $1+\delta$ (corresponding to a doped hole density $\delta$ relative to half-filling) is recovered, in accordance with the Luttinger theorem [see Fig. \ref{fig:schematics}(b)]. This scenario offers a qualitative understanding of the rapid increase observed in the normal-state Hall number near the critical doping of YBCO and Nd-LSCO \cite{Taillefer-annuphys-2019}. Nevertheless, a smooth crossover of the Fermi surface volume from $\delta$ to $1+\delta$ is possible in the SM phase via a mixture of two fluids: (i) normal metal with $T$-quadratic resistivity by partial boson condensation and a large Fermi surface volume ($1+\delta$), and (ii) strange metal with $T$-linear resistivity without boson condensation and with a small Fermi surface volume ($\delta$), as was observed in Ref. \cite{hussey-incoherent-nature-2021}.


\section{Discussions}
The umklapp scattering is usually expected to be important for electrical resistivity by relaxing electron momentum (charge current). Here, we propose an alternative relaxation mechanism without umklapp scattering via local charge Kondo fluctuations where electrical charge current decays into charge-neutral spinon current and local charge fluctuations despite forward scattering \cite{pepin-KB-PRL}. 
Due to the heavy $\xi$-band, we find that the sub-leading corrections to the scattering rate are negligible compared to the leading Planckian scattering rate mentioned above, within the temperature range of our interest: $1/\tau_{\xi,\rm{gauge}} \sim (T/\chi_d)^{4/3} \varepsilon_{F,\xi} \propto a  (T/D)^{4/3}$  with $a/D \sim O(\delta^2) \ll 1$ and  specific heat coefficient $\gamma(T) \sim (b/D) \cdot (T/D)^{-2/3}$ with $b \sim O(\delta^2) \ll 1$ contributed from U(1) gauge field \cite{Nagaosa-1990-PRL}, see Supplementary Notes 5. Meanwhile, due to the local nature of the slave boson in our approach, the \textit{T}-linear scattering rate contributed from U(1) gauge field found in Ref. \cite{Nagaosa-1990-PRL} is absent here. The Planckain metal phase we find here can be further stablized in the generalized version of our model in large-\textit{N} and multi-channel limit where fluctuations from gauge-field and higher order processes are suppressed. 
The Planckian phase in transport and thermodynamical coefficients within our theoretical framework show distinct qualitative features: the former is insensitive to but the latter strongly depend on the microscopic couplings. This suggests possible breakup of low-energy excitations in this phase into charged particles (electrons) and the charge-neutral spinons; the former dominates in transport while the latter in thermodynamics, reminiscent of the Kondo breakdown QCP in heavy-fermion system \cite{Custers2003Nature}.

To reach the Planckian bound, it was argued in Ref. \cite{Hartnoll-mackenzie-rmp-planckian} that a strong inelastic scattering with local energy relaxation $\Delta E \sim k_B T$ is required so that the equilibration time $\tau_{eq} \sim \hbar/\Delta E \sim \hbar/k_B T$ reaches the Planckian time scale. The charge fluctuating effective Kondo term in our model offers a realization of such strong inelastic scattering channel with a local energy relaxation rate $1/\tau \sim k_B T/\hbar$, where electron loses its energy to local slave boson by coupling to a spinon band. 

It was pointed out that the strong disorder coupled to the electron interactions is crucial to realize the Planckian strange metal state, as shown in the SKY model \cite{Patel-PRL-SM,Patel-2023-SYK-Sci}. Within our two-band heavy-fermion slave-boson $t$-$J$ model, local fluctuating non-condensed slave boson plays the role of disorder embedded in electron interaction in the fluctuating Kondo hybridization. MFL and SM properties are thus  expected, reminiscent of the SKY models. Via the effective Kondo term, the fluctuating slave-boson also generates impurity-like scattering, leading to a residual scattering rate at $T = 0$ due to the breaking of the translational symmetry of the original Hamiltonian.   
A formal link of our model to the SYK-like models deserves further investigations. 

\section{Conclusions}
We provide a microscopic mechanism for the Planckian metal phase in the Kondo-Heisenberg formulation of the slave-boson \textit{t-J} model based on the dynamical charge Kondo breakdown near the localized-to-delocalized phase transition in the form of critical charge (Kondo) fluctuations. The slave-boson $t$-$J$ model was first mapped onto an effective Kondo-Heisenberg model. Via perturbative RG analysis on this effective model,  we realized an extended quantum critical Planckian metal phase. Within the one-loop RG, we identified a quantum critical point separating pseudogap, $d$-wave superconducting, a normal Fermi liquid and a strange metal phases.  In particular, we find a stable strange metal phase in our RG phase diagram close to this critical point exhibiting $T$-linear scattering rate and $\omega/T$-quantum critical scaling and the Planckian behavior where $\alpha_P \sim O(1)$ that is independent of coupling constant, in excellent agreement with the optical conductivity and magnetoresistance measurements in various overdoped cuprates. Our theoretical predictions on the specific heat coefficient, effective mass enhancement, and single particle spectral function in the strange metal state agree well with the experimental observations and hence offers a microscopic mechanism for the marginal Fermi liquid ansatz. Our theoretical framework offer a promising route to reveal the mystery of quantum critical strange metal phase and its link to the generic phase diagram of high-$T_c$ cuprate superconductors.


\section{Methods}
\textit{The slave-boson approach to the $t$-$J$ model.} The model we consider here is the $t$-$J$ model $ H	=H_{t}+H_{J}$,
where 
\begin{align}
    H_{t}& =-t\sum_{\langle i,j\rangle,\sigma}c_{i\sigma}^{\dagger}c_{j\sigma}-\mu\sum_{i\sigma}c_{i\sigma}^{\dagger}c_{i\sigma}, \nn	H_J & = J_{H}\sum_{\langle i,j\rangle}\bm{S}_{i}\cdot\bm{S}_{j}.
\end{align}
 $\bm{S}_{i}=\frac{1}{2}c_{i\sigma}^{\dagger}\bm{\sigma}_{\sigma\sigma^\prime}c_{i\sigma^\prime}$ represents the (dimensionless) spin operator at site $i$, $t$ is the hopping strength, and $\mu$ denotes the chemical potential of the $c$-band. Here, $\langle i,j\rangle$ means only the nearest-neighboring sites are considered. The Heisenberg coupling $J_{H}\sim t^{2}/U>0$. For low dopings, the doubly-occupied state is not energetically favorable. To project out the doubly-occupied state from the Hilbert space, we exploit the slave-boson technique, which splits a conduction electron as a product of a charge-neutral spinon $f_{i\sigma}$ and a spinless charged boson $b_i$, i.e., $c_{i\sigma}^{\dagger}\rightarrow f_{i\sigma}^{\dagger}b_{i}$ subject to the single-occupancy constraint for the $b_i$ and $f_{i\sigma}$ operators on each site \cite{Coleman-SB,Kotliar-PRL-dwave,Punk-SBtJ-PRB}.

 Near half-filling, the doubly-occupied state is not energetically favorable, thus will be ruled out from the Hilbert space. Therefore, a no double-occupancy constraint has to be enforced on every site, i.e. $\sum_{\sigma}c_{i\sigma}^{\dagger}c_{i\sigma}\leq 1$. In terms of the fermion operators, the inner product of two spin operators takes the form of
 \begin{align}
     \bm{S}_{i}\cdot\bm{S}_{j}=&-\frac{1}{4}\sum_{\sigma\sigma^\prime}\left(\tilde{\sigma}c_{i\sigma}^{\dagger}c_{j,-\sigma}^{\dagger}\right)\left(\tilde{\sigma^\prime}c_{j,-\sigma^\prime}c_{i\sigma^\prime}\right) \nn
     &+\frac{1}{4}\sum_{\sigma \sigma^\prime}c_{i\sigma}^{\dagger}c_{j\sigma^\prime}^{\dagger}c_{j\sigma}c_{i\sigma^\prime},
 \end{align}
where $\tilde{\sigma}\equiv \text{sgn}(\sigma)$.


Near half-filling, the doubly-occupied state is not energetically favorable, thus will be ruled out. In this circumstance, we may apply the slave-boson $c_{i\sigma}^{\dagger}\rightarrow f_{i\sigma}^{\dagger}b_{i}$ to project out the doubly occupied state from the Hilbert space, leading to the following single-occupancy constraint for the b and f operators on each site: $ b_{i}^{\dagger}b_{i}+\sum_{\sigma}f_{i\sigma}^{\dagger}f_{i\sigma}=1.$  This constrain can be incorporated into the whole Hamiltonian via applying the Lagrange multiplier, 
\begin{align}
    \sum_{i}\lambda_{i}\left(b_{i}^{\dagger}b_{i}+\sum_{\sigma}f_{i\sigma}^{\dagger}f_{i\sigma}-1\right).    
\end{align}
In the slave-boson representation, the kinetic term $H_{t}$ and the chemical potential term $H_{\mu}$ can be expressed as ($t>0$ being the strength of hopping)
\begin{align}
    H_{t}+H_{\mu}=-t\sum_{\langle i,j\rangle,\sigma}\left(f_{i\sigma}^{\dagger}b_{i}\right)b_{j}^{\dagger}f_{j\sigma}-\mu\sum_{i,\sigma}f_{i\sigma}^{\dagger}f_{i\sigma}.
\end{align}

The Heisenberg term $H_J$ can be expressed in the forms of particle-particle (p-p) and particle-hole (p-h) channels,
\begin{align}
     & \text{p-p} \to \left(-\frac{J_{H}}{2}\right)\sum_{\langle i,j\rangle}\sum_{\alpha\beta\gamma\delta}\left(\varepsilon_{\alpha\beta}f_{i\alpha}^{\dagger}f_{j\beta}^{\dagger}\right)\left(\varepsilon^{\gamma\delta}f_{j\delta}f_{i\gamma}\right)+const., \nn
    & \text{p-h} \to \left(-\frac{J_{H}}{2}\right)\sum_{\langle i,j\rangle}\sum_{\sigma,\sigma^\prime}\left(f_{i\sigma}^{\dagger}f_{j\sigma}\right)\left(f_{j\sigma^\prime}^{\dagger}f_{i\sigma^\prime}\right),
\end{align}
where $\varepsilon=i\sigma_{y}$ is an anti-symmetric tensor. In our mean-field analysis, we consider both of them.

\section{Acknowledgments}
We acknowledge discussions with Cenke Xu, G. Kotliar, A. Georges, J. Schmalian, E. Berg, T. Devereaux, S. A. Kivelson, P. A. Lee, T. K. Lee, P. Coleman, N. Hussey, R. Greene, B. Michon, D. Van Der Marel. This work is supported by the National Science and Technology Council (NSTC) Grants 110-2112-M-A49-018-MY3, the National Center for Theoretical Sciences of Taiwan, Republic of China (C.- H.C.). Y.-Y.C. acknowledges the financial support from The 2023 Postdoctoral Scholar Program of Academia Sinica, Taiwan. C.-H. C. acknowledges the hospitality of Aspen Center for Physics, USA and Kavli Institute for Theoretical Physics, UCSB, USA where part of the work was done. 


%
\newpage
\onecolumngrid

\setcounter{section}{0}
\setcounter{equation}{0}
\setcounter{figure}{0}
\renewcommand\thesection{Supplementary Note \arabic{section}}
\renewcommand{\theequation}{S\arabic{equation}}
\renewcommand{\figurename}{Supplementary Figure}

\section{Hubbard-Stratonovich (HS) transformation, mean-field theory and the effective action for the U(1) FL$^*$ phase}
The Hamiltonian of the Kondo-Heisenberg slave-boson approach to the $t$-$J$ model reads
    \begin{align}
        H=&-t\sum_{\langle i,j\rangle\sigma}\left(f_{i\sigma}^{\dagger}b_{i}\right)b_{j}^{\dagger}f_{j\sigma}-\mu\sum_{i\sigma}f_{i\sigma}^{\dagger}f_{i\sigma}+\sum_{i}\lambda_{i}\left(b_{i}^{\dagger}b_{i}+\sum_{\sigma}f_{i\sigma}^{\dagger}f_{i\sigma}-1\right) \nn
        & -J_{H}\sum_{\langle i,j\rangle}\sum_{\sigma\sigma^\prime}\left(\tilde{\sigma}f_{i\sigma}^{\dagger}f_{j,-\sigma}^{\dagger}\right)\left(\tilde{\sigma^\prime}f_{j,-\sigma^\prime}f_{i\sigma^\prime}\right)-J_{H}\sum_{\langle i,j\rangle}\sum_{\sigma\sigma^\prime}\left(f_{i\sigma}^{\dagger}f_{j\sigma}\right)\left(f_{j\sigma^\prime}^{\dagger}f_{i\sigma^\prime}\right) ,
        \label{eq:H-SB}
    \end{align}
where $\sigma,\sigma^\prime$ refer to the spin-1/2 indices. Following Ref. \cite{Punk-SBtJ-PRB}, we perform the HS transformation on the $H_{t}$ term. To do so, we first define the following operator
\begin{align}
\xi_{ij,\sigma}^{\dagger}=\left(f_{i\sigma}^{\dagger}b_{j}^{\dagger}+f_{j\sigma}^{\dagger}b_{i}^{\dagger}\right)/\sqrt{2},
\label{eq:def-HS-xi_ij}
\end{align}
and re-express the hopping term $H_{t}$ in terms of $\xi_{ij,\sigma}$. The $\xi_{ij,\sigma}$ field is defined on the middle of the bond connecting sites $i$ and $j$ such that
\begin{align}
    \xi_{ij,\sigma}=\xi_{\sigma}\left(\frac{\bm{r}_{i}+\bm{r}_{j}}{2}\right)=\xi_{ji,\sigma}\np
\end{align}
On a two-dimensional square lattice, its Fourier component can be expressed as
\begin{align}
    \xi_{\bm{k}\sigma}^{a}=\frac{1}{\sqrt{N_{s}}}\sum_{i}\xi_{i,i+\hat{e}_{a},\sigma}e^{-i\bm{k}\cdot(\bm{r}_{i}+\hat{e}_{a}/2)},
\end{align}
where $\hat{e}_{a}=\bm{r}_{j}-\bm{r}_{i}$ with $a = x,y$ being the label of the nearest neighbor sites for a 2D square lattice and $N_s$ the number of sites. It is straightforward to show that the $H_{t}$ term can be expressed as a quadratic form in $\xi_{ij,\sigma}$:
\begin{align}
     \sum_{\langle i,j\rangle,\sigma}\xi_{ij,\sigma}^{\dagger}\xi_{ij,\sigma}	=\sum_{\langle i,j\rangle,\sigma}n_{f}^{i}n_{b}^{j}-H_{t}/t \Rightarrow  H_{t}=-t\sum_{\langle i,j\rangle,\sigma}\xi_{ij,\sigma}^{\dagger}\xi_{ij,\sigma}.
\end{align}
The $n_{f}^{i}n_{b}^{j}\sim(1-n_{b}^{i})n_{b}^{j}$ term shown above represents the holon-holon density interaction. In our theory, as we consider the low-doping regime, this interaction can be neglected. Upon performing the HS transformation, the hoping term takes the form of
\begin{align}
    H_{t}	\to  -\frac{t}{\sqrt{2}}\sum_{\langle i,j\rangle,\sigma}\left[\left(f_{i\sigma}^{\dagger}b_{j}^{\dagger}+f_{j\sigma}^{\dagger}b_{i}^{\dagger}\right)\xi_{ij,\sigma}+H.c.\right] 
    +t\sum_{\langle i,j\rangle\sigma}\xi_{ij,\sigma}^{\dagger}\xi_{ij,\sigma},
    \label{eq:Ht-HS-realspace}
\end{align}
which is analogous to the Kondo interaction of a Kondo lattice model. In the momentum space, the hoping term in the form of Eq. (\ref{eq:Ht-HS-realspace}) becomes
\begin{align}
   H_{t} \to \frac{2t}{\sqrt{N_{s}}}\sum_{\bm{k}\bm{k}^{\prime}\sigma a}\left(f_{\bm{k}\sigma}^{\dagger}b_{\bm{k}^{\prime}}^{\dagger} y_{\bm{k}-\bm{k}^\prime}^{a}\xi_{\bm{k}^{\prime}+\bm{k},\sigma}^{a}+H.c.\right)+t\sum_{\bm{k}\sigma a}\xi_{\bm{k}\sigma}^{a\,\dagger}\xi_{\bm{k}\sigma}^{a}
\end{align}
with
\begin{align}
   y_{\bm{k}-\bm{p}}^{a}&\equiv e^{i(\bm{k}-\bm{p})\cdot\hat{e}_{a}/2}.
\end{align}

To decompose the Heisenberg term ($H_J$), we introduce two HS fields, $\chi_{ij}$ and $\Delta_{ij}$, corresponding to the spinon bond field (particle-hole channel) and  spinon pairing field (particle-particle), respectively, 
\begin{align}
    H_{J}	 \to\left(-J_{H}\right)\sum_{\langle i,j\rangle}\left(\sum_{\sigma}\chi_{ij}f_{i\sigma}^{\dagger}f_{j\sigma}+H.c.-\left|\chi_{ij}\right|^{2}\right) +J_{H}\sum_{\langle i,j\rangle}\sum_{\sigma,\sigma^\prime}\left(\Delta_{ij}\varepsilon_{\sigma\sigma^\prime}f_{i\sigma}^{\dagger}f_{j\sigma^\prime}^{\dagger}+H.c.\right)  +J_{H}\sum_{\langle i,j\rangle}\left|\Delta_{ij}\right|^{2}.
\end{align}
In the following analysis, we assume a finite mean-field value of spinon bond field $\left\langle \chi_{ij}\right\rangle =\chi \neq 0$, which is uniformly distributed on the whole square lattice. The spinon pairing field $\Delta_{ij}$ can be introduced via $\Delta_{i,i+\hat{e}_{a}}\to\eta_{a}\Delta+J_{H}\varphi_{i,i+\hat{e}_{a}}$ where $\Delta \equiv \langle \Delta_{ij}\rangle$ denotes the mean-field value of the spinon pairing bond field and $\varphi_{i,i+\hat{e}_{a}}$ represents the fluctuation field of $\Delta$. Both the spinon bond field and the spinon  pairing field are defined on the middle of the links, similar to the $\xi$ field. Here, $\eta_{a}$ is a bond-dependent index which characterizes the underlying symmetry of lattice. We choose the $d$-wave pairing in our theory, corresponding to $\eta_{x}=1=-\eta_{y}$. The full mean-field Hamiltonian of $H_{J}$ reads
\begin{align}
    H_{J}^{MF}	=\sum_{\bm{k},\sigma}h_{\bm{k}}f_{\bm{k}\sigma}^{\dagger}f_{\bm{k}\sigma}+\sum_{\bm{k},\sigma}\left(\Delta\tilde{\sigma}f_{\bm{k}\sigma}^{\dagger}f_{-\bm{k},-\sigma}^{\dagger}\sum_{a}\eta_{a}\cos k_{a}+H.c.\right) + \frac{2N_{s}\left|\chi\right|^{2}}{J_{H}}+\frac{2N_{s}\left|\Delta\right|^{2}}{J_{H}},
\end{align}
where $h_{\bm{k}}=-2J_{H}\chi\left(\cos k_{x}+\cos k_{y}\right)$ denotes the spinon dispersion.  The fluctuating part of $H_{J}$, denoted as $H_{J}^{fluc}$ is
\begin{align}
    H_{J}^{fluc}\to J_{H}\sum_{\langle i,j\rangle,\alpha}\left(\varphi_{i,i+\hat{e}_{a}}\tilde{\alpha}f_{i\alpha}^{\dagger}f_{i+\hat{e}_{a},-\alpha}^{\dagger}+H.c.\right)+J_{H}\sum_{i,a}\left|\varphi_{i,i+\hat{e}_{a}}\right|^{2}.
\end{align}
\added{The Fourier transform of the fluctuating  spinon bond field $\varphi_{\bm{k}}^{a}$ is defined as}
\begin{align}
    \varphi_{\bm{k}}^{a}=\frac{1}{\sqrt{N_{s}}}\sum_{i}\varphi_{i,i+\hat{e}_{a}}e^{-i\bm{k}\cdot(\bm{r}_{i}+\hat{e}_{a}/2)}.
\end{align}

\added{The fluctuating part of $H_{J}$  can thus be expressed as}
\begin{align}
    H_{J}^{fluc}\to\frac{2J_{H}}{\sqrt{N_{s}}}\sum_{\bm{k}\bm{p}}\left[f_{\bm{k}\uparrow}^{\dagger}f_{\bm{p}\downarrow}^{\dagger}\sum_{a}\cos\left(\frac{k_{a}-p_{a}}{2}\right)\varphi_{\bm{k}+\bm{p}}^{a}+H.c.\right]+J_{H}\sum_{\bm{k}a} \left(\varphi_{\bm{k}}^{a}\right)^\dagger \varphi_{\bm{k}}^{a}.
\end{align}

 The $H_{t}$ and $H_{J}$ shown above both contain a nonlinear (momentum-dependent) form factors, $y^{a}_{\bm{k-p}}$ and $\cos \left(\frac{k_a-p_a}{2}\right)$, respectively. To simplify our calculations, we first perform Taylor expansion of these momentum-dependent form factors as a polynomial in momentum, i.e. $f(k) = 1+ f^\prime(k=0) k^2 + (2!)^{-1}f^{\pp}(k=0)k^2 + O(k^3)$. By power-counting the scaling dimensions of these Taylor-expanded coupling constants, $t$ and \textit{J}, we keep only the zeroth-order (constant) terms of them and neglect all the higher-order momentum terms. This approximation is justified since the higher-order momentum-dependent couplings are much more irrelevant than the constant terms under renormalization group (RG) transformation. We therefore use these approximate forms of $H_t$ and $H_J$ in our following analysis. They read 
\begin{align}
    H_{t}	\to \frac{g}{\sqrt{N_{s}}}\sum_{\bm{k}\bm{p}\sigma a}\left(f_{\bm{k}\sigma}^{\dagger}b_{\bm{p}}^{\dagger}\xi_{\bm{k}+\bm{p},\sigma}^{a}+H.c.\right)+\frac{g}{2}\sum_{\bm{k}\sigma a} \left(\xi_{\bm{k}\sigma}^{a}\right)^\dagger \xi_{\bm{k}\sigma}^{a},\nn
	H_{J}^{fluc}	\to \frac{J}{\sqrt{N_{s}}}\sum_{\bm{k}\bm{p}a}\left(\varphi_{\bm{k}+\bm{p}}^{a}f_{\bm{k}\uparrow}^{\dagger}f_{\bm{p}\downarrow}^{\dagger}+H.c.\right)+\frac{J}{2}\sum_{\bm{k}a}\left(\varphi_{\bm{k}}^{a}\right)^\dagger \varphi_{\bm{k}}^{a},
\end{align}
with the newly defined couplings
\begin{align}
    g\equiv2t,\quad J\equiv2J_{H}.
\end{align}
\added{In the following, we will use the simplified $H_{J}$ and $H_{t}$ terms for further evaluations.}

\added{The electron hoping $t$, in general, depends on doping. To extract the doping dependence of hoping, we may rescale the whole Hamiltonian with the half bandwidth of the $f$-spinon}, i.e.  i.e., $H \to H/D$ with $D=4 J_H  \chi \sim 4J_H$. In the absence of boson condensate,  $\langle b \rangle =0$, we expect the hoping parameter $t$ in $H_t$ is significantly suppressed. Below, we estimate the amount of the suppressed hoping term  upon perturbatively taking the expectation value of $H_t$, i.e. $\langle H_t \rangle$. Due to $\langle b \rangle =0$, the expectation value of $H_t$ in the unperturbed limit vanishes, i.e.  $\langle H_t \rangle = 0$; however, the second-order  contribution $\langle H_t^2 \rangle_0$ survives:    
\begin{align}
    \langle H_{t}^2\rangle_{0}&=\frac{2t^{2}}{D^{2}}\sum_{\langle i,j\rangle,\sigma}\sum_{\langle k,l\rangle,\sigma^\prime}\langle\xi_{ij,\sigma}^{\dagger}\left(b_{j}f_{i\sigma}+b_{i}f_{j\sigma}\right)\left(f_{k\sigma^\prime}^{\dagger}b_{l}^{\dagger}+f_{l\sigma^\prime}^{\dagger}b_{k}^{\dagger}\right)\xi_{kl,\sigma^\prime}\rangle_{0} \nn
    = & \frac{2t^{2}}{D^{2}}\sum_{\langle i,j\rangle,\sigma}\sum_{\langle k,l\rangle,\sigma^\prime}\left\langle \xi_{ij,\sigma}^{\dagger}\xi_{kl,\sigma^\prime}\right\rangle \left[\begin{array}{c}
\langle b_{j}b_{l}^{\dagger}\rangle_{0}\langle f_{i\sigma}f_{k\sigma^\prime}^{\dagger}\rangle_{0}+\langle b_{j}b_{k}^{\dagger}\rangle_{0}\langle f_{i\sigma}f_{l\sigma^\prime}^{\dagger}\rangle_{0}\\
+\langle b_{i}b_{l}^{\dagger}\rangle_{0}\langle f_{j\sigma}f_{k\sigma^\prime}^{\dagger}\rangle_{0}+\langle b_{i}b_{k}^{\dagger}\rangle_{0}\langle f_{j\sigma}f_{l\sigma^\prime}^{\dagger}\rangle_{0}
\end{array}\right] \nn
= & \frac{2t^{2}}{D^{2}}\sum_{\langle i,j\rangle,\sigma}\sum_{\langle k,l\rangle,\sigma^\prime}\left\langle \xi_{ij,\sigma}^{\dagger}\xi_{kl,\sigma^\prime}\right\rangle \left[\begin{array}{c}
\delta_{jl}\delta_{ik}\delta_{\sigma\sigma^\prime}\left(1-\langle f_{i\sigma}f_{i\sigma}^{\dagger}\rangle_{0}\right)+\delta_{jk}\delta_{il}\delta_{\sigma\sigma^\prime}\left(1-\langle f_{i\sigma}f_{i\sigma}^{\dagger}\rangle_{0}\right)\\
+\delta_{il}\delta_{jk}\delta_{\sigma\sigma^\prime}\left(1-\langle f_{j\sigma}f_{j\sigma}^{\dagger}\rangle_{0}\right)+\delta_{jl}\delta_{ik}\delta_{\sigma\sigma^\prime}\left(1-\langle f_{j\sigma}f_{j\sigma}^{\dagger}\rangle_{0}\right)
\end{array}\right] \nn
= & \frac{4t^{2}}{D^{2}}\sum_{\langle i,j\rangle,\sigma}\left\langle \xi_{ij,\sigma}^{\dagger}\xi_{ij,\sigma}\right\rangle _{0}\left[\langle f_{i,-\sigma}f_{i,-\sigma}^{\dagger}\rangle_{0}+\langle f_{j,-\sigma}f_{j,-\sigma}^{\dagger}\rangle_{0}\right].
\end{align}

\begin{figure}
    \centering
    \includegraphics[width=0.5\textwidth]{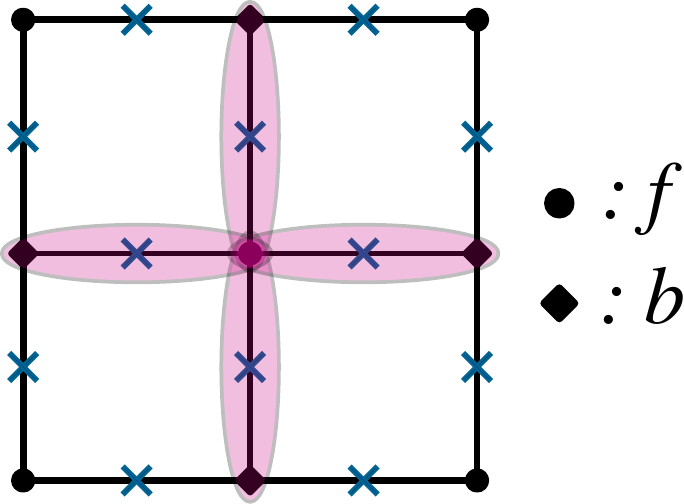}
    \caption{\added{This plot illustrates the schematic of a $\xi$ field (pink eclipse) on a square lattice, with its position (blue cross) being defined in between a spinon field $f$ (black circle) and slave boson $b$ (black diamond). The lattice constant of the unit cell formed by the $\xi$ field is $1/\sqrt{2}$ times of the original square lattice.}}
    \label{fig:xi-bond}
\end{figure}

\added{To proceed $\left\langle H_t^2 \right\rangle_0, \,\left\langle \xi_{ij,\sigma}^\dagger \xi_{ij,\sigma}\right\rangle_0$ is treated in the homogeneous and uniform limit, $ \left\langle \xi_{ij,\sigma}^\dagger \xi_{ij,\sigma}\right\rangle_0 \longrightarrow \overline{\left\langle \xi_{ij,\sigma}^\dagger \xi_{ij,\sigma}\right\rangle_0}$ where  $\overline{\left\langle \xi_{ij,\sigma}^\dagger \xi_{ij,\sigma}\right\rangle_0}$ takes the average value, independent of spins and sites. }
The value of $\overline{\left\langle \xi_{ij,\sigma}^{\dagger}\xi_{ij,\sigma}\right\rangle _{0}}$ can be obtained from the doping constraint. In the uniform limit, the doping constraint over a unit cell (consisting of two $ \xi$-bonds per spin flavor, see Supplementary Figure \ref{fig:xi-bond}) reads
\begin{align}
    \frac{2}{N_{s}}\sum_{\langle i,j\rangle\sigma}  \overline{\left\langle \xi_{ij,\sigma}^{\dagger}\xi_{ij,\sigma}\right\rangle _{0}}=\frac{2}{N_{s}}\times4N_{s}\times  \overline{\left\langle \xi_{ij,\sigma}^{\dagger}\xi_{ij,\sigma}\right\rangle _{0}}=\delta,
\end{align}
leading to $\overline{\left\langle \xi_{ij,\sigma}^{\dagger}\xi_{ij,\sigma}\right\rangle _{0}}=\frac{\delta}{8}$. The above result gives rise to
\begin{align}
    \langle H_{t}^2\rangle_{0}&=\frac{4t^{2}}{D^{2}}\times\frac{\delta}{8}\sum_{\langle i,j\rangle,\sigma}\left[\langle f_{i,-\sigma}f_{i,-\sigma}^{\dagger}\rangle_{0}+\langle f_{j,-\sigma}f_{j,-\sigma}^{\dagger}\rangle_{0}\right].
\end{align}
The original hoping $t$ is thus modified by a prefactor proportional to $\sqrt{\delta}$: 
\begin{align}
    \frac{\langle H_{t}^2\rangle_{0}}{8N_{s}}=\frac{t^{2}\delta}{4D^{2}}\Rightarrow t\to t_{eff}=\frac{t\sqrt{\delta}}{2}.
\end{align}
The suppression of the hoping parameter by the non-condensed slave boson mentioned above is reminiscent of the reduction (by a factor of $\delta$ instead) in hoping term within the renormalized mean-field theory approach, i.e. $t \to t_{eff} = t \times g_t$ with $g_t \sim 2\delta/(1+\delta)^2$ (see, e.g., Ref. \cite{Sun-doping-KM-model}).

The RG-relevant action of slave-boson approach to the \textit{t}-\textit{J} model for the U(1) FL$^*$ phase ($\Delta = 0$ and $\langle b \rangle = 0$ but $\chi \neq 0$) \cite{Punk-SBtJ-PRB} is described as
\begin{align}
    S=	& -\sum_{k\sigma}f_{k\sigma}^{\dagger}\left(i\omega-\varepsilon_{\bm{k}}\right)f_{k\sigma}^{\dagger}-\sum_{k}b_{k}^{\dagger}\left(i\omega-\lambda\right)b_{k} +\frac{g}{2}\sum_{k\sigma a}\left(\xi_{k\sigma}^{a}\right)^{\dagger}\xi_{k\sigma}^{a}  +\frac{J}{2}\sum_{ka}\left(\varphi_{k}^{a}\right)^{\dagger}\varphi_{k}^{a} \nn
	& +\frac{g}{\sqrt{\beta N_{s}}}\sum_{kp\sigma a}\left(f_{k\sigma}^{\dagger}b_{p}^{\dagger}\xi_{k+p,\sigma}^{a}+H.c.\right) +\frac{J}{\sqrt{\beta N_{s}}}\sum_{kpa}\left[f_{k\uparrow}^{\dagger}f_{p\downarrow}^{\dagger}\varphi_{k+p}^{a}+H.c.\right]
	+\rm{const.}.
 \label{eq:action}
\end{align}
Here, the simplified notations $k \equiv (\omega_n, \, \bm{k})$ and $p \equiv (\nu_n, \, \bm{p})$ are used to represent both Matsubara frequency ($\omega_n, \, \nu_n$) and momentum ($\bm{k}, \, \bm{p}$) simultaneously. Based on the action $S$ of Eq. (\ref{eq:action}) above, we define the  bare  (without RPA corrections) Green's functions of various fields as
\begin{align}
    &\mathcal{G}_{f}(k)=\frac{1}{i\omega_{n}-\varepsilon_{\bm{k}}}, \quad
    \mathcal{G}_{b}(k)=\frac{1}{i\omega_{n}-\lambda},\nn
    & \mathcal{G}_{\xi}(k)=\frac{2}{-g},\quad\mathcal{G}_{\varphi}(k)=\frac{2}{-J}.
\end{align}
They are diagrammatically represented in Fig. 2(a) of the main text. Note that, in  \ref{app: dynamics of fields} below, we will use RPA to  perturbatively generate dispersion and dynamics of the $\xi$ field. Thus, the form of its Green's function will change accordingly.

\section{Dynamics of various fields}
\label{app: dynamics of fields}
\added{In this section, we perturbatively estimate the dispersion and dynamics of the slave boson $b$, the spinon-holon bond field $\xi$  and the spinon pairing field $\varphi$.}

At the bare level, the $\xi$ fermion does not have dynamics. We \added{generate its dynamics and band dispersion by }  perturbatively calculating the leading non-trivial self-energy correction, $\Sigma_{\xi}$, via Random Phase Approximation (RPA) approach where the dynamics and dispersion of the $\xi$ fermion are generated by integrating out the higher energy modes in its self-energy \cite{Chang-2018-SM,YYC-SSC-2019}. We first evaluate $\Sigma_{\xi}$: The Green's function of the $\xi$ fermion to the second order in hoping term is given by
\begin{align}
    \mathcal{G}_{\xi}^{(2)}(\bm{k},\omega_{n})	& =-\frac{g^{2}}{\beta}\sum_{p,q,\sigma}\sum_{p^{\prime},q^{\prime},\sigma^{\prime}}\langle\xi_{\bm{k}\sigma}(\omega_{n})\xi_{p+q,\sigma}^{\dagger}b_{q}f_{p\sigma}f_{p^{\prime}\sigma^{\prime}}^{\dagger}b_{q^{\prime}}^{\dagger}\xi_{p^{\prime}+q^{\prime},\sigma^{\prime}}\xi_{\bm{k}\sigma}^{\dagger}(\omega_{n})\rangle_{0} \nn
	& =-\frac{g^{2}}{\beta}\sum_{p,q,\sigma}\sum_{p^{\prime},q^{\prime},\sigma^{\prime}}\left\langle \xi_{\bm{k}\sigma}(\omega_{n})\xi_{p+q,\sigma}^{\dagger}f_{p\sigma}f_{p^{\prime}\sigma^{\prime}}^{\dagger}\xi_{p^{\prime}+q^{\prime},\sigma^{\prime}}\xi_{\bm{k}\sigma}^{\dagger}(\omega_{n})\right\rangle _{0}\left\langle b_{q}b_{q^{\prime}}^{\dagger}\right\rangle _{0} \nn
	& =-\frac{g^{2}}{\beta}\sum_{p}\mathcal{G_{\xi}}^{2}(\bm{k},\omega_{n})\mathcal{G}_{f}(\bm{p},p_{n})\mathcal{G}_{b}(\bm{k-p},\omega_{n}-q_{n}).
\end{align}
Thus, via the Dyson equation, $G_{\xi}^{-1}=\mathcal{G}_{\xi}^{-1}-\Sigma_{\xi}$, the full $G_{\xi}$ up to the second order approximation, $  G_{\xi}(\bm{k},\omega_{n})	 \approx\mathcal{G}_{\xi}(\bm{k},\omega_{n})+\mathcal{G}_{\xi}^{(2)}(\bm{k},\omega_{n}) $, leads to
	\begin{align}
	    G_{\xi}^{-1}(\bm{k},\omega_{n})=\mathcal{G}_{\xi}^{-1}(\bm{k},\omega_{n})- \Sigma_{\xi}(\bm{k},i\omega_{n}),
	\end{align}
    where we obtain 
    \begin{align}
        \Sigma_{\xi}(\bm{k},\omega_{n})=-\frac{g^{2}}{\beta}\sum_{\bm{p},p_{n}}\mathcal{G}_{f}(\bm{p},p_{n})\mathcal{G}_{b}(\bm{k}-\bm{p},\omega_{n}-p_{n}).
    \end{align}
To acquire the dispersion of the $\xi$ fermion, we only focus on real part of $\Sigma_{\xi}$. We integrate out the higher energy modes of the conduction electron, leading to 
\begin{align}
    \Sigma_{\xi}(\bm{k},\omega_{n})	& =-\frac{g^{2}}{\beta}\sum_{\bm{p},p_{n}}\mathcal{G}_{f}(\bm{p},p_{n})\mathcal{G}_{b}(\bm{k}-\bm{p},\omega_{n}-p_{n}) \nn
	& =\sum_{\bm{p}}\left(\frac{g^{2}}{i\omega_{n}-\lambda-\varepsilon_{\bm{p}}}\right)+\sum_{\bm{p}}^{\sim}\left(\frac{g^{2}}{-i\omega_{n}+\varepsilon_{\bm{p}}+\lambda}\right),
\end{align}
where, in the above equation, $\tilde{\sum}_{\bm{p}}$ means summing over the relatively higher energy (quasi-momenta) states whose energy levels are slightly below the Fermi surface.

Letting $i\omega_{n}\to\omega+i\epsilon$ with $\epsilon>0$ being an infinitesimal number. The real part of first term is
\begin{align}
    \sum_{\bm{p}}\left(\frac{g^{2}}{\omega-\lambda-\varepsilon_{\bm{p}}}\right)= & -g^{2}\rho_{0}\left(\int_{-D}^{-D+\varepsilon_{\bm{k}}}+\int_{D-\varepsilon_{\bm{k}}}^{D}\right)\frac{d\varepsilon}{-\omega+\lambda+\varepsilon} \nn
	 \approx & \left(-g^{2}\rho_{0}\right)\left[\ln\left(1+\frac{\omega-\lambda-\varepsilon_{\bm{k}}}{D}\right)-\ln\left(1+\frac{-\omega+\lambda-\varepsilon_{\bm{k}}}{D}\right)\right] \nn
	 \approx &  \frac{-2g^{2}\rho_{0}}{D}\left(\omega-\lambda\right),
\end{align}
and the real part of the second term is given by
\begin{align}
     \sum_{\bm{p}}^{\sim}\left(\frac{g^{2}}{-\omega+\varepsilon_{\bm{p}}+\lambda}\right) 	
    = & g^{2}\rho_{0}\int_{-D}^{-D+\varepsilon_{\bm{k}}}\frac{d\varepsilon}{-\omega+\varepsilon+\lambda} \nn
    =& g^{2}\rho_{0}\ln\left(\frac{-\omega+\lambda-D+\varepsilon_{\bm{k}}}{-\omega+\lambda-D}\right) \nn
	 \approx & \frac{g^{2}\rho_{0}}{D}\left(\omega-\lambda-\varepsilon_{\bm{k}}\right).
\end{align}
The real part of $\Sigma_{\xi}(\omega+i\epsilon,\bm{k})$  then reads
\begin{align}
    \Sigma_{\xi}(\omega+i\epsilon,\bm{k})\approx-\frac{g^{2}\rho_{0}}{D}\varepsilon_{\bm{k}}-\frac{g^{2}\rho_{0}}{D}\left(\omega-\lambda\right).
\end{align}

Next, we evaluate the imaginary part of $\Sigma_{\xi}$, denoted as $\Sigma_{\xi}^{\prime\prime}$. Using $(x\pm i\epsilon)^{-1}=\mathcal{P}(\frac{1}{x})\mp i\pi\delta(x)$ and assuming $\left|\omega-\lambda\right|\leq D$, we have 
\begin{align}
     \Sigma_{\xi}^{\prime\prime}	(\omega\pm i\epsilon,\bm{k})   & =   \sum_{\bm{p}}\left(\frac{g^{2}}{\omega-\lambda-\varepsilon_{\bm{p}}\pm i\epsilon}\right)^{\prime\prime}+\sum_{\bm{p}}^{\sim}\left(\frac{g^{2}}{-\omega+\varepsilon_{\bm{p}}+\lambda\pm i\epsilon}\right)^{\prime\prime} \nn
	& =\left(\mp\pi g^{2}\rho_{0}\right)\int_{-D}^{D}d\varepsilon\delta\left(\omega-\lambda-\varepsilon\right)\pm\pi g^{2}\rho_{0}\int_{-D}^{0}\delta\left(-\omega+\varepsilon+\lambda\right) \nn
	& =\begin{cases}
\mp\pi g^{2}\rho_{0} & \text{if \ensuremath{\lambda\leq\omega}\ensuremath{\leq D+\lambda}}\\
0 & \text{otherwise}
\end{cases} 
\end{align}

Since $\lambda$ is positive, it implies that $\omega$ should also be positive. In the Matsubara frequency space, we have $\Sigma_{\xi}(\bm{k},i\omega_{n})$
\begin{align}
     \Sigma_{\xi}(\bm{k},\omega_{n}) 	
    = &   \sum_{\bm{p}}\left(\frac{g^{2}}{i\omega_{n}-\lambda-\varepsilon_{\bm{p}}}\right)+\sum_{\bm{p}}^{\sim}\left(\frac{g^{2}}{-i\omega_{n}+\varepsilon_{\bm{p}}+\lambda}\right) \nn
	 = & g^{2}\rho_{0}\int_{-D}^{D}\frac{d\varepsilon}{i\omega_{n}-\lambda-\varepsilon}+g^{2}\rho_{0}\int_{-D}^{0}\frac{d\varepsilon}{-i\omega_{n}+\varepsilon+\lambda} \nn 
	= & g^{2}\rho_{0}\ln\left(\frac{i\omega_{n}-\lambda+D}{i\omega_{n}-\lambda-D}\right)+g^{2}\rho_{0}\ln\left(\frac{i\omega_{n}-\lambda}{i\omega_{n}-\lambda+D}\right) \nn
	 = & g^{2}\rho_{0}\ln\left(\frac{i\omega_{n}+D}{i\omega_{n}-D}\right)+g^{2}\rho_{0}\ln\left(\frac{i\omega_{n}-\lambda}{i\omega_{n}+D}\right) \nn
	 = & -i\pi g^{2}\rho_{0}\text{sgn}(\omega_{n})+\frac{1}{2}g^{2}\rho_{0}\ln\left(\frac{\omega_{n}^{2}+\lambda^{2}}{\omega_{n}^{2}+D^{2}}\right) \nn
	 & \qquad +ig^{2}\rho_{0}\left[\theta_{1}(\omega_{n})-\theta_{2}(\omega_{n})\right],
	\end{align}
	where 
	\begin{align}
	    \theta_{1}(\omega_{n})	=\tan^{-1}\left(\frac{\omega_{n}}{-\lambda}\right), \,\,
\theta_{2}(\omega_{n})	=\tan^{-1}\left(\frac{\omega_{n}}{D}\right).
	\end{align}
Thus, we obtain the imaginary part of $\Sigma_{\xi}$ 
\begin{align}
    \Sigma_{\xi}^{\prime\prime}(\bm{k},\omega_{n})	
     =& -\pi g^{2}\rho_{0}\text{sgn}(\omega_{n})+g^{2}\rho_{0}\left[\tan^{-1}\left(\frac{\omega_{n}}{-\lambda}\right)-\tan^{-1}\left(\frac{\omega_{n}}{D}\right)\right] \nn
	 =& -\pi g^{2}\rho_{0}\text{sgn}(\omega_{n})-g^{2}\rho_{0}\left[\frac{\omega_{n}}{\lambda}+\frac{\omega_{n}}{D}\right] \nn
	 \approx & -\pi g^{2}\rho_{0}\text{sgn}(\omega_{n}) \; ,
\end{align}
where, in the last line of the above equation, we have assumed $\lambda, \, D\gg \omega_n$. Combing $\Sigma_{\xi}^{\prime}$ and $\Sigma_{\xi}^{\prime\prime}$, the dressed retarded Green's function of the $\xi$ field up to the leading non-trivial self-energy correction is given by 
\begin{align}
    G_{\xi}^{R}(\omega+i\epsilon,\bm{k}) & 	=\frac{1}{-t-\Sigma_{\xi}^{\prime}(\omega+i\epsilon,\bm{k})-i\Sigma_{\xi}^{\prime\prime}(\omega+i\epsilon,\bm{k})} \nn
	& \equiv\frac{1}{\omega_{\xi}-\xi_{\bm{k}}+i\Gamma_{\xi}},\quad\text{with }\text{\ensuremath{\lambda}\ensuremath{\leq\omega}\ensuremath{\leq D},}
\end{align}
where the quasi-particle width $\Gamma_{\xi}$, the generated  frequency $\omega_{\xi}$, and the dispersion $\xi_{\bm{k}}$ are given by 
\begin{align}
    &  \Gamma_{\xi}\equiv\pi g^{2}\rho_{0}, \nn
    &  \omega_{\xi}\equiv\frac{g^{2}\rho_{0}}{D}\omega\equiv\zeta\omega, \nn
    & \xi_{\bm{k}}= -\zeta\varepsilon_{\bm{k}}+\zeta\lambda+t
\end{align}
Here, $\zeta\equiv\frac{g^{2}\rho_{0}}{D}$, $\rho_{0}\sim1/D$ and $\varepsilon_{\bm{k}} = h_{\bm{k}} -\mu_f \approx v |\bm{k}| $ with $\mu_f \equiv \mu - \lambda $ being the effective chemical potential for $f$ spinon. Here, the $f$-spinon band is approximated by a linear-in-momentum dispersion near the effective chemical potential $\mu_f$. As we treat $g$ as a perturbation, we expect $\zeta=(g/D)^{2}\ll1$. In addition, as the $\xi$ fermion is dispersionless at the bare level, its bandwidth of the renormalized band structure falls in the range of  $-\zeta D \leq \zeta \varepsilon_{\bm{k}}\leq\zeta D$ for $\bm{k}$ to be satisfactory with $-D\leq\varepsilon_{\bm{k}}\leq D$. The value of $\zeta D=g^{2}/D\sim t\zeta\ll t$. This leads to $\xi_{\bm{k}}>0$.

We now generate and estimate  the dispersion of the slave boson $b$. At the bare level, the slave boson $b$ shows  only flat band with an atomic energy level $\lambda$. Here, we estimate the dispersion of the slave boson by computing its effective hoping through second-order perturbation in $H_t$. The effective hoping of slave boson can be obtained by contracting the $\xi$ and $f$ fermions of the following 
\begin{align}
    & g^2 \sum_{ i,j,\sigma}\sum_{ i^\prime,j^\prime,\sigma^\prime} \left\langle  f_{i\sigma}^{\dagger} \xi_{ij,\sigma}\xi^\dagger_{i^\prime j^\prime,\sigma^\prime}f_{i^\prime \sigma^\prime} \right\rangle b_{j}^{\dagger}b_{j^\prime} 
    \sim  g^2 \sum_{ i,j,\sigma}\sum_{ i^\prime,j^\prime,\sigma^\prime} \left\langle  f_{i\sigma}^{\dagger} f_{i^\prime \sigma^\prime} \right\rangle \left\langle \xi_{ij,\sigma}\xi^\dagger_{i^\prime j^\prime,\sigma^\prime} \right\rangle b_{j}^{\dagger}b_{j^\prime}.
\end{align}
\added{Taking $\left\langle  f_{i\sigma}^{\dagger} f_{i^\prime \sigma^\prime} \right\rangle \sim \chi $ and $\left\langle \xi_{ij,\sigma}\xi^\dagger_{i^\prime j^\prime,\sigma^\prime} \right\rangle  \sim g^2/D^2 $, the dispersion of the slave boson field thus take the form $\varepsilon_b (\bm{k}) = \lambda + t_b (\cos k_x +\cos k_y)$ with the effective hoping  $t_b \propto g^4 \chi = t^4 \delta^2 \chi$ (in unit of \textit{D}).  For small hole-doping $\delta$ and a weakly perturbative coupling $t$ where $t/D \approx 3/4$ with $D \approx 4J_H$, $t_b$ can be estimated as $t_b \sim t^4 \delta^2 \sim (3/4)^4 \delta^2 $ (in unit of \textit{D}). We can fix the value of $\lambda$ such that $D>\lambda \gg |t_b|$, and therefore the effecitve hoping $t_b$ term of the generated slave boson band is negligible.}


We now evaluate the leading-order self-energy correction to generate dynamics of the $\varphi$ field. The self-energy $\Sigma_{\varphi}$ proportional to $J^{2}$ is given by
\begin{align}
    \Sigma_{\varphi}\left(\bm{k},\omega_{n}\right)	 =& \left(-\frac{J^{2}}{\beta}\right)\sum_{q_{n},\bm{q}}\mathcal{G}_{f}\left(\bm{k}-\bm{q},\omega_{n}-q_{n}\right)\mathcal{G}_{f}\left(\bm{q},q_{n}\right) \nn
	= & \left(-\frac{J^{2}}{\beta}\right)\sum_{q_{n},\bm{q}}\frac{1}{iq_{n}-\varepsilon_{\bm{q}}}\frac{1}{i\omega_{n}-iq_{n}-\varepsilon_{\bm{k-q}}} \nn
	= & J^{2}\sum_{\bm{q}}\left[\frac{n_{F}(\varepsilon_{\bm{q}})}{-i\omega_{n}+\varepsilon_{\bm{q}}+\varepsilon_{\bm{k-q}}}+\frac{n_{F}(i\omega_{n}-\varepsilon_{\bm{k-q}})}{i\omega_{n}-\varepsilon_{\bm{q}}-\varepsilon_{\bm{k-q}}}\right] \nn
	= & J^{2}\sum_{\bm{q}}\frac{1-n_{F}(\varepsilon_{\bm{k}-\bm{q}})-n_{F}(\varepsilon_{\bm{q}})}{i\omega_{n}-\varepsilon_{\bm{q}}-\varepsilon_{\bm{k-q}}}.
\end{align}
By setting the external momentum $\bm{k}=0$, we have
\begin{align}
    \Sigma_{\varphi}\left(\omega_{n}\right)	=J^{2}\sum_{\bm{q}}\frac{1-2n_{F}(\varepsilon_{\bm{q}})}{i\omega_{n}-2\varepsilon_{\bm{q}}}=J^{2}\sum_{\bm{p}}\frac{1}{i\omega_{n}-2\varepsilon_{\bm{q}}}-2J^{2}\sum_{\bm{p}}^{\sim}\frac{1}{i\omega_{n}-2\varepsilon_{\bm{q}}}.
\end{align}
In the real frequency representation, the real part of $\Sigma_{\varphi}$ is
\begin{align}
    \Sigma_{\varphi}^{\prime}\left(\bm{k},\,\omega + i\epsilon\right)	=& J^{2}\rho_{0}\left(\int_{D-\varepsilon_{\bm{k}}}^{D}+\int_{-D}^{-D+\varepsilon_{\bm{k}}}\right)\frac{d\varepsilon}{\omega-2\varepsilon}-2J^{2}\rho_{0}\int_{-D}^{-D+\varepsilon_{\bm{k}}}\frac{d\varepsilon}{\omega-2\varepsilon} \nn
	= & \left(-\frac{J^{2}\rho_{0}}{2}\right)\left(\int_{2D-2\varepsilon_{\bm{k}}}^{2D}+\int_{-2D}^{-2D+2\varepsilon_{\bm{k}}}\right)\frac{du}{u-\omega}+\left(J^{2}\rho_{0}\right)\int_{-2D}^{-2D+2\varepsilon_{\bm{k}}}\frac{du}{u-\omega} \nn
	= &\left(-\frac{J^{2}\rho_{0}}{2}\right)\ln\left|\frac{-\omega+2D}{-\omega+2D-2\varepsilon_{\bm{k}}}\right|+\left(-\frac{J^{2}\rho_{0}}{2}\right)\ln\left|\frac{\omega+2D-2\varepsilon_{\bm{k}}}{\omega+2D}\right|
	+J^{2}\rho_{0}\ln\left|\frac{\omega+2D-2\varepsilon_{\bm{k}}}{\omega+2D}\right| \nn	
	= & \frac{J^{2}\rho_{0}}{2}\ln\left|\frac{-\omega+2D-2\varepsilon_{\bm{k}}}{-\omega+2D}\right|+\frac{J^{2}\rho_{0}}{2}\ln\left|\frac{\omega+2D-2\varepsilon_{\bm{k}}}{\omega+2D}\right| \nn
	= & \frac{J^{2}\rho_{0}}{2}\ln\left[\frac{\left(2D-2\varepsilon_{\bm{k}}\right)^{2}-\omega^{2}}{(2D)^{2}-\omega^{2}}\right],
\end{align}
where the logarithmic term can be approximated as
\begin{align}
    \ln\frac{\left(2D-2\varepsilon_{\bm{k}}\right)^{2}-\omega^{2}}{(2D)^{2}-\omega^{2}}	\approx & \ln\frac{\left(2D-2\varepsilon_{\bm{k}}\right)^{2}-\omega^{2}}{(2D)^{2}} \nn
	\approx &  \ln\left[\frac{1}{\left(2D\right)^{2}}\left[(2D)^{2}-8D\varepsilon_{\bm{k}}-\omega^{2}\right]\right] \nn
	= & \ln\left[1-\frac{2\varepsilon_{\bm{k}}}{D}-\frac{\omega^{2}}{4D^{2}}\right] \nn
	= & -\frac{2\varepsilon_{\bm{k}}}{D}-\frac{\omega^{2}}{4D^{2}}.
\end{align}
The imaginary part of $\Sigma_{\varphi}$ vanishes in the wide-band and low-energy (frequency) limit, $D \gg \omega$:  
\begin{align}
\Sigma_{\varphi}^{\prime\prime}\left(\bm{k},\omega+i\epsilon\right)	=-\pi J^{2}\rho_{0}\left(\int_{D-\varepsilon_{\mathbf{k}}}^{D}+\int_{-D}^{-D+\varepsilon_{\mathbf{k}}}\right)d\varepsilon\delta(\omega-2\varepsilon)+2\pi J^{2}\rho_{0}\int_{-D}^{-D+\varepsilon_{\mathbf{k}}}d\varepsilon\delta(\omega-2\varepsilon)
	= 0.
\end{align}

Combining $\Sigma_{\varphi}^{\prime}$ and $\Sigma_{\varphi}^{\prime\prime}$, we have 
\begin{align}
    \Sigma_{\varphi}=\left(-\frac{J^{2}\rho_{0}}{2}\right)\left(\frac{2\varepsilon_{\bm{k}}}{D}+\frac{\omega^{2}}{4D^{2}}\right)\equiv-\varepsilon_{\varphi}(\bm{k})-\frac{\omega_{\varphi}^{2}}{D}.
\end{align}
with
 \begin{align}
     \omega_{\varphi}=\omega\sqrt{\frac{\kappa}{8}}\quad,\quad\varepsilon_{\varphi}(\bm{k})=\kappa\varepsilon_{\bm{k}}\quad,\text{with }\kappa\equiv\frac{J^{2}\rho_{0}}{D}=\frac{J^{2}}{D^{2}}.
 \end{align}
The renormalized Green's function $G_{\varphi}$ is given by
\begin{align}
    G_{\varphi}^{-1}=\mathcal{G}_{\varphi}^{-1}-\Sigma_{\varphi}=-J+\varepsilon_{\varphi}(\bm{k})+\frac{\omega_{\varphi}^{2}}{D}\approx-J+\frac{\omega_{\varphi}^{2}}{D}
\end{align}
Since $J/t \ll 1$,  the generated frequency and dispersion which are of the order of $J^2/D^2$ can be thus neglected.

\section{Renormalization group (RG) analysis}
\label{app:RG}
In this section, we provide details of the RG analysis of our slave-boson approach to the Kondo-Heisenberg-like \textit{t-J} model. The  RG procedure is primarily based on Ref. \cite{Yamamoto-RG-PRB, qimiao-prb-local-fluc,lijun-prb-loca-fluc-bose-fermi} .

\subsection{Scaling dimension}
The RG procedure starts from the scaling of the wave vector $\bm{k}$ and the frequency $\omega$, where 
\begin{align}
    k_i^{\prime}=e^{\ell}k_i,\quad\omega^{\prime}=e^{z \ell}\omega,
\end{align}
where $\ell$ is called the scaling factor, $i$ above denotes the component of $\bm{k}$, and $z$ represents the dynamical exponent. The bare scaling dimension for wave vector $k_i$ is chosen to be $[k_i]=1$. To preserve the Luttinger's theorem, the Fermi velocity has to be kept scale invariant. Since the dispersion of the $f$ spinon band is linear-in-momentum, we thus have
\begin{align}
    z=[\omega]=[\varepsilon_{\bm{k}}]=1.
\end{align}
The scaling dimension for the measure $d^{d}k$ for fermions with scale-invariant Fermi velocity is given by \cite{Yamamoto-RG-PRB}
\begin{align}
    [d^{d}k]=1.
\end{align}

The scale invariance for the action of the kinetic term of the $f$ spinon, $S_{0}^{f}$, implies that the bare scaling dimension of the  $f$ spinon operator is
\begin{align}
    1+1+1+2[f_{k\sigma}]=0\to[f_{k\sigma}]=-\frac{3}{2},
\end{align}
where $k \equiv (\omega,\, \bm{k})$. Since the bare slave boson $b_k$ is local, its scaling dimension can be thus obtained as
\begin{align}
[b_k]=-\left(\frac{d+z}{2}\right).
\end{align}
The non-interacting part of the composite fermion field $\xi^a_{k\sigma}$ is given by
\begin{align}
    S_{0}^{\xi}\sim\sum_{\sigma a}\int d\omega d^d k \left(\omega_{\xi}-\xi_{\bm{k}}\right)\left( \xi^a_{k\sigma}\right)^{\dagger}\xi^a_{k\sigma} \nc
\end{align}
where the renormalized frequency $\omega_{\xi}=\frac{g^{2}\rho_{0}}{D}\omega$ and renormalized dispersion $\xi_{\bm{k}}=-\frac{g^{2}\rho_{0}}{D}\varepsilon_{\bm{k}}+\text{const.}$. While performing the scaling procedure for $\xi_{\bm{k}}$ and $\omega_{\xi}$, we only scale $\varepsilon_{\bm{k}}$ and $\omega$. The prefactor $\frac{g^{2}\rho_{0}}{D}$ is not involved in the  scaling. This is equivalent to fixing $\frac{g^{2}\rho_{0}}{D}$ at its fixed point value, i.e. $\frac{g^{2}\rho_{0}}{D}\to \frac{(g^*)^{2}\rho_{0}}{D} $. We thus have the bare scaling dimension of the $\xi_{k\sigma}$, 
\begin{align}
    [\xi^a_{k\sigma}]=-\frac{3}{2} \np
\end{align}
In our theory, since the dispersion for the RVB pairing bond field $\varphi_{k}$ by perturbative correction is sub-leading,  we approximate the $\varphi^a_{k}$ field to be local. Hence, its bare scaling dimension is identical to that of the slave boson $b$, namely, 
\begin{align}
    [\varphi^a_k] = -\left(\frac{d+z}{2}\right).
\end{align}
The scaling dimension for various couplings can be computed in a similar approach. The action for $H_{t}$, denoted as $S_{t}$, takes the following form, 
\begin{align}
    S_{t}\sim \bar{g}\int d\omega d\nu\int d^{d}kd^{d}p\sum_{a\sigma}\left[f_{k\sigma}^{\dagger}b_{p}^{\dagger}\bar{\xi}_{k+p,\sigma}^{a}+H.c.\right].
    \label{eq:S_t-action}
\end{align}
with $k=(\omega,\bm{k})$ and $p=(\nu,\bm{p})$. From Ref. \cite{Yamamoto-RG-PRB}, for a special case $z=1$, the calculation of the scaling dimension for a boson-fermion coupling term can be performed via the usual way without invoking the patch scheme. The scaling dimension of $\bar{g}$ can be obtained similarly by simple power-counting, $ 0=[\bar{g}]+2z+1+d+[f_{k\sigma}]+[\xi_{k\sigma}]+[b_k]$. Consequently, we obtain
 \begin{align}
    [\bar{g}]	=\frac{4-d-3z}{2}=-\frac{1}{2} \np
 \end{align}
Note that, in Eq. (\ref{eq:S_t-action}), $\bm{k}$ denote the wave vector $\bm{k}$ for the $f$ spinon while $\bm{p}$ for the slave boson. Therefore, by Ref. \cite{Yamamoto-RG-PRB}, we take $[d^dk] = 1$ and $[d^d p] = d$ in the above derivation for $[\bar{g}]	$. Following a similar approach, the bare scaling dimension of the Heisenberg coupling is given by
\begin{align}
    [J]=\frac{4-d-3z}{2}=-\frac{1}{2} \np
\end{align}

\subsection{Self-energy corrections}
The self-energy correction to the bare Green's function of a given field begins with the following relation between the dressed (renormalized) Green's function $G$ of that field: 
\begin{align}
    G=\mathbb{G}\mathcal{G} \nc
\end{align}
with $\mathcal{G}$ being the bare Green's function and $\mathbb{G}$ being the corrections to $\mathcal{G}$ by perturbation expansion. From the Dyson equation, we can also show that $G^{-1}=\mathcal{G}^{-1}-\Sigma=\mathcal{G}^{-1}\left(1-\mathcal{G}\Sigma\right)$ with $\Sigma$ here being denoted as the self-energy correction, implying that $ \mathbb{G}=\frac{1}{1-\mathcal{G}\Sigma}$. For weak coupling theory, we can expand $\mathbb{G}$ as a power series, 
\begin{align}
    \mathbb{G}\approx1+\mathcal{G}\Sigma+O(\mathcal{G}\Sigma)^{2}.
\end{align}
Note that the self-energy correction $\Sigma_{\xi}^{g}$ for the $\xi$ field in the RG analysis requires an additional negative sign compared to $\Sigma_{\xi}^{g}$ from RPA. This difference arises from the minus sign  appearing in the $k$-dependent term in the dispersion $\xi_{\bm{k}}$ for the $\xi$ field computed from RPA. As a result, $\Sigma_{\xi}^{g}$ is changed to
\begin{align}
    \Sigma_{\xi}^{g} (\omega_n)&=\left(\frac{\bar{g}^{2}}{N_{s}\beta}\right)\sum_{\bm{p},p_{m}}\frac{1}{ip_{m}-\varepsilon_{\bm{p}}}\cdot\frac{1}{i\omega_{n}-ip_{m}-\lambda} \nn
    & = \left(-\frac{\bar{g}^{2}}{N_{s}}\right)\sum_{\bm{p}}\left[\frac{n_{F}(\varepsilon_{\bm{p}})}{-i\omega_{n}+\varepsilon_{\bm{p}}+\lambda}+\frac{n_{F}(i\omega_{n}-\lambda)}{i\omega_{n}-\lambda-\varepsilon_{\bm{p}}}\right].
    \label{eq:Sigma_xi_RG}
\end{align}
Taking $-i\omega_{n}=-\omega-\lambda$, we have
\begin{align}
    \Sigma_{\xi}^{g} (\omega)  = \left(-\bar{g}^{2}\rho_{0}\right)\int_{-D}^{0}\left(\frac{d\varepsilon}{-\omega+\varepsilon}\right)+\text{others}  = \bar{g}^{2}\rho_{0}\ln\frac{D}{\omega}+\text{others} \nc
\end{align}
 where ``others'' in the above equation represents the terms with no logarithmic correction.

 The self-energy of the $f$ spinon, $\Sigma_{f}^{g}$,  contributed from perturbation of $H_t$ is given by
\begin{align}
    \Sigma_{f}^{g} (\bm{k}, \omega_n) &=\left(-\frac{\bar{g}^{2}}{N_{s}\beta}\right) \sum_{p}\overline{\mathcal{G}}_{\xi}\left(k+p\right)\mathcal{G}_{b}\left( p \right) \nn
    & =\left(-\frac{\bar{g}^{2}}{N_{s}\beta}\right)\sum_{p_{m},\bm{p}}\frac{1}{i\omega_{n}+ip_{m}-\zeta^{-1}\xi_{\bm{k}+\bm{p}}}\cdot\frac{1}{ip_{m}-\lambda} \nn
    & = \frac{\bar{g}^{2}}{N_{s}}\sum_{\bm{p}}\frac{n_{F}(\zeta^{-1}\xi_{\bm{p}})}{i\omega_{n}+\varepsilon_{\bm{p}}+\zeta^{-1}\varUpsilon+\lambda},
\end{align}
where $\varUpsilon\equiv-\zeta\lambda-t+\mu_{\xi}$. After substituting $i\omega_{n} + \lambda \to \omega$, we obtain
\begin{align}
    \Sigma_{f}^{g} (\omega) = & \bar{g}^{2}\rho_{0}\int_{-\zeta^{-1}\varUpsilon}^{D}\frac{d\varepsilon}{\omega+\varepsilon} \nn
     = & \bar{g}^{2}\rho_{0}\ln\left(\frac{D}{\omega-\zeta^{-1}\varUpsilon}\right).
\end{align}
From the result in RPA section, we know that the self-energy of the slave boson contributes no logarithmic divergence, $\Sigma^g_b \propto (\ln D)^0$. 

Now, we evaluate the self-energy correction of various fields from the perturbation of the Heisenberg term, $H_J$. The self-energy of the \textit{f} spinon is given by
\begin{align}
    \Sigma_{f}^{J}(\bm{k},\omega_{n})=\frac{J^{2}}{N_{s}\beta}\sum_{q_{m},\bm{q}}\mathcal{G}_{f,\downarrow\downarrow}(\bm{q},q_{m})\mathcal{\bar{G}}_{\varphi}(\bm{k+q},\omega_{n}+q_{m}).
\end{align}
Using the $\varphi$ propagator 
\begin{align}
    \bar{\mathcal{G}}_{\varphi}(\omega_{n},\bm{k})=-1,
\end{align}
$\Sigma_{f}^{J}$ reads
\begin{align}
    \Sigma_{f}^{J}(\bm{k},\omega_{n})	& =\left(-\frac{J^{2}}{N_{s}\beta}\right)\sum_{q_{m},\bm{q}}\frac{1}{ip_{m}-\varepsilon_{\bm{q}}}
	=\left(-J^{2}\rho_{0}\right)\int_{-D}^{D}d\varepsilon n_{F}(\varepsilon) \nn
 & =-J^{2}\rho_{0}\int_{-D}^{0}d\varepsilon
	=-J^{2}\rho_{0}D.
\end{align}
We find no logarithmic correction from $\Sigma_{f}^{J}$, i.e. $\Sigma_{f}^{J}  \sim\left(\ln D\right)^{0}$. Additionally, no imaginary part for $\Sigma_{f}^{J}$ exists.

We now evaluate the self-energy correction $\Sigma_{\varphi}$ for the fluctuating $\varphi$, which is given by
\begin{align}
    \Sigma_{\varphi}^{J}\left(\bm{k},\omega_{n}\right)	& =\left(\frac{J^{2}}{N_{s}\beta}\right)\sum_{q_{n},\bm{q}}\mathcal{G}_{f}\left(\bm{k}-\bm{q},\omega_{n}-q_{n}\right)\mathcal{G}_{f}\left(\bm{q},q_{n}\right) \nn
	&=\frac{J^{2}}{N_{s}\beta}\sum_{q_{n},\bm{q}}\frac{1}{iq_{n}-\varepsilon_{\bm{q}}}\frac{1}{i\omega_{n}-iq_{n}-\varepsilon_{\bm{k-q}}} \nn
	& =\left(\frac{-J^{2}}{N_{s}}\right)\sum_{\bm{q}}\frac{1-n_{F}(\varepsilon_{\bm{k}-\bm{q}})-n_{F}(\varepsilon_{\bm{q}})}{i\omega_{n}-\varepsilon_{\bm{q}}-\varepsilon_{\bm{k-q}}}.
\end{align}
Since our purpose is the find to logarithmic correction due to the internal momentum integral, we thus can set the external momentum to zero. This leads to 
 \begin{align}
      \Sigma_{\varphi}^{J}\left(\omega_{n}\right)=\left(\frac{-J^{2}}{N_{s}}\right)\sum_{\bm{q}}\frac{1-2n_{F}(\varepsilon_{\bm{q}})}{i\omega_{n}-2\varepsilon_{\bm{q}}}=\left(2J^{2}\rho_{0}\right)\int_{-D}^{0}\frac{d\varepsilon}{i\omega_{n}-2\varepsilon}.
 \end{align}
 Taking $i\omega_{n}\to\omega+i\epsilon$ and using $(x\pm i\epsilon)^{-1}=\mathcal{P}(\frac{1}{x})\mp i\pi\delta(x)$, the self-energy correction becomes
\begin{align}
    \Sigma_{\varphi}^{J}\left(\omega\right)	& =\left(2J^{2}\rho_{0}\right)\int_{-D}^{0}\frac{d\varepsilon}{\omega-2\varepsilon+i\epsilon}=\left(-J^{2}\rho_{0}\right)\int_{-2D}^{0}\frac{d\varepsilon^{\prime}}{\varepsilon^{\prime}-\omega-i\epsilon} \nn
	& =\left(-J^{2}\rho_{0}\right)\int_{-2D}^{0}d\varepsilon^{\prime}\left[\mathcal{P}(\frac{1}{\varepsilon^{\prime}-\omega})+i\pi\delta(\varepsilon^{\prime}-\omega)\right].
\end{align}
We find the real part for $\Sigma_{\varphi}^{J}$ is given by
\begin{align}
   \left[  \Sigma_{\varphi}^{J}\left(\omega\right) \right]^\prime &	=\left(-J^{2}\rho_{0}\right)\int_{-2D}^{0}d\varepsilon^{\prime}\mathcal{P}(\frac{1}{\varepsilon^{\prime}-\omega}) \nn
    & =\left(-J^{2}\rho_{0}\right)\ln\frac{\omega}{2D+\omega}\approx\left(-J^{2}\rho_{0}\right)\ln\frac{\omega}{2D} \nn
	& \sim J^{2}\rho_{0}\ln\frac{D}{\omega}, 
\end{align}
which contributes a logarithmic correction that is relevant for further RG analysis. While the imaginary part can be evaluated as
\begin{align}
    \left[\Sigma_{\varphi}^{J}\left(\omega\right) \right]^{\prime\prime}=\left(-J^{2}\pi\rho_{0}\right)\int_{-2D}^{0}d\varepsilon^{\prime}\delta(\varepsilon^{\prime}-\omega)=-J^{2}\pi\rho_{0}\Theta(-\omega).
\end{align}

\subsection{Vertex corrections}
Vertex correction is defined as the correction for the bare vertex function due to interactions. If we denote the full vertex function as $\Pi$ and the bare one as $\Pi^{0}$. The vertex correction, represented as $U$, is defined through
\begin{align}
    \Pi = U \Pi^{0}.
    \label{eq:def-vertex-correction}
\end{align}
While employing perturbation expansion, $U$ can be written as $U=1+U^{(1)}+U^{(2)}+U^{(3)}+\cdots$ with $U^{(n)}$ being the vertex correction from the $n$-th order perturbation.

It is straightforward to show that there is no vertex corrections up to one-loop order contributed from the hopping  term $H_{t}$. The only relevant one-loop vertex correction is coming from the Heisenberg term $H_J$. The vertex function of $H_J$, denoted as $\Pi_{J}$, is defined as
\begin{align}
    \Pi_{J}\left(k,p\right)=\left\langle \varphi_{k+p}f_{k\uparrow}^{\dagger}f_{p\downarrow}^{\dagger}\right\rangle ,
\end{align}
where $k=(\bm{k},k_{n}),\, p=(\bm{p},p_{n})$. Expanding $\Pi_{J}$ above, it is easy to show that no zero-th order term is available. The first non-vanishing contribution is from the first-order expansion in $J$, given by
\begin{align}
    \Pi_{J}^{(1)}\left(k,p\right)&=J\sum_{k^{\prime},p^{\prime}}\left\langle \varphi_{k+p}f_{k\uparrow}^{\dagger}f_{p\downarrow}^{\dagger}\varphi_{k^{\prime}+p^{\prime}}^{\dagger}f_{p^{\prime}\downarrow}f_{k^{\prime}\uparrow}\right\rangle _{0} \nn
    & =  \left(-J\right)\mathcal{\bar{G}}_{\varphi}\left(k+p\right)\mathcal{G}_{f}^{\downarrow\downarrow}(p)\mathcal{G}_{f}^{\uparrow\uparrow}(k).
    \label{eq:bare-vertex-Pi_J}
\end{align}
Thus, $\Pi_{J}^{(1)}\left(k,p\right)=\Pi_{J}^{0}$  is considered as the bare vertex function for the fluctuating Heisenberg term $H_J$. It is easy to demonstrate that $\Pi^{(2)}\propto J^{2}$ is zero. Thus, the lowest non-vanishing correction to the bare vertex function $\Pi_{J}^{(1)}$ comes from $\Pi^{(3)}\propto J^3$, which takes the following form
\begin{align}
    \Pi_{J}^{(3)}(k,p)  = & \left(-\frac{3J^{3}}{3!\beta}\right)\sum_{k^{\prime},p^{\prime}}\sum_{k^{\prime\prime},p^{\prime\prime}}\sum_{k^{\prime\prime\prime},p^{\prime\prime\prime}}\left\langle \varphi_{k+p}f_{k\uparrow}^{\dagger}f_{p\downarrow}^{\dagger}\varphi_{k^{\prime}+p^{\prime}}^{\dagger}f_{p^{\prime}\downarrow}f_{k^{\prime}\uparrow}\varphi_{k^{\prime\prime}+p^{\prime\prime}}^{\dagger}f_{p^{\prime\prime}\downarrow}f_{k^{\prime\prime}\uparrow}\varphi_{k^{\prime\prime\prime}+p^{\prime\prime\prime}}f_{k^{\prime\prime\prime}\uparrow}^{\dagger}f_{p^{\prime\prime\prime}\downarrow}^{\dagger}\right\rangle_{0} \nn
    = & \left(\frac{J^{3}}{\beta}\right)\sum_{p^{\prime}}\sum_{k^{\prime\prime}}\left\langle \varphi_{k+p}\varphi_{k^{\prime\prime}+p^{\prime}}\right\rangle _{0}\left\langle \varphi_{k+p^{\prime}}^{\dagger}\varphi_{k^{\prime\prime}+p}^{\dagger}\right\rangle _{0}\left\langle f_{k\uparrow}f_{k\uparrow}^{\dagger}\right\rangle _{0}\left\langle f_{p\downarrow}f_{p\downarrow}^{\dagger}\right\rangle _{0}\left\langle f_{k^{\prime\prime}\uparrow}f_{k^{\prime\prime}\uparrow}^{\dagger}\right\rangle _{0}\left\langle f_{p^{\prime}\downarrow}f_{p^{\prime}\downarrow}^{\dagger} \right\rangle_{0}.
    \label{eq:Pi_J_3-1}
\end{align}
From Eqs. (\ref{eq:def-vertex-correction}), (\ref{eq:bare-vertex-Pi_J}) and (\ref{eq:Pi_J_3-1}), the vertex correction of $ \Pi_{J}^{(1)}$ is related to the third-order perturbation,  denoted as $U_{J}^{(3)}(k,p)$, through $\Pi_{J}^{(3)}(k,p) = U_{J}^{(3)}(k,p)\Pi_{J}^{(1)}\left(k,p\right)$, with
\begin{align}
U_{J}^{(3)}(k,p)=\left(-\frac{J^{2}}{\beta}\right) &\sum_{p^{\prime}}\mathcal{\bar{G}}_{\varphi}\left(k+p^{\prime}\right)\mathcal{G}_{f}^{\uparrow\uparrow}(-k-p-p^{\prime}) \mathcal{G}_{f}^{\downarrow\downarrow}(p^{\prime}). 
\end{align}
Using $\mathcal{\bar{G}}_{\varphi}=-1$, we have
\begin{align}
    U_{J}^{(3)}(k,p)&=\frac{J^{2}}{\beta}\sum_{p^{\prime}}\mathcal{G}_{f}^{\uparrow\uparrow}(-k-p-p^{\prime})\mathcal{G}_{f}^{\downarrow\downarrow}(p^{\prime}) \nn
    & = \frac{J^{2}}{\beta}\sum_{p^{\prime}}\mathcal{G}_{f}^{\uparrow\uparrow}(-p^{\prime})\mathcal{G}_{f}^{\downarrow\downarrow}(p^{\prime}) \nn
    & =\frac{J^{2}}{\beta}\sum_{ip_{n}^{\prime},\bm{p}^{\prime}}\frac{1}{ip_{n}^{\prime}-\varepsilon_{\bm{p}^{\prime}}}\cdot\frac{1}{-ip_{n}^{\prime}-\varepsilon_{\bm{p}^{\prime}}} \nn
    & =\left(-J^{2}\rho_{0}\right)\int_{-D}^{D}d\varepsilon\frac{n_{F}(\varepsilon)-n_{F}(-\varepsilon)}{2\varepsilon} \nn
    & =\left(-J^{2}\rho_{0}\right)\ln\frac{D}{r}\Big|_{r\to0}+\text{others}.
\end{align}
We obtain a $\ln D$ term from the vertex correction of $H_J$.

In summary, the self-energy corrections of various fields from $H_{t}$ and $H_{J}$ are given by
\begin{align}
    \Sigma_{\xi}^{g}(\omega) & \sim\bar{g}^{2}\rho_{0}\ln\frac{D}{\omega}, \nn
\Sigma_{b}^{g}(\omega)	& \sim(\ln D)^{0}, \nn
\Sigma_{f}^{g}(\omega)	& \sim\bar{g}^{2}\rho_{0}\ln\left(\frac{D}{\omega-\zeta^{-1}\varUpsilon}\right), \nn
\Sigma_{\varphi}^{J}(\omega) & \sim J^{2}\rho_{0}\ln\frac{D}{\omega}.
\end{align}
This leads to the following form of field renormalization $ \mathbb{G}$:
\begin{align}
    \mathbb{G}_{f}	& \approx 1+ \left( \bar{g}^{2}\rho_{0}\right)\mathcal{G}_{f}\ln\left(\frac{D}{\omega-\zeta^{-1} \varUpsilon} \right) 
     \sim  1+ \left(\bar{g}^{2}\rho_{0}\right)\frac{1}{\omega}\ln\left(\frac{D}{\omega-\zeta^{-1}\varUpsilon}\right) \nc \nn
\mathbb{G}_{\xi} & \approx 1+ \bar{g}^{2}\rho_{0} \mathcal{\bar{G}}_{\xi}\ln\frac{D}{\omega} \sim 1+\bar{g}^{2}\rho_{0}\frac{1}{\omega}\ln\frac{D}{\omega}\nc \nn
\mathbb{G}_{\varphi} & \approx 1+ J^{2} \rho_{0} \mathcal{\bar{G}}_{\varphi}\ln\frac{D}{\omega} \sim 1+ J^{2}\rho_{0}\frac{1}{\omega}\ln\frac{D}{\omega} \nc \nn
\mathbb{G}_{b} & = 1 \np 
\end{align}
Here, we have replaced the bare Green's function with $\mathcal{G}\sim1/\omega$. In addition, we find the perturbative expansion to the vertex function of $H_J$ and $H_t$ contributes the following vertex corrections:
\begin{align}
     U_{J}(k,p)	& =1-J^{2}\rho_{0}\ln\frac{D}{r}\Big|_{r\to0}\nc \nn
     U_{g}(k,p)	& =1 \np
\end{align}

\subsection{The renormalization $z$-factors  and the RG $\beta$ functions}
To compute the RG $\beta$ functions, we start by reducing the cutoff energy from $D\to D^{\prime}=yD$ with $0<y<1$. This generates the relations between the wave-function (Green's function) and vertex corrections between $D$ and $D^\prime$, which shows that \cite{qimiao-prb-local-fluc,lijun-prb-loca-fluc-bose-fermi}
\begin{align}
    \mathbb{G}(\omega,D^{\prime},J_{i}^{\prime}) & =z_{\psi}\text{\ensuremath{\mathbb{G}}}(\omega,D,J_{i})\nc \nn U(\omega,D^{\prime},J_{i}^{\prime}) & =z_{U}^{-1}U(\omega,D,J_{i})
    \label{eq:G-U-renormal}
\end{align}
with $z_\psi$ and $z_U$ being denoted as the renormalization factor for the Green's function for the $\psi$ field and vertex correction $U$, respectively. The renormalization factor for the $\xi$ field is given by
\begin{align}
    z_{\xi}	& = \left(1+\bar{g}^{2}\rho_{0}\frac{1}{\omega}\ln\frac{D}{\omega}\right)^{-1}\left(1+\bar{g}^{2}\rho_{0}\frac{1}{\omega}\ln\frac{D^{\prime}}{\omega}\right) \nn
	        & =1-\frac{\bar{g}^{2}\rho_{0}}{\omega}\ln \ell \nc
\end{align}
where $\ln \ell \equiv\ln\frac{D}{D^{\prime}}$. As $\omega$ does not undergo a RG scaling [see Eq. (\ref{eq:G-U-renormal})], we can set, for convenience, $\omega=D\to1/\omega=1/D=\rho_{0}$. This yields 
\begin{align}
    z_{\xi}=1-\bar{g}^{2}\rho_{0}^{2}\ln \ell \np
\end{align}
Following a similar approach, we obtain the other two $z$ factors of fields,
\begin{align}
    z_{f} & =1-\left(\bar{g}\rho_{0}\right)^{2}\ln \ell \nc \nn
    z_{\varphi} & =1-\left(J\rho_{0}\right)^{2}\ln \ell \np 
\end{align}
The $z$ factors for the couplings are
\begin{align}
    & z_g =1 \nnc 
    &z_{J}=1+J^{2}\rho_{0}^{2}\ln \ell \np
\end{align}

\subsection{The RG scaling equations ($\beta$ functions)}
The renormalized couplings constants are defined as
\begin{align}
    \bar{g}^{\prime}&=z_{f}^{-\frac{1}{2}}z_{b}^{-\frac{1}{2}}z_{\xi}^{-\frac{1} {2}}z_{g}\bar{g} \nnc
J^{\prime} & =z_{f}^{-1}z_{\varphi}^{-\frac{1}{2}}z_{J}J \np
\end{align}
Expanding $\bar{g}^\prime$ to the leading order in $\bar{g}\rho_0$, we obtain
\begin{align}
    \bar{g}^{\prime}&	=\left(1-\left(\bar{g}\rho_{0}\right)^{2}\ln \ell\right)^{-\frac{1}{2}}\left(1-\bar{g}^{2}\rho_{0}^{2}\ln \ell\right)^{-\frac{1}{2}}\bar{g} \nn
	 & =\left(1+\frac{1}{2}\left(\bar{g}\rho_{0}\right)^{2}\ln \ell\right)\left(1+\frac{1}{2}\left(\bar{g}\rho_{0}\right)^{2}\ln \ell\right)\bar{g} \nn
	& =\bar{g}+\bar{g}\left(\bar{g}\rho_{0}\right)^{2}\ln \ell \np
\end{align}
Similarly, $J^\prime$ shows
\begin{align}
    J^{\prime}	& =\left[1-\left(\bar{g}\rho_{0}\right)^{2}\ln \ell\right]^{-1} \left[1-\left(J\rho_{0}\right)^{2}\ln \ell\right]^{-\frac{1}{2}}\left[1+J^{2}\rho_{0}^{2}\ln \ell \right]J \nn
	  & =\left[1+\left(\bar{g}\rho_{0}\right)^{2}\ln \ell \right] \left[1+\frac{1}{2}\left(J\rho_{0}\right)^{2}\ln \ell \right]\left(1+J^{2}\rho_{0}^{2}\ln \ell \right)J \nn
	  & =J+J\left[\left(\bar{g}\rho_{0}\right)^{2}+\frac{3}{2}\left(J\rho_{0}\right)^{2}\right]\ln \ell \np
\end{align}
Multiplying both sides of $J^{\prime}$ and $\bar{g}^{\prime}$ by $\rho_{0}$, we have 
\begin{align}
    \left(J\rho_{0}\right)^{\prime}	& =\bar{g}\rho_{0}+\left(\bar{g}\rho_{0}\right)^{3}\ln \ell \nnc 
    (\bar{g}\rho_{0})^{\prime}	& =J\rho_{0}+\left[\left(J\rho_{0}\right)\left(\bar{g}\rho_{0}\right)^{2}+\frac{3}{2}\left(J\rho_{0}\right)^{3}\right]\ln \ell \np
\end{align}
These lead to the following RG $\beta$ functions:
\begin{align}
    \frac{d(\bar{g}\rho_{0})}{dl}	 & =-\left(\frac{d-z}{2}\right)(\bar{g}\rho_{0}) + \left(\bar{g}\rho_{0}\right)^{3} \nnc
\frac{d(J\rho_{0})}{dl}	& =-\left(\frac{d-z}{2}\right)(J\rho_{0})+\left(J\rho_{0}\right)\left(\bar{g}\rho_{0}\right)^{2}+\frac{3}{2}\left(J\rho_{0}\right)^{3} \np
\end{align}
The fixed points occur at  $\left(\left(\bar{g}\rho_{0}\right)^{2},\quad\left(J\rho_{0}\right)^{2}\right)=\left(0,\quad0\right),\quad Q:\left(0,\quad\frac{\epsilon}{3}\right),\quad P:\left(\frac{\epsilon}{2},\quad0\right)$
where $\epsilon \equiv d-z$. In our case, we set $d=2, \, z=1$ and thus  $\epsilon=1$. Note that the RG scheme is still reliable for $\epsilon=1$ though it makes the $\epsilon$ expansion divergent \cite{goldenfeld-book-RG}. 

\subsection{Rescale the coupling constants}
In this section, we will show how the coupling constants are  rescaled near the critical point $C_r$. Near the fixed point $P$, we fix $\bar{g}\rho_0$ at its fixed point value, namely $\bar{g}\rho_0 =\bar{g}^* \rho_0 = \sqrt{\epsilon/2} $. As $\bar{g}\rho_0$ is now fixed at a constant, it is equivalent of making the $\bar{g}\rho_0$  coupling marginal since it cannot flow anymore. Making $\bar{g}\rho_0$ marginal can be done by $H_t \sim \left( \bar{g}/e^{-\epsilon \ell/2}\right) e^{-\epsilon \ell/2} f^\dagger b^\dagger \xi  = \bar{g}^\prime f^{\dagger\,\prime} b^{\dagger \prime} \xi^\prime  $ with $\bar{g}^\prime = \bar{g} /e^{-\epsilon \ell/2} $ and $f^\prime = e^{-\epsilon \ell/2} f$ and $b^\prime = b$ and $\xi^\prime = \xi$. This leads to a shift in the bare scaling dimensions $[\bar{g}^\prime] = [\bar{g}]+\epsilon/2 = 0$ and $[f^\prime]=[f] +\epsilon/4$. We attribute the remaining exponential terms into $b$ and $\xi$ and change their scaling dimensions accordingly. In the meantime, due to the change of the scaling dimension of $f$, the scaling dimension of $J$ needs to be changed to $[J^\prime] = [J] - \epsilon/2$. To lighten the notation, here we will use the same symbols for the couplings after the rescaling as we did before the rescaling. Fixing $\bar{g}\rho_{0} = \left(\bar{g}\rho_{0}\right)^* = \sqrt{\epsilon/2}$  yields the following RG equation for $J\rho_0$ near $P$:
\begin{align}
    \frac{d(J\rho_{0})}{dl}&=\left(-\frac{\epsilon}{2} -\frac{\epsilon}{2}\right)\left(J\rho_{0}\right)+\left(J\rho_{0}\right)\left[\left(\bar{g}\rho_{0}\right)^*\right]^{2}+\frac{3}{2}\left(J\rho_{0}\right)^{3} \nn
    =& -\frac{\epsilon}{2} \left(J\rho_{0}\right)+\frac{3}{2}\left(J\rho_{0}\right)^{3},
\end{align}
which yields the same fixed point value for $ J\rho_0$, namely $\left[\left( J\rho_0\right)^*\right]^2  = \epsilon/3 $.

Following the similar approach, we evaluate the RG flow of $\bar{g}\rho_0$ near $Q$ by fixing $J\rho_0 =J^* \rho_0 = \sqrt{\epsilon/3}$. We make $J$ marginal by $[J^\prime] = [J] + \epsilon/2$ and $[\varphi^\prime] =[\varphi]-\epsilon/2 $. The RG equation for $\bar{g}$ does not change after this transformation. The RG flow diagram  of the rescaled coupling constants is shown in Fig. 3(a) of the main text.

\section{Specific heat coefficient}
\label{app:specificheat}

\begin{figure}
    \centering
    \includegraphics[width=0.45\textwidth]{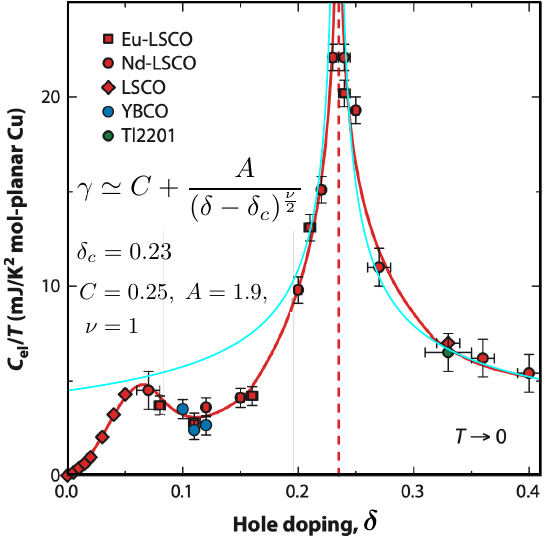}
    \caption{Power-law divergence of the electronic specific heat coefficient $C_V/T$ for various samples around the optimal doping. Experimental data are reproduced from Ref. \cite{michon-nature-2019-QC-cuprate}. The blue line corresponds to the power-law fit for which we find a coefficient $b \approx 0.5$ in good agreement with our theoretical prediction [cf. Eq. (\ref{eq:C_V^f-pwr})].}
    \label{fig:specificheat-powerlaw}
\end{figure}

In this section, we estimate the specific heat coefficient contributed solely from the half-filled \textit{f}-spinon band. The entropy of a non-interacting many-fermion system with Hamiltonian $H_f=\sum_{\bm{k}\sigma}\varepsilon_{\bm{k}}f_{\bm{k}\sigma}^{\dagger}f_{\bm{k}\sigma}$ with  dispersion $\varepsilon_{\bm{k}}=-2 J_H \chi \left(\cos k_{x}+\cos k_{y}\right)$ is given by
\begin{align}
    S_f(T)= & -2\sum_{\bm{k}}  \left[n_{F} (\varepsilon_{\bm{k}})\ln n_F (\varepsilon_{\bm{k}}) 
     +\left(1-n_F \left(\varepsilon_{\bm{k}} \right)\right)\ln\left(1-n_{F} (\varepsilon_{\bm{k}})\right)\right] \nn
    = &-2\int_{-D(T)}^{0}  d\varepsilon\rho(\varepsilon,T)\left[n_F(\varepsilon)\ln n_F(\varepsilon)
     +\left(1-n_F(\varepsilon)\right)\ln\left(1-n_F(\varepsilon)\right)\right],
\end{align}
 where $n_{F} (\varepsilon_{\bm{k}})=1/\left(e^{\varepsilon_{\bm{k}}/T}+1\right)$ and the temperature-dependent half-bandwidth $D(T) = 4 J_H \chi(T)$. The Boltzmann constant is set to be $k_{B}=1$, while the factor of $2$ in $S_f$ comes from the spins. 
 
For the low-temperature regime, we can approximate $D(T) \approx D(T=0) \approx 4 J_H \chi(T = 0 ) = 4 J_H \chi$ with $\chi \equiv \chi(T=0)$, leading to $\rho(\varepsilon,T)\to\rho(\varepsilon,T=0)\equiv\rho_{0}(\varepsilon)$. The entropy becomes approximately
\begin{align}
    S_f(T)\approx-2\int_{-D_{0}}^{0}d\varepsilon\rho_{0}(\varepsilon)\left[n_F(\varepsilon)\ln n_F(\varepsilon)
    +\left(1-n_F(\varepsilon)\right)\ln\left(1-n_F(\varepsilon)\right)\right].
\end{align}

At half-filling, the density of states for the tight-binding dispersion $\varepsilon_{\bm{k}}$ on a  2D square lattice  exhibits a logarithmic divergence in the low-energy regime, i.e., $ \rho_{0}(\varepsilon) \sim - \left(\frac{1}{\Lambda} \right)\ln \left(\varepsilon / \varepsilon_0 \right)$ as $\varepsilon \to 0$, where the prefactor $\Lambda$, carrying the same unit of energy, is proportional to $J$, and $\varepsilon_0$ represents a constant energy scale. Consequently, the entropy in the low-energy regime is dictated by the logarithmically divergent density of states, given by
\begin{align}
    S_f(T)	\sim &  \frac{2}{\Lambda}\int_{-D}^{0}d\varepsilon\ln \left(\varepsilon / \varepsilon_0 \right) \left[n_F(\varepsilon)\ln n_F(\varepsilon)   +\left(1-n_F(\varepsilon)\right)\ln\left(1-n_F(\varepsilon)\right)\right] \vspace{10pt} \nn
		\approx & \frac{2}{\Lambda} \int_{-D/T}^{0}\left(Tdx\right)\ln\left( \frac{Tx}{\varepsilon_0}\right)\left[\frac{1}{e^{x}+1}\ln\left(\frac{1}{e^{x}+1}\right)+\left(1-\frac{1}{e^{x}+1}\right)\ln\left(1-\frac{1}{e^{x}+1}\right)\right] \vspace{10pt} \nn
	= & \frac{2}{\Lambda} \int_{-\infty}^{0}\left(Tdx\right)\left[\ln \left( \frac{T}{\varepsilon_0}\right)+\ln x\right]F(x),
\end{align}
where in the last line of the above equation, we take the approximation $D/T\to\infty $  for $T \to 0$, and where we define 
\begin{align}
    F(x)\equiv & \frac{1}{e^{x}+1}\ln\left(\frac{1}{e^{x}+1}\right)  +\left(1-\frac{1}{e^{x}+1}\right)\ln\left(1-\frac{1}{e^{x}+1}\right).
\end{align}
For the low-temperature limit, the entropy can be compactly expressed as
\begin{align}
    S_f(T)	=  \frac{\alpha^\prime T}{\Lambda}\ln \left( \frac{T}{\varepsilon_0}\right)+ \frac{\beta^\prime T}{\Lambda}
    \label{eq:S-low-T}
\end{align}
 with $\alpha^\prime, \, \beta^\prime$ being constant prefactors:
 \begin{align}
     \alpha^\prime \equiv 2\int_{-\infty}^{0}F(x)dx,\quad \beta^\prime \equiv2\int_{-\infty}^{0} dx F(x)\ln x.
 \end{align}

The contribution of the non-interacting $f$ spinon to the specific heat coefficient at the low-temperature regime can be thus obtained by taking the temperature derivative of $S_f$ in Eq. (\ref{eq:S-low-T}), $\frac{C^f_V}{T} = \frac{\partial S_f(T)}{\partial T}$, showing a logarithmic-in-\textit{T} divergence
\begin{align}
    \frac{C^f_V}{T} = \frac{\alpha^\prime}{\Lambda} \ln \left( \frac{\gamma^\prime T}{\varepsilon_0}\right)
\end{align}
with $\gamma^\prime = e^{\frac{\alpha^\prime + \beta^\prime}{\alpha^\prime}}$.

At low temperatures and near the QCP, the behavior of the specific heat coefficient $\frac{C^f_V}{T}$ has to be carefully estimated by considering the coupling constant renormalization. Here,  $\frac{C^f_V}{T}$ only depends on the Heisenberg coupling; thus, under the coupling constant renormalization $J \to J(\ell)$, we have $\frac{C^f_V}{T} \sim J^{-1}(\ell_0)$ at some fixed renormalization scale $\ell = \ell_0$. The renormalization of the Heisenberg coupling can be expressed as $J(\ell_0) = J e^{-\epsilon \ell_0 /2}$.  Using the relation of $e^{\ell_0} = \eta \sim |\delta - \delta_c|^{-\nu}$ with $\eta$ being the correlation length and $\nu = 1/z = 1$ the correlation length exponent in our theory,  $\frac{C^f_V}{T}$ near the QCP behaves as
\begin{align}
     \frac{C^f_V}{T} \sim |\delta -\delta_c|^{-1/2}.
     \label{eq:C_V^f-pwr}
\end{align}
The low-temperature behavior of  $\frac{C^f_V}{T} $ contributed from the half-filled $f$-spinon band exhibits a power-divergence as a function of $\delta -\delta_c$ with power-law exponent equal to $0.5$. A similar behavior has been experimentally observed in the cuprate superconductors \cite{michon-nature-2019-QC-cuprate, Taillefer-annuphys-2019}, see Supplementary Figure \ref{fig:specificheat-powerlaw}. Note that due to the divergence of the specific heat coefficient in Eq. (\ref{eq:C_V^f-pwr}), the \textit{T}-logarithmic marginal Fermi liquid behavior of the specific heat coefficient contributed from \textit{f}-spinon band dominates over that contributed from the fluctuating charge (Kondo) term.  
\section{Transport properties}
\subsection{Total electrical resistivity by Ioffe-Larkin's composition rule}
Below, we use the Ioffe-Larkin composition rule \cite{larkin-rule-PRB} to compute the total resistivity (or conductivity) of the Kondo-Heisenberg approach to the slave boson \textit{t-J} model [Eq. (1) of the main text]. To apply the Ioffe-Larkin composition rule in the calculation of the total resistivity, we need to analyze the gauge transformation and include both the  intrinsic U(1) gauge field and external gauge field in the form of vector potential of external magnetic field into this model. The model, described by the Lagrangian density below
\begin{align}
    \mathcal{L}(\tau)	& =\sum_{i}b_{i}^{\dagger}\left(\partial_{\tau}+i\lambda_{i}\right)b_{i}+\sum_{i\sigma}f_{i\sigma}^{\dagger}\left(\partial_{\tau}+i\lambda_{i}-\mu\right)f_{i\sigma}+t\sum_{\langle i,j\rangle,\sigma}\left[\left(f_{i\sigma}^{\dagger}b_{j}^{\dagger}+f_{j\sigma}^{\dagger}b_{i}^{\dagger}\right)\xi_{ij,\sigma}+H.c.\right] \nn
	& \qquad+\sum_{\langle i,j\rangle\sigma}\left(-\chi_{ij}f_{i\sigma}^{\dagger}f_{j\sigma}+H.c.\right)+\sum_{\langle i,j\rangle,\alpha}\left(\Delta_{ij}\tilde{\sigma}f_{i\sigma}^{\dagger}f_{j,-\sigma}^{\dagger}+H.c.\right)+{\rm others},
\end{align}
is invariant under the following gauge transformations
\begin{align}
b_{i}	&\to b_{i}e^{i\theta_{i}},\nn
f_{i\sigma}	& \to f_{i\sigma}e^{i\theta_{i}},\nn
\xi_{ij,\sigma}	& \to\xi_{ij,\sigma}e^{i(\theta_{i}+\theta_{j})},\nn
\Delta_{ij}	&\to \Delta_{ij}e^{i(\theta_{i}+\theta_{j})}, \nn
\chi_{ij}	&\to \chi_{ij}e^{i(-\theta_{i}+\theta_{j})}, \nn
\lambda_{i}	&\to \lambda_{i}-\partial_{\tau}\theta_{i}.
\end{align}

We fix $i\lambda_{i}$ and $\chi_{ij}$ at their saddle point solution such that  $i\lambda_{i}$ and $\chi_{ij}$ in our model are replaced by the mean-field values i.e. $i\lambda_{i}\to\lambda$ and $\langle\chi_{ij}\rangle \to \chi$. The fluctuations beyond the mean-field $\lambda$ and $\chi$ can be included in the effective Lagrangian by writing $i\lambda_{i} =\lambda+ia_{0}(\bm{r}_{i},\tau)$ and $\chi_{ij}=\chi e^{ia_{ij}}$, where $a_{0}(\bm{r}_{i},\tau)$ and $a_{ij}$ are the fluctuating fields for $\lambda$ and $\chi$ ($a_{ij}$ is the phase fluctuation for $\chi$). With the inclusion  of the fluctuations in the amplitude of  $\lambda$, the non-interacting parts for $f$ and $b$ can be rewritten as	
\begin{align}
& \sum_{i}b_{i}^{\dagger}\left(\partial_{\tau}+i\lambda_{i}\right)b_{i}+\sum_{i\sigma}f_{i\sigma}^{\dagger}\left(\partial_{\tau}+i\lambda_{i}-\mu\right)f_{i\sigma} \nn
\longrightarrow &	\sum_{i}b_{i}^{\dagger}\left[\partial_{\tau}+\lambda+ia_{0}(\bm{r}_{i},\tau)\right]b_{i}+\sum_{i\sigma}f_{i\sigma}^{\dagger}\left[\partial_{\tau}+\lambda-\mu+ia_{0}(\bm{r}_{i},\tau)\right]f_{i\sigma}.
\end{align}
Note that in our further derivation of the conductivity (or the resistivity),  we will drop the non-interacting Lagrangian density of the $b$ field since its dynamics and dispersion are negligibly small. In addition, when the phase fluctuation $a_{ij}$ is included, the fermion hopping term becomes  $-\sum_{\langle i,j\rangle\sigma}\chi_{ij}f_{i\sigma}^{\dagger}f_{j\sigma}\to-\sum_{\langle i,j\rangle\sigma}\chi e^{ia_{ij}}f_{i\sigma}^{\dagger}f_{j\sigma}$, which, in the continuum limit, can be expressed approximately in the following familiar form, 
\begin{align}
    -\frac{1}{2m_{f}^{*}}\sum_{\sigma}\int d^{2}rf_{\sigma}^{\dagger}(\bm{r},\tau)\left[-i\nabla_{\bm{r}}+\bm{a}(\bm{r},\tau)\right]^{2}f_{\sigma}(\bm{r},\tau).
\end{align}
Within the saddle point solutions, the gauge transformations now becomes $a_{0}(\bm{r}_{i},\tau)\to a_{0}(\bm{r}_{i},\tau)-\partial_{\tau}\theta_{i}$ and $a_{ij}\to a_{ij}-\theta_{i}+\theta_{j}$. Thus, we observe that the fluctuating field $a_0(\bm{r},\tau)$ governs the gauge transformation for the scalar potential, while the phase field $a_{ij}$ governs the gauge transformation for the vector potential. The latter becomes apparent when we take the continuum limit of our model.

In the continuum limit, the phase field $a_{ij}$ can be approximate in the form of \cite{Nagaosa-1990-PRL,Nagaosa-2006-RMP}
\begin{align}
    a_{ij}(\tau)=(\bm{r}_{i}-\bm{r}_{j})\cdot\bm{a}\left(\frac{\bm{r}_{i}+\bm{r}_{j}}{2},\tau\right) \sim\int d\bm{r}\cdot\bm{a}(\bm{r},\tau),
\end{align}
which defines the gauge field $\bm{a}(\bm{r},\tau)$. The original form of the gauge transformation for $a_{ij}$ has now been modified to $\bm{a}(\bm{r},\tau) \to \bm{a}(\bm{r},\tau) -\nabla \theta (\bm{r},\tau)$.

The $\xi$ field should also couple to the gauge field since we introduce dynamics and dispersion to the $\xi$ field via Random Phase Approximation (RPA). This introduces an additional contribution to the conductivity via the Ioffe-Larkin composition rules, as described below. To demonstrate how the $\xi$ field couples with the gauge fields, we need to integrate out the $f$ spinon and the slave boson $b$, which generate terms of the form $\sim\xi_{ij,\sigma}^{\dagger}\xi_{jk,\sigma}\chi_{kl}\chi_{li}$ \cite{Punk-SBtJ-PRB}. With a finite mean-field value for $\chi$, the $\xi$ field will couple to the phase field of $\chi_{ij}$, namely $a_{ij}$. In the continuum limit, this coupling of the $\xi$ field to the phase field is described by the following Lagrangian density  \cite{Punk-SBtJ-PRB}, 
\begin{align}
    \sum_{\sigma}\int d^{2}r\xi_{\sigma}^{\dagger}(\bm{r},\tau)\left[\partial_{\tau}-2ia_{0}(\bm{r},\tau)\right]\xi_{\sigma}(\bm{r},\tau)-\frac{1}{2m_{\xi}^{*}}\sum_{\sigma}\int d^{2}r\xi_{\sigma}^{\dagger}(\bm{r},\tau)\left[-i\nabla_{\bm{r}}+2\bm{a}(\bm{r},\tau)\right]^{2}\xi_{\sigma}(\bm{r},\tau).
\end{align}
This term implies that $\xi$ is a charge-2$e$ fermion field. A similar term also exists for the pairing bond field $\Delta_{ij}$. However, we will neglect this contribution in our further analysis of the conductivity since the dynamics and dispersion of $\Delta_{ij}$ have been shown to be subdominant.

Next, we consider the effect of an external electromagnetic field coupling to the system. The coupling of the system to the electromagnetic field can be described by Peierls' substitution,
\begin{align}
    tc_{i\sigma}^{\dagger}c_{j\sigma}\to te^{-ie\int_{i}^{j}d\bm{r}\cdot\bm{A}_{em}}c_{i\sigma}^{\dagger}c_{j\sigma}=t\sum_{\langle i,j\rangle,\sigma}\left[e^{-ie\int_{i}^{j}d\bm{r}\cdot\bm{A}_{em}}\left(f_{i\sigma}^{\dagger}b_{j}^{\dagger}+f_{j\sigma}^{\dagger}b_{i}^{\dagger}\right)\xi_{ij,\sigma}+H.c.\right].
\end{align}
We can couple the external vector potential $\bm{A}_{em}$ either to $f$, $b$, or $\xi$, but not to all simultaneously. Here, we choose $\bm{A}_{em}$ to couple to the $\xi$ field. The non-interacting part of the Lagrangian for the $\xi$ field shown above should be replaced by
\begin{align}
    \sum_{\sigma}\int d^{2}r\xi_{\sigma}^{\dagger}(\bm{r},\tau) & \left[\partial_{\tau}-2ia_{0}(\bm{r},\tau)-iA_{em,0}(\bm{r},\tau)\right]  \xi_{\sigma}(\bm{r},\tau) \nn
    &-\frac{1}{2m_{\xi}^{*}}\sum_{\sigma}\int d^{2}r\xi_{\sigma}^{\dagger}(\bm{r},\tau)\left[-i\nabla_{\bm{r}}+2\bm{a}(\bm{r},\tau)+\bm{A}_{em}(\bm{r},\tau)\right]^{2}\xi_{\sigma}(\bm{r},\tau).
\end{align}
The non-interacting effective Lagrangian, which includes all fields relevant for the following conductivity analysis coupled to the gauge fields, can be summarized as	
\begin{align}
& \sum_{i\sigma}f_{i\sigma}^{\dagger}\left[\partial_{\tau}+\lambda-\mu+ia_{0}(\bm{r}_{i},\tau)\right]f_{i\sigma}-\frac{1}{2m_{f}^{*}}\sum_{\sigma}\int d^{2}rf_{\sigma}^{\dagger}(\bm{r},\tau)\left[-i\nabla_{\bm{r}}+\bm{a}(\bm{r},\tau)\right]^{2}f_{\sigma}(\bm{r},\tau)\nn
	&\quad\qquad+\sum_{\sigma}\int d^{2}r\xi_{\sigma}^{\dagger}(\bm{r},\tau)\left[\partial_{\tau}-2ia_{0}(\bm{r},\tau)\right]\xi_{\sigma}(\bm{r},\tau)-\frac{1}{2m_{\xi}^{*}}\sum_{\sigma}\int d^{2}r\xi_{\sigma}^{\dagger}(\bm{r},\tau)\left[-i\nabla_{\bm{r}}+2\bm{a}(\bm{r},\tau)\right]^{2}\xi_{\sigma}(\bm{r},\tau).
\end{align}

Following the rules by Ioffe and Larkin \cite{larkin-rule-PRB}, we integrate out the $f$ and $\xi$ fermion fields to obtain the effective action of the gauge fields,
\begin{align}
    S_{eff}	& \sim\sum_{\omega_{n},\bm{k}}\sum_{\mu=0,\perp}\Pi_{\mu}^{f}(\omega_{n},\bm{k})a_{\mu}(\omega_{n},\bm{k})a_{\mu}(-\omega_{n},-\bm{k}) \nn
 &\qquad\qquad+\sum_{\omega_{n},\bm{k}}\sum_{\mu=0,\perp}\Pi_{\mu}^{\xi}(\omega_{n},\bm{k})\left[2a_{\mu}(\omega_{n},\bm{k})+A_{em,\mu}(\omega_{n},\bm{k})\right]\left[2a_{\mu}(-\omega_{n},-\bm{k})+A_{em,\mu}(-\omega_{n},-\bm{k})\right] \nn
	&=\sum_{\omega_{n},\bm{k}}\sum_{\mu=0,\perp}\left[\Pi_{\mu}^{f}(\omega_{n},\bm{k})+4\Pi_{\mu}^{\xi}(\omega_{n},\bm{k})\right]\left|a_{\mu}(\omega_{n},\bm{k})\right|^{2}+\sum_{\omega_{n},\bm{k}}\sum_{\mu=0,\perp}\Pi_{\mu}^{\xi}(\omega_{n},\bm{k})\left|A_{em,\mu}(\omega_{n},\bm{k})\right|^{2} \nn
	& \qquad\qquad+\sum_{\omega_{n},\bm{k}}\sum_{\mu=0,\perp}4\Pi_{\mu}^{\xi}(\omega_{n},\bm{k})A_{em,\mu}(\omega_{n},\bm{k})a_{\mu}(-\omega_{n},-\bm{k})
\end{align}
with $\Pi_{\alpha}^{f}\left(\Pi_{\alpha}^{\xi}\right)$ being the current-current correlation function for the $f \, (\xi)$ fermion. And $\mu \in 0,\, \perp$ denotes the scalar and transverse components. The transverse component is directly related to the conductivity. When further integrating out $a_{\mu}(\omega_{n},\bm{k})$, the resulting effective action for $\bm{A}_{em,\mu}$ describes the response of the system to the external electromagnetic field,
\begin{align}
S_{eff}\sim\sum_{\omega_{n},\bm{k}}\sum_{\mu=0,\perp}\Pi_{\mu}^{{\rm tot}}(\omega_{n},\bm{k})A_{em,\mu}(\omega_{n},\bm{k})A_{em,\mu}(-\omega_{n},-\bm{k}),
\end{align}
where $\Pi_{\mu}^{{\rm tot}}$ represents the correlation function of the total current, given by 
\begin{align}
    \Pi_{\mu}^{{\rm tot}} = \frac{\Pi_{\mu}^{\xi}\Pi_{\mu}^{f}}{\Pi_{\mu}^{f}+4\Pi_{\mu}^{\xi}}.
\end{align}
Since the transverse component of current-current correlation function is related to the conductivity through $\Pi_{\mu}^{f}(\omega_{n}, \bm{k}=0) = \sigma_{f}|\omega_{n}|$ (same for the $\xi$ field), the total conductivity is thus approximately  given by $\sigma_{{\rm tot}} = \frac{\sigma_{f}\sigma_{\xi}}{\sigma_{f} + 4\sigma_{\xi}}$, implying that the total resistivity is approximately given by
 \begin{align}
     \rho_{{\rm tot}} \approx 4\rho_{f}+\rho_{\xi}.
 \end{align}
From Drude's model, the electrical resistivity is proportional to the effective mass, we thus obtain $\rho_{{\rm tot}}\approx\rho_{\xi}$ since the ratio of the effective masses for the $f$ spinon and $\xi$ field is
\begin{align}
    m_{\xi}^{*}/m_{f}^{*}\sim\frac{\frac{1}{\zeta\chi}}{\frac{1}{\chi}} = \frac{1}{\zeta} \propto \frac{(D/t)^2}{\delta}  \gg 1 \nc 
\end{align}
for doping $0.2<\delta <0.3$ and $t/D \approx 3/4$. 
 
 The resistivity from the $\xi$ fermion can be further spitted into two contributions: one from the scattering of the $\xi$ fermions by $H_t$, denoted as $\rho_{\xi,g}$, and the other from scattering of the $\xi$ fermions with gauge fields, denoted as $\rho_{\xi,\rm{gauge}}$. Calculation of $\rho_{\xi,g}$ is provided in the section below. \added{The electron scattering rate $1/\tau_{\xi, \rm{gauge}}$ contributed from the gauge field, and the corresponding conductivity $\sigma_{\xi, \rm{gauge}}$ and resistivity $\rho_{\xi, \rm{gauge}}$ related by $\rho_{\xi, \rm{gauge}} = 1/\sigma_{\xi, \rm{gauge}} \propto 1/\tau_{\xi, \rm{gauge}}$ within a similar slave-boson approach has been estimated in Ref. \cite{Nagaosa-1990-PRL}: } 
 \begin{align}
     \tau^{-1}_{\xi,\rm{gauge}} &\sim \left(\frac{T}{\chi_d}\right)^{4/3} \varepsilon_{F,\xi} 
      \sim  \left( \frac{T}{J} \right)^{4/3}  \frac{t^2 \delta^2}{D} ,
 \end{align}
 where $\chi_d \sim (m_f^*)^{-1} \sim J$ and the Fermi energy for the \added{$\xi$ band $\varepsilon_{F,\xi} \propto \delta \times (g^2/D^2) \times D \sim t^2 \delta^2 / D$.} Note that the estimate in Ref. \cite{Nagaosa-1990-PRL} is applicable here for the $\xi$-band since the coupling of the gauge-field to the fermion band  ($\xi$-band here) and the gauge-field propagator are of the same form as that in our model. The scattering rate $1/\tau_{\xi, \rm{gauge}}$ receives a sub-leading $T$-super-linear power-law correction $1/\tau_{\xi, \rm{gauge}} \sim \bar{a} \times (T/D)^{4/3}$  with a negligible prefactor  $\bar{a}/D \sim (4t^2/D^2) \times \delta^2 \gg 1$. For the relevant temperature range $0<T<300$ K, it is apparent that $1/\tau_{\xi,\rm{gauge}}$ is negligible compared to the leading Planckian scattering rate $1/\tau = \alpha_P k_B T/\hbar$ with $\alpha_P \sim O(1) $ from the hopping (effective Kondo) term:
\begin{align}
    \frac{\tau_{\xi, \rm{gauge}}^{-1}}{\tau_{\xi,g}^{-1}}\sim  \left( \frac{T}{D}\right)^{1/3} \left(\frac{t}{D}\right)^2  \delta^2 \ll 1 
\end{align}
for $0<T<300$ K, $D \sim  1$ eV, $\delta = 0.2 \sim 0.3$ and  $1/\tau_{\xi, g}  \sim  T $ (set $k_B = \hbar = 1$ here).


\subsection{Evaluation of $\Sigma_{f}$ and its scattering rate}
To evaluate the scattering rate for the $f$ spinon, we start by calculating its self-energy contributed from the perturbation of the hopping term, $H_t$, denoted as $\Sigma_{f}$:
\begin{align}
    \Sigma_{f}^{g}(\bm{k},\omega_{n})&=\left(-\frac{g^{2}}{N_{s}\beta}\right)\sum_{p_{m}}\sum_{\bm{p}}\mathcal{G}_{\xi}\left(p\right)\mathcal{G}_{b}\left(p-k\right) \nn
    &  =-\frac{g^{2}}{N_{s}\beta}\sum_{p_m}\sum_{\bm{p}}\frac{1}{ip_{\xi}-\xi_{\bm{p}}-i\Sigma_{\xi}^{\prime\prime}(ip_{m})}
     \times \frac{1}{ip_{m}-i\omega_{n}-\lambda}.
    \label{eq:Sigma_g_f-scattering-rate}
\end{align}
Note that, due to the negative sign in front of the (k-dependent) dispersion $\xi_{\bm{k}}$, the imaginary part of the self-energy $\Sigma_{\xi}^{\prime\prime}$ for the $\xi$ field in the above equation will require an additional minus sign, which differs from $\Sigma_{\xi}$ evaluated via RPA. Similar situation also occurs for the calculation of $\Sigma_{\xi}$ in the RG section. Thus, we take $\Sigma_{\xi}^{\prime\prime}(ip_{m})=\pi g^{2}\rho_{0}\text{sgn}(p_{m})=\pi D\zeta\text{sgn}(p_{m})$. Due to the sign function in $\Sigma_{\xi}^{\prime\prime}$, Eq. (\ref{eq:Sigma_g_f-scattering-rate}) can be expressed as  $\Sigma^g_{f}(\bm{k},i\omega_{n})=\Sigma_{f}^{g,>} (\bm{k},i\omega_{n}) +\Sigma_{f}^{g,<} (\bm{k},i\omega_{n})$ where $\Sigma_{f}^{g,\gtrless}$ corresponds to the $p_{m} \gtrless 0$ branch in the Matsubara sum:
    \begin{align}
        \Sigma_{f}^{g,\gtrless} (\bm{k},\omega_{n}) 	& \equiv-\frac{g^{2}}{\beta}\sum_{p_{m}\gtrless0}\sum_{\bm{p}}\frac{1}{ip_{\xi}-\xi_{\bm{p}}\mp i\pi g^{2}\rho_{0}}\cdot\frac{1}{ip_{m}-i\omega_{n}-\lambda} \nn
	     & =-\frac{g^{2}}{\zeta\beta}\sum_{p_{m}\gtrless0}\sum_{\bm{p}}\frac{1}{ip_{m}+\left(\varepsilon_{\bm{p}}-\bar{\mu}_{\xi}\right)\mp iD\pi}\cdot\frac{1}{ip_{m}-i\omega_{n}-\lambda}.
      \label{eq:Sigma_f-RL}
    \end{align}

To analytically evaluate Eq. (\ref{eq:Sigma_f-RL}), we need to analyze the following contour integral, defined as 
\begin{align}
    \mathcal{S}_{F}^{\gtrless}	& =-\frac{g^{2}}{\zeta}\oint_{C_{>}}\frac{dz}{2\pi i}\,\frac{n_{F}(z)}{z+\left(\varepsilon_{\bm{p}}-\bar{\mu}_{\xi}\right)\mp iD\pi}  \times \frac{1}{z-i\omega_{n}-\lambda},
    \label{eq:S_F-RL}
\end{align}
where $C_{>}$ encircles the upper hemisphere of the $z$ plane counterclockwise where $p_{m}>0$. There are two approaches to evaluate this contour integral either by residue theorem and line integral along $C_{>}$. On one hand, $\mathcal{S}_{F}^{>}$ can be  evaluated by using the residue theorem, namely
    \begin{align}
        \mathcal{S}_{F}^{>}	= &   \frac{g^{2}}{\zeta}\Bigg[\frac{1}{\beta}\sum_{p_{m}>0}\frac{1}{ip_{m}+\left(\varepsilon_{\bm{p}}-\bar{\mu}_{\xi}\right)-iD\pi}\cdot\frac{1}{ip_{m}-i\omega_{n}-\lambda}-\frac{n_{F}(i\omega_{n}+\lambda)\Theta(\omega_{n})}{i\omega_{n}+\left(\varepsilon_{\bm{p}}-\bar{\mu}_{\xi}\right)+\lambda-iD\pi} \nn
	 &\qquad\qquad\qquad+\frac{n_{F}(-\varepsilon_{\bm{p}}+\bar{\mu}_{\xi}+iD\pi)}{i\omega_{n}+\varepsilon_{\bm{p}}-\bar{\mu}_{\xi}+\lambda-iD\pi}\Bigg] \nn
	= & \frac{g^{2}}{\zeta}\Bigg[\frac{1}{\beta}\sum_{p_{m}>0}\frac{1}{ip_{m}+\left(\varepsilon_{\bm{p}}-\bar{\mu}_{\xi}\right)-iD\pi}\cdot\frac{1}{ip_{m}-i\omega_{n}-\lambda}+\frac{n_{B}(\lambda)}{i\omega_{n}+\left(\varepsilon_{\bm{p}}-\bar{\mu}_{\xi}\right)+\lambda-iD\pi} \nn
	&\qquad\qquad\qquad+\frac{n_{F}(-\varepsilon_{\bm{p}}+\bar{\mu}_{\xi}+iD\pi)}{i\omega_{n}+\varepsilon_{\bm{p}}-\bar{\mu}_{\xi}+\lambda-iD\pi}\Bigg] \nn
	 = & \frac{g^{2}}{\zeta}\Bigg[\frac{1}{\beta}\sum_{p_{m}>0}\frac{1}{ip_{m}+\left(\varepsilon_{\bm{p}}-\bar{\mu}_{\xi}\right)-iD\pi}\cdot\frac{1}{ip_{m}-i\omega_{n}-\lambda}+\frac{n_{F}(-\varepsilon_{\bm{p}}+\bar{\mu}_{\xi}+iD\pi)}{i\omega_{n}+\varepsilon_{\bm{p}}-\bar{\mu}_{\xi}+\lambda-iD\pi}\Bigg].
    \end{align}
On the other hand, $\mathcal{S}_{F}^{>}$ can be alternatively analyzed via directly doing the line integral along  $C_{>}$ on the complex plane: first perform the integral along the real axis and then along the upper semicircle counterclockwise. The integrand  takes the approximate form $\sim\frac{1}{e^{\beta z}+1}\frac{1}{z^{2}}$. For the integral over $-\infty\leq|z|\leq 0$, integral along $[-\infty,0]$ gives $\mathcal{S}_{F}^{>}$ to be zero since the integrand is approximated as $\frac{1}{e^{\beta z}+1}\frac{1}{z^{2}}\to\frac{1}{z^{2}}$ when $T\to0$. While integrating over $0\leq|z|\leq\infty$, the integrand is approximately given by $\frac{1}{e^{\beta z}+1}\frac{1}{z^{2}}\to\frac{1}{e^{\beta z}}\frac{1}{z^{2}}$. It decays even faster than the previous one and thus the integral over this range is also zero. The second part of this integral is also zero as we take the radius of the semicircle to infinity. From the second approach, we have $\mathcal{S}_{F}^{>}=0$.

Combing these, we have
\begin{align}
    \left(-\frac{g^{2}}{\zeta\beta}\right) \sum_{p_{m}>0}\frac{1}{ip_{m}+\left(\varepsilon_{\bm{p}}-\bar{\mu}_{\xi}\right)-iD\pi}\times \frac{1}{ip_{m}-i\omega_{n}-\lambda} 	= \left(\frac{g^{2}}{\zeta}\right)\frac{n_{F}(-\varepsilon_{\bm{p}}+\bar{\mu}_{\xi}+iD\pi)}{i\omega_{n}+\varepsilon_{\bm{p}}-\bar{\mu}_{\xi}+\lambda-iD\pi}.
\end{align}
Thus, $\Sigma_{f}^{g,>}(\bm{k},\omega_{n}>0)$ reads
\begin{align}
     \Sigma_{f}^{g,>}(\bm{k},\omega_{n}>0) 
    & =-\frac{g^{2}}{\zeta\beta}\sum_{p_{m}>0}\sum_{\bm{p}}\frac{1}{ip_{m}+\left(\varepsilon_{\bm{p}}-\bar{\mu}_{\xi}\right)+iD\pi} 
    \frac{1}{ip_{m}-i\omega_{n}-\lambda} \nn
	& =\frac{g^{2}}{\zeta}\sum_{\bm{p}}\frac{n_{F}(-\varepsilon_{\bm{p}}+\bar{\mu}_{\xi}+iD\pi)}{i\omega_{n}+\varepsilon_{\bm{p}}-\bar{\mu}_{\xi}+\lambda-iD\pi}\nn
	& =\frac{g^{2}\rho_{0}}{\zeta}\int_{-D+\bar{\mu}_{\xi}}^{D+\bar{\mu}_{\xi}}d\varepsilon^{\prime}\frac{n_{F}(\varepsilon^{\prime}+iD\pi)}{i\omega_{n}-\varepsilon^{\prime}+\lambda-iD\pi},
\end{align}
 where we have make the following change  $-\varepsilon+\bar{\mu}_{\xi}=\varepsilon^{\prime}$. By absorbing $\lambda$ into $\omega$ and performing the analytic continuation, namely $i\omega_{n}\to\omega-\lambda+i\epsilon$ with $\epsilon \to 0^+ > 0 $, we arrive $(-D+\bar{\mu}_{\xi}<0)$ 
\begin{align}
    \Sigma_{f}^{g,>}(\omega>0)	=\frac{g^{2}\rho_{0}}{\zeta}\int_{-D+\bar{\mu}_{\xi}}^{D+\bar{\mu}_{\xi}}d\varepsilon^{\prime}\frac{n_{F}(\varepsilon^{\prime}+iD\pi)}{\omega-\varepsilon^{\prime}-iD\pi}.
\end{align}
	For $T\to0$, the Fermi function exhibits the following asymptotic result, 
\begin{align}
    n_{F}(\varepsilon^{\prime}+iD\pi)	=\begin{cases}
1, & \text{for }-D^{\prime}\leq\varepsilon\leq0\\
0, & \text{for }-D+\bar{\mu}_{\xi}\leq\varepsilon\leq D+\bar{\mu}_{\xi}.
\end{cases}
\end{align}
This gives rise to 
\begin{align} 
\left[\Sigma_{f}^{g,>}(\omega>0)\right]^{\prime\prime}  = & \frac{g^{2}D\pi\rho_{0}}{\zeta}\int_{-D+\bar{\mu}_{\xi}}^{0}\frac{d\varepsilon^{\prime}}{\left(\omega-\varepsilon^{\prime}\right)^{2}+(D\pi)^{2}} \nn
	= & \left(-\frac{g^{2}\rho_{0}}{\zeta}\right)\left[\tan^{-1}\left(\frac{\omega}{\pi D}\right)-\tan^{-1}\left(\frac{D-\bar{\mu}_{\xi}+\omega}{\pi D}\right)\right] \nn
	\approx & \left(-\frac{1}{\rho_{0}}\right)\left[\frac{\omega}{\pi D}-\tan^{-1}\left(\frac{1}{\pi}\right)\right],
\end{align}
where the assumption $D\gg\omega,\, \bar{\mu}_{\xi} \, \lambda$ has been made in the last line of the above equation. Finally, we reach
\begin{align}
    \left[\Sigma_{f}^{g,>}(\omega>0)\right]^{\prime\prime}=\alpha-\varsigma\omega.
\end{align}
with $\alpha=D\tan^{-1}\left(\frac{1}{\pi}\right)\approx D/3$ (where $\tan^{-1}\left(\frac{1}{\pi}\right) \approx 1/3$) and $\varsigma=\frac{1}{\pi}$.
Next, we turn to the $p_{m}<0$ branch of $\Sigma_f$. Following a similar approach, we need to analyze the following contour integral by the residue theorem,
\begin{align}
    \mathcal{S}_{F}^{<}&=-\frac{g^{2}}{\zeta}\oint_{C_{<}}\frac{dz}{2\pi i}\,\frac{n_{F}(z)}{z+\left(\varepsilon_{\bm{p}}-\bar{\mu}_{\xi}\right)+iD\pi}\cdot\frac{1}{z-i\omega_{n}-\lambda},
\end{align}
where $C_{<}$ encircles the lower hemisphere where $p_{m}<0$ of the complex $z$-plane counterclockwise. Below, we skip all the immediate steps and directly jump to the results. The final result of $\Sigma_{f}^{g,<}$ can be simply obtained by replacing $iD\pi$ with $-iD\pi$ in $\Sigma_{f}^{g,>}$. We finally have
\begin{align}
   \Sigma_{f}^{g,<}(\omega_{n}<0)&=\frac{g^{2}\rho_{0}}{\zeta}\int_{-D+\bar{\mu}_{\xi}}^{D+\bar{\mu}_{\xi}}d\varepsilon^{\prime}\frac{n_{F}(\varepsilon^{\prime}-iD\pi)}{i\omega_{n}-\varepsilon^{\prime}+\lambda+iD\pi},
\end{align}
leading to 
\begin{align}
   \left[\Sigma_{f}^{<\prime\prime}(\omega<0)\right]^{\prime\prime}&=\left(-\frac{g^{2}D\pi\rho_{0}}{\zeta}\right)\int_{-D+\bar{\mu}_{\xi}}^{0}\frac{d\varepsilon^{\prime}}{\left(\omega-\varepsilon^{\prime}\right)^{2}+(D\pi)^{2}}.
\end{align}
We therefore conclude 
\begin{align}
    \left[\Sigma_{f}^{g,<}(\omega<0)\right]^{\prime\prime} =-\left(\alpha-\varsigma\omega\right).
\end{align}

In summary, combining the results of $\Sigma_{f}^{g,\gtrless}$ above, the imaginary part of $\Sigma_f^g$ reads
\begin{align}
    \left[\Sigma_{f}^{g}(\omega)\right]^{\prime\prime}=-\text{sgn(\ensuremath{\omega})}\left(\alpha-\varsigma\omega\right).
    \label{eq:sigma_f-img-scatteringrate}
\end{align}
Due to the constant term in $\left[\Sigma_{f}^{g}(\omega)\right]^{\prime\prime}$ of Eq. (\ref{eq:sigma_f-img-scatteringrate}), we note that, although our Kondo-Heisenberg-like lattice model exhibits translational symmetry, various fields that are local at the bare level still exhibit local characteristics even after perturbation corrections. This implies that the translational symmetry of this model is still broken, analogous to the single-impurity problem. 

\subsection{Scattering rate for the spinon-holon bound field $\xi$}
In this section, we calculate the resistivity contributing from the $\xi$ field: The self-energy of the $\xi$ field is given by 
\begin{align}
    \Sigma_{\xi}(ik_{n})	& =\left(\frac{\bar{g}^{2}}{N_{s}\beta}\right)\sum_{p}G_{f}(p)\mathcal{G}_{b}(k-p) \nn
	& =\left(\frac{\bar{g}^{2}}{\beta}\right)\sum_{ip_{m}}\left(\frac{1}{N_{s}}\sum_{\bm{p}}G_{f}(p)\right)\mathcal{G}_{b}(k-p). 
 \label{eq:Sigma_xi}
\end{align}
Similarly, we also need to introduce an additional minus sign into $\Sigma_{\xi}$ here, as we did during the calculation of $\Sigma_{\xi}$ in the RG analysis, as depicted in Eq. (\ref{eq:Sigma_xi_RG}).  Here, $\Sigma_{\xi}$ can be justified to be local by iteration due to the local $b$ boson.

 To evaluate $\Sigma_{\xi}$, we first need to evaluate the dressed Green's function of the $f$ spinon, defined as $G_{f}(ip_{m}) = N^{-1}_s \sum_{\bm{p}} G_f (p)$. This can be explicitly computed as
 \begin{align}
     G_{f}(p_{m})	 = & \frac{1}{N_{s}}\sum_{\bm{p}}\frac{1}{ip_{m}-\varepsilon_{\bm{p}}-\Sigma_{f}(p_{m})} \nn
	= & \left(-\rho_{0}\right)\ln\frac{D-ip_{m}+\Sigma_{f}(p_{m})}{-D-ip_{m}+\Sigma_{f}(p_{m})} \nn 
	= & \left(-\rho_{0}\right)\ln\frac{D+\Sigma_{f}^{\prime}(p_{m})-ip_{m}+i\Sigma_{f}^{\prime\prime}(p_{m})}{-D+\Sigma_{f}^{\prime}(p_{m})-ip_{m}+i\Sigma_{f}^{\prime\prime}(p_{m})}\nn
	= & \left(-\rho_{0}\right)\ln\left[\frac{\sqrt{\left(1+\Sigma_{f}^{\prime}/D\right)^{2}+\left(p_{m}-\Sigma_{f}^{\prime\prime}\right)^{2}/D^{2}}}{\sqrt{\left(1-\Sigma_{f}^{\prime}/D\right)^{2}+\left(p_{m}-\Sigma_{f}^{\prime\prime}\right)^{2}/D^{2}}}e^{i(\theta_{1}-\theta_{2})}\right] \nn
	= & \left(-\rho_{0}\right)\ln\frac{\sqrt{\left(1+\Sigma_{f}^{\prime}/D\right)^{2}+\left(p_{m}-\Sigma_{f}^{\prime\prime}\right)^{2}/D^{2}}}{\sqrt{\left(1-\Sigma_{f}^{\prime}/D\right)^{2}+\left(p_{m}-\Sigma_{f}^{\prime\prime}\right)^{2}/D^{2}}} -i\rho_{0}(\theta_{1}-\theta_{2})
 \end{align}
 with
    \begin{align}
        &\theta_{1}=\tan^{-1}\left(\frac{-(p_{m}-\Sigma_{f}^{\prime\prime})}{D+\Sigma_{f}^{\prime}(p_{m})}\right), \nn
        &\theta_{2}=\tan^{-1}\left(\frac{-(p_{m}-\Sigma_{f}^{\prime\prime})}{-D+\Sigma_{f}^{\prime}(p_{m})}\right).
    \end{align}
where $\Sigma_{f}=\Sigma_{f}^{J}+\Sigma_{f}^{g}$ can be justified to be local. In the wide-band limit where $D\gg|p_{m}|,\,\Sigma_{f}^{\prime},\,\Sigma_{f}^{\prime\prime}$,  we have 
\begin{align}
    & \theta_{1}	\approx\tan^{-1}\left(\frac{-(p_{m}-\Sigma_{f}^{\prime\prime})}{D}\right)=\begin{cases}
2\pi, & p_{m}-\Sigma_{f}^{\prime\prime}>0\\
0 & p_{m}-\Sigma_{f}^{\prime\prime}<0
\end{cases}, \nn
\quad
&  \theta_{2}	\approx\tan^{-1}\left(\frac{-(p_{m}-\Sigma_{f}^{\prime\prime})}{-D}\right)=\begin{cases}
\pi, & p_{m}-\Sigma_{f}^{\prime\prime}>0,\\
\pi & p_{m}-\Sigma_{f}^{\prime\prime}<0.
\end{cases}
\end{align}
Thus, in the wide-band limit, we have $\theta_{1}-\theta_{2}=\pi\text{sgn}(p_{m}-\Sigma_{f}^{\prime\prime})$. Now we deal with the leading order approximation of $G_{f}$ in terms of small $\frac{(p_{m}-\Sigma_{f}^{\prime\prime})}{D}$.

For $\theta_{1}$ and $p_{m}-\Sigma_{f}^{\prime\prime}>0$, we may write $\theta_{1}=2\pi-\delta\quad\text{with }0<\delta\ll 1$. We thus have
\begin{align}
    \tan(2\pi-\delta)=\left(\frac{-(p_{m}-\Sigma_{f}^{\prime\prime})}{D}\right).
\end{align}
Using $\tan(A+B)=\left(\tan A+\tan B\right)/(1-\tan A\tan B)$, we reach $\tan(2\pi-\delta)=-\tan\delta\approx-\delta$, leading to $\delta=\frac{(p_{m}-\Sigma_{f}^{\prime\prime})}{D}$. We therefore obtain
\begin{align}
    \theta_{1}=2\pi-\frac{(p_{m}-\Sigma_{f}^{\prime\prime})}{D},\quad\text{for \ensuremath{p_{m}}-\ensuremath{\Sigma_{f}^{\prime\prime}}}>0.
\end{align}
On the other hand, for $\theta_{1}$ and $p_{m}-\Sigma_{f}^{\prime\prime}<0$, we may write $\theta_{1}=\delta\quad\text{with }0<\delta\ll 1$. We find 
\begin{align}
    \theta_{1}=\delta=-\frac{(p_{m}-\Sigma_{f}^{\prime\prime})}{D},\quad\text{for }\ensuremath{p_{m}}-\Sigma_{f}^{\prime\prime} <0.
\end{align}
Gathering the above results, we have
\begin{align}
    \theta_{1}\approx
    \begin{cases}
2\pi-\frac{p_{m}-\Sigma_{f}^{\prime\prime}}{D}, & \text{for } p_{m}-\Sigma_{f}^{\prime\prime}>0\\
-\frac{p_{m}-\Sigma_{f}^{\prime\prime}}{D} & \text{for } p_{m}-\Sigma_{f}^{\prime\prime}<0
\end{cases}.
\end{align}
Following  a similar approach, to evaluate $\theta_2$, we can express
\begin{align}
    \theta_{2}\approx
    \begin{cases}
\pi+\delta, & p_{m}-\Sigma_{f}^{\prime\prime}>0,\\
\pi-\delta & p_{m}-\Sigma_{f}^{\prime\prime}<0.
\end{cases}
\end{align}
Applying a similar way to find $\delta$, we reach
\begin{align}
    \theta_{2}=\pi+\frac{(p_{m}-\Sigma_{f}^{\prime\prime})}{D}.
\end{align}
Thus, to the leading order in $\frac{p_{m}-\Sigma_{f}^{\prime\prime}}{D},$ 
\begin{align}
     \theta_{1}-\theta_{2}=\pi\text{sgn}(p_{m}-\Sigma_{f}^{\prime\prime})-\frac{2(p_{m}-\Sigma_{f}^{\prime\prime})}{D}.
\end{align}
The real part of $G_{f}$ can be approximated as
\begin{align}
    G_{f}^{\prime}	(p_{m}) & =\left(-\rho_{0}\right)\ln\frac{\sqrt{\left(1+\Sigma_{f}^{\prime}(p_{m})/D\right)^{2}+\left(p_{m}-\Sigma_{f}^{\prime\prime}(p_{m})\right)^{2}/D^{2}}}{\sqrt{\left(1-\Sigma_{f}^{\prime}(p_{m})/D\right)^{2}+\left(p_{m}-\Sigma_{f}^{\prime\prime}(p_{m})\right)^{2}/D^{2}}} \nn
    & \approx\left(-\rho_{0}\right)\ln\frac{\sqrt{1+2\Sigma_{f}^{\prime}(p_{m})/D}}{\sqrt{1-2\Sigma_{f}^{\prime}(p_{m})/D}} \nn
   & \approx-\frac{2\rho_{0}}{D}\Sigma_{f}^{\prime}(p_{m}),
\end{align}
leading to 
 \begin{align}
     G_{f}(p_{m})	& \approx -\frac{2\rho_{0}}{D}\Sigma_{f}^{\prime}(p_{m})-i\pi \rho_{0}\text{sgn}\left[p_{m}-\Sigma_{f}^{\prime\prime}(p_{m})\right]+i\frac{2\rho_{0}}{D}\left[p_{m}-\Sigma_{f}^{\prime\prime}(p_{m})\right] \nn
	 & = -i\pi \rho_{0} \text{sgn} \left[p_{m}-\Sigma_{f}^{\prime\prime}(p_{m})\right]-\frac{2\rho_{0}}{D}\Sigma_{f}(p_{m})+\frac{2\rho_{0}}{D}ip_{m}.
  \label{eq:Gf-final}
 \end{align}
Utilizing the wide-band limit once more, where $|p_{m}|\ll D$, and the absence of an imaginary part for $\Sigma_{f}^{J}$, we have $\left(\Sigma_{f} (ip_{m})\right)^{\prime\prime} = \left(\Sigma_{f}^{g}(ip_{m})\right)^{\prime\prime}=\alpha-\varsigma\left|p_{m}\right|$, the sign function in $G_{f}(ip_{m})$ above can be approximated as 
\begin{align}
    \text{sgn}\left[p_{m}-\left(\Sigma_{f}^{g}(p_{m})\right)^{\prime\prime}\right]	\approx\text{sgn}\left(-\alpha\right)=-1,
\end{align}
since $\alpha \gg |p_{m}|>0$. Note that we do not include the contribution from the impurity scattering $A$ here in $\Sigma_{f}^{\prime\prime}$ since $A$ will be put by hand in $\Sigma_{f}^{\prime\prime}$ and thus will only be included at the end of the calculation when we want to evaluate the scattering rate for the $f$ spinon. Thus, $A$ should not be included in the subsequent calculation for $\Sigma_{\xi}$.  Applying $\rho_{0}=1/D$, $G_{f}$ in Eq. (\ref{eq:Gf-final}) becomes 
\begin{align}
    G_{f}(p_{m})&	= i\pi \rho_{0}-\frac{2\rho_{0}}{D}\Sigma_{f}(p_{m})+\frac{2\rho_{0}}{D}ip_{m} \nn
	& =i\pi \rho_{0}-\frac{2\rho_{0}}{D}\Sigma_{f}^{\prime}(p_{m})-i\frac{2\rho_{0}}{D}\Sigma_{f}^{\prime\prime}(p_{m})+\frac{2\rho_{0}}{D}ip_{m} \nn
	& =i\pi \rho_{0}-\frac{2\rho_{0}}{D}\Sigma_{f}^{\prime}(p_{m})-i\frac{2\rho_{0}}{D}\left(\frac{D}{3}-\frac{|p_{m}|}{\pi}\right)+\frac{2\rho_{0}}{D}ip_{m} \nn
	& \approx i\frac{5}{2}\rho_{0}-\frac{2\rho_{0}}{D}\Sigma_{f}^{\prime}(p_{m})+\frac{2\rho_{0}}{\pi D}\text{sgn(\ensuremath{p_{m}})}ip_{m}+\frac{2\rho_{0}}{D}ip_{m}.
\end{align}
where we have taken the approximation for $\pi-\frac{2}{3}\approx\frac{5}{2}$ in the last line of the above equation. Note that the last term in the above equation $\frac{2\rho_{0}}{D}ip_{m}$ is linearly proportional to frequency; thus we shall see that this contribution will vanish when we perform the integration over frequency.

Plugging the result of $G_{f}$ into $\Sigma_{\xi}$ of Eq. (\ref{eq:Sigma_xi}), we have
\begin{align}
  \Sigma_{\xi}(k_{n})&=\frac{\bar{g}^{2}}{\beta}\sum_{ip_{m}}\frac{i\frac{5}{2}\rho_{0}-\frac{2\rho_{0}}{D}\Sigma_{f}^{\prime}(ip_{m})+\frac{2\rho_{0}}{\pi D}\text{sgn(\ensuremath{p_{m}})}ip_{m}+\frac{2\rho_{0}}{D}ip_{m}}{ik_{n}-ip_{m}-\lambda} \nn
  & =-\frac{\bar{g}^{2}}{\beta}\sum_{ip_{m}}\frac{i\frac{5}{2}\rho_{0}-\frac{2\rho_{0}}{D}\Sigma_{f}^{\prime}(ip_{m})+\frac{2\rho_{0}}{D}(ip_{m})}{ip_{m}-ik_{n}+\lambda}-\frac{2\rho_{0}\bar{g}^{2}}{\pi D\beta}\sum_{ip_{m}}\frac{ (i p_{m}) \text{sgn}(p_{m})}{ip_{m}-ik_{n}+\lambda} \nn
    &=-\bar{g}^{2}n_{F}(ik_{n}-\lambda)\left[i\frac{5}{2}\rho_{0}-\frac{2\rho_{0}}{D}\Sigma_{f}^{\prime}(ik_{n}-\lambda)+\frac{2\rho_{0}}{D}\left(ik_{n}-\lambda\right)\right]-\frac{2\bar{g}^{2}\rho_{0}}{\pi D}\left(\frac{1}{\beta}\sum_{ip_{m}}\frac{ (ip_{m}) \text{sgn}(p_{m}) }{ip_{m}-ik_{n}+\lambda}\right) \nn
    & =-\bar{g}^{2}\left[i\frac{5}{2}\rho_{0}-\frac{2\rho_{0}}{D}\Sigma_{f}^{\prime}(ik_{n}-\lambda)+\frac{2\rho_{0}}{D}\left(ik_{n}-\lambda\right)\right]-\frac{2\bar{g}^{2}\rho_{0}}{\pi D}\left(\frac{1}{\beta}\sum_{p_{m}>0}\frac{ip_{m}}{ip_{m}-ik_{n}+\lambda}-\frac{1}{\beta}\sum_{p_{m}<0}\frac{ip_{m}}{ip_{m}-ik_{n}+\lambda}\right).
\end{align}
The summation over the $p_{m}>0$ and $p_{m}<0$ branches will be evaluated separately. For the $p_{m}>0$, we have 
\begin{align}
    \frac{1}{\beta}\sum_{p_{m}>0}\frac{ip_{m}}{ip_{m}-ik_{n}+\lambda}=(ik_{n}-\lambda)\Theta(k_{n})n_{F}(ik_{n}-\lambda)-\int_{-\infty}^{\infty}\frac{dx}{2\pi i}\cdot\frac{xn_{F}(x)}{x-ik_{n}+\lambda}.
    \label{eq:p-greater}
\end{align}
Following a similar approach, we have, for the $p_{m}<0$ branch,
\begin{align}
    -\frac{1}{\beta}\sum_{p_{m}<0}\frac{ip_{m}}{ip_{m}-ik_{n}+\lambda}=-(ik_{n}-\lambda)\Theta(-k_{n})n_{F}(ik_{n}-\lambda)+\int_{\infty}^{-\infty}\frac{dx}{2\pi i}\cdot\frac{xn_{F}(x)}{x-ik_{n}+\lambda}.
    \label{eq:p-lesser}
\end{align}
Combining Eqs. (\ref{eq:p-greater}) and (\ref{eq:p-lesser}), when $T\to0$, we obtain 
\begin{align}
    \frac{1}{\beta}\sum_{ip_{m}}\frac{\text{sgn(\ensuremath{p_{m}})}ip_{m}}{ip_{m}-ik_{n}+\lambda}=(ik_{n}-\lambda)\text{sgn}(k_{n})-\int_{-\infty}^{\infty}\frac{dx}{\pi i}\cdot\frac{xn_{F}(x)}{x-ik_{n}+\lambda},
\end{align}
leading to
\begin{align}
    \Sigma_{\xi}(k_{n})	& =-\bar{g}^{2}\left[i\frac{5}{2}\rho_{0}-\frac{2\rho_{0}}{D}\Sigma_{f}^{\prime}(ik_{n}-\lambda)+\frac{2\rho_{0}}{D}\left(ik_{n}-\lambda\right)\right]+\frac{2\rho_{0}\bar{g}^{2}}{\pi D}\left[i|k_{n}|-\lambda\text{sgn}(k_{n})\right] \nn
	& \qquad \qquad -\frac{2\rho_{0}\bar{g}^{2}}{\pi D}\int_{-\infty}^{\infty}\frac{dx}{\pi i}\cdot\frac{xn_{F}(x)}{x-ik_{n}+\lambda}.
\end{align}
In addition, using the results of $\left(\Sigma_{f}^{J} \right)^\prime=-J^{2}\rho_{0}D$ and $\left(\Sigma_{f}^{g}(k_{n}) \right)^\prime=\bar{g}^{2}\rho_{0}\ln\left(\frac{D}{k_{n}-\zeta^{-1}\varUpsilon}\right)$, we obtain the real and imaginary part for $\Sigma_{\xi}$, given by
\begin{align}
    \Sigma_{\xi}^{\prime}(k_{n})&=\frac{2\rho_{0}^{2}\bar{g}^{4}}{D}\ln\left(\frac{D}{k_{n}-\zeta^{-1}\varUpsilon}\right)-\frac{2\rho_{0}\bar{g}^{2}}{D}J^{2}+\frac{2\rho_{0}\lambda\bar{g}^{2}}{D}+\frac{2\rho_{0}\lambda\bar{g}^{2}}{\pi D}\text{sgn}(k_{n})+\frac{2\rho_{0}\bar{g}^{2}}{\pi D}\text{Re}\left(\int_{-\infty}^{\infty}\frac{dx}{\pi i}\cdot\frac{xn_{F}(x)}{x-ik_{n}+\lambda}\right),
\end{align}
and
\begin{align}
  \Sigma_{\xi}^{\prime\prime}(\omega+i\delta)=-\frac{5}{2}\bar{g}^{2}\rho_{0}-\frac{2\rho_{0}\bar{g}^{2}}{\pi^{2}D}\omega_{0}-\frac{2\rho_{0}\bar{g}^{2}}{\pi D}\left|\omega\right|-\frac{2\rho_{0}\bar{g}^{2}}{\pi^{2}D}\left(\omega-\lambda\right)\ln\frac{|\omega-\lambda|}{\omega_{0}}.
\end{align}
For $\lambda \gg \omega$, we may simply approximate $\ln\frac{|\omega-\lambda|}{\omega_{0}} \approx \ln\frac{\lambda}{\omega_{0}}$. In addition, the linear-in-$\omega$ term in the last term of $\Sigma_\xi^{\prime\prime}$ above can be shown to vanish when we calculate the electrical conductivity below (due to the symmetric boundary for the frequency integral). We therefore can neglect this term. Now,  $\Sigma_{\xi}^{\prime\prime}(\omega+i\delta)$ becomes
\begin{align}
    \Sigma_{\xi}^{\prime\prime}(\omega+i\delta) = -A - \varsigma |\omega|
    \label{eq:Im-Sigma-xi-final}
\end{align}
with $A = \frac{5}{2}\bar{g}^{2}\rho_{0} + \frac{2\rho_{0}\bar{g}^{2}}{\pi^{2}D}\omega_{0}-\frac{2\rho_{0}\bar{g}^{2}}{\pi^{2}D}\lambda \ln\frac{\lambda}{\omega_{0}} $ and $
\varsigma = \frac{2}{\pi}$ a constant here.

\subsection{The Planckian metal in DC-resistivity}

Having the result of scattering rate, we further compute  the electrical resistivity of our model in the DC limit (zero frequency) directly via Boltzmann formula under the assumption of temperature-independent effective mass and carrier concentration. Using  Eq. (\ref{eq:Im-Sigma-xi-final}) in the above section and the the relation $\Sigma_c = (1/2)\Sigma_\xi$, the relaxation time for the conduction $c$ electron shows
\begin{align}
   \frac{\hbar}{\tau_c(\omega)} = -2\Sigma_c^{\prime\prime} (\omega)=A + \varsigma |\omega|
\end{align} 
 The conductivity can be obtained as
 \begin{align}
     \sigma(T)	& =\left(-\frac{ne^{2}}{m^{\star}}\right)\int\tau_c(\omega)\frac{\partial n_{F}(\omega)}{\partial\omega}d\omega \nn
	 & \approx\left(\frac{ne^{2}\hbar}{m^{\star}}\right)\left[\int_{0}^{\infty}\frac{1}{A}\left(1+\frac{\varsigma\omega}{A}\right)\frac{\partial n_{F}(\omega)}{\partial\omega}d\omega  -\int_{-\infty}^{0}\frac{1}{A}\left(1+\frac{\varsigma\omega}{A}\right)\frac{\partial n_{F}(\omega)}{\partial\omega}d\omega\right] 
	 \nn
	& =\frac{ne^{2}\hbar}{m^{\star}A}-\left(\frac{ne^{2}\hbar\varsigma Y}{m^{\star}A^{2}}\right)k_{B}T,
 \end{align}
where $Y\equiv \int_{-\infty}^{\infty}\frac{|x|e^{x}}{(e^{x}+1)^{2}}dx \approx 1.39$ is a constant. Thus, the resistivity at sufficiently low temperatures can be approximated as 
\begin{align}
    \rho(T)	& =\frac{1}{\sigma(T)}=\frac{m^{\star}A}{ne^{2}\hbar}\left(\frac{1}{1+\left(\frac{\varsigma Y}{A}\right)k_{B}T}\right)
	\nn
	& \approx\frac{m^{\star}A}{ne^{2}\hbar}+\frac{m^{\star}\varsigma Y}{ne^{2}\hbar}k_{B}T.
\end{align}

The linear-in-temperature term of resistivity $\rho_{L}(T)$ shows 
\begin{align}
    \rho_{L}(T)=\frac{m^{\star}}{ne^{2}\hbar}(\varsigma Yk_{B}T)\equiv\frac{m^{\star}}{ne^{2}\hbar}\left(\alpha_{P}k_{B}T\right),
\end{align}
where coefficient $\alpha_{P}$ defined above is $\alpha_{P}=\varsigma Y=1.39 \times \frac{2}{\pi}\approx 0.9$, indicating the Planckian scattering rate.

We find that the \textit{T}-linear resistivity comes as a consequence of the isotropic electronic states. Though the self energy diagram, leading to the electron scattering rate, depends on $g^2$, this coupling constant dependence in electron scattering rate is eventually canceled by the same factor in the Green's function of the composite fermionic spinon-holon state. As a result, the system remains a Planckian state in transport in this phase without knowing that the coupling constant $t$ is flowing to an irrelevant fixed point at U(1) FL$^*$. Similar cancellation in transport rate has been found in the SYK model \cite{Patel-PRL-SM,Patel-2023-SYK-Sci}.

\subsection{Estimation of scattering rate with slightly anisotropic Fermi surface}
In this section, we estimate the change of the scattering rate in the Planckian strange metal state if the Fermi surface includes a small anisotropic part, as shown in Supplementary Figure \ref{fig:aniso-fs}.  We find that the contribution to the scattering rate from the small anisotropic part of the Fermi surface is negligible.  

For small anisotropy of the dispersion (or the Fermi surface), we assume that the anisotropic band structure starts to deviate significantly relative to the isotropic one at an energy $-\delta D < 0$ below the Fermi energy, where $\delta D$ is assumed to obey $\delta D/ D \ll 1$ to satisfy the small anisotropy condition. The small anisotropic part is also affected by a shift, $\delta k_F$, in the Fermi wave vector $k_F$.  Due to the additional anisotropic portions, the density of states at the Fermi surface is subjected to a small change,
\begin{align}
    \rho_0 \to \rho^\prime  (\varepsilon) = \rho_0 + \delta \rho (\varepsilon).
\end{align}
Here, $\delta \rho (\varepsilon) \propto \delta k_F$ denotes the change in the density of states at the Fermi surface with respect to $\rho_0$.  $\Sigma_{f}^{>}$ with tiny anisotropic Fermi surface can then be expressed as
\begin{align}
  \left[ \Sigma_{f}^{>}(\omega_{n}) \right]^\prime& =\frac{g^{2}}{\zeta} \sum_{\bm{p}} \frac{n_{F}(\varepsilon^{\prime}_{\bm{p}}+iD\pi)}{i\omega_{n}-\varepsilon^{\prime}_{\bm{p}}+\lambda-iD\pi} \nn
   & = \frac{g^{2}}{\zeta}\int_{-D+\bar{\mu}_{\xi}}^{0}  \frac{\rho^\prime (\varepsilon)d\varepsilon}{i\omega_{n}-\varepsilon+\lambda-iD\pi},
\end{align}
where, in general, the density of states $\rho (\varepsilon)$ contains angular dependence, 
\begin{align}
    \rho (\varepsilon) = 4\times \int_{\varepsilon = \varepsilon_{\bm{k}}} \frac{d\theta_{\bm{k}}}{(2\pi)^2} \frac{1}{\left| \nabla_{\bm{k}}\varepsilon_{\bm{k}} \right|}
\end{align}
with $\theta_{\bm{k}}$ being the angle between $k_x$ and $k_y$ for the two-dimensional Brillouin zone. The change of $\Sigma_{f}^{>}$ in the presence of an anisotropic Fermi surface relative to the isotropic one takes the form,
\begin{align}
    \delta \Sigma_{f}^{>} (\omega_n) & =\left[ \Sigma_{f}^{>}(\omega_{n}) \right]^\prime - \Sigma_{f}^{>}(\omega_{n})  \nn
    & \approx  \frac{g^{2}}{\zeta}\int_{-\delta D}^{0}  \frac{\rho^\prime (\varepsilon) d\varepsilon }{i\omega_{n}-\varepsilon+\lambda-iD\pi} -  \frac{g^{2}\rho_0}{\zeta}\int_{-\delta D}^{0}  \frac{d\varepsilon}{i\omega_{n}-\varepsilon+\lambda-iD\pi} \nn
    & = \frac{g^{2}}{\zeta}\int_{-\delta D}^{0}  \frac{\delta \rho (\varepsilon) d\varepsilon }{i\omega_{n}-\varepsilon+\lambda-iD\pi}.
\end{align}
For small $\delta D$ and performing the analytic continuation, we approximate $ \delta \Sigma_{f}^{>} (\omega +i\epsilon)$  as
\begin{align}
    \delta \Sigma_{f}^{>} (\omega +i\epsilon) \approx  \frac{g^{2}}{\zeta}  \frac{\delta \rho (0) \delta D }{\omega+\lambda-iD\pi} ,
\end{align}
and its imaginary part is given by
\begin{align}
    \left[\delta \Sigma_{f}^{>} (\omega +i\epsilon)\right]^{\prime\prime} =  \frac{g^{2} }{\zeta \pi}  \delta \rho(0) \frac{\delta D}{D}.  
\end{align}
We find that the change of $\Sigma_{f}^{>}$ due to an anisotropic Fermi surface is proportional to $ \frac{\delta D}{D}$, thus negligibly small. Following a similar approach, however, it can be demonstrated that the correction for $\Sigma_{f}^{<}$ includes an additional minus sign compared to $\Sigma_{f}^{>}$. Therefore, the complete leading-order  correction to $\left[ \Sigma_{f}\right]^{\prime\prime}$ due to a slightly anisotropic Fermi surface vanishes.

\begin{figure}[ht]
    \centering
    \includegraphics[width=0.6\textwidth]{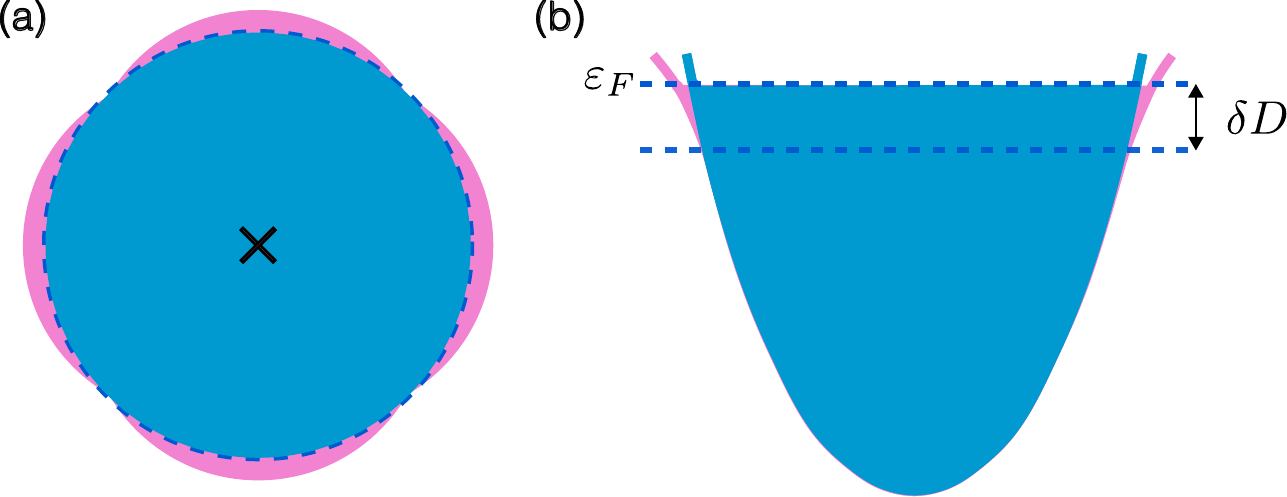}
    \caption{(a) A band with slightly anisotropic dispersion is composed of an isotropic region (blue), with the blue dashed line representing the isotropic Fermi surface,  and four crescent-shaped anisotropic portions (pink areas). (b) Schematic plots of the band with isotropic Fermi surface (blue, associated with the Fermi surface of blue region in (a)) and the slightly anistropic one (pink, associated with the Fermi surface of pink region in (a)). The band with slightly anisotropic Fermi surface starts to deviate from the blue one at energy $-\delta D$ below the Fermi energy $\varepsilon_F$.}
    \label{fig:aniso-fs}
\end{figure}


\subsection{Contribution from vertex correction to scattering rate}

\begin{figure}[ht]
    \centering
    \includegraphics[width=0.8\textwidth]{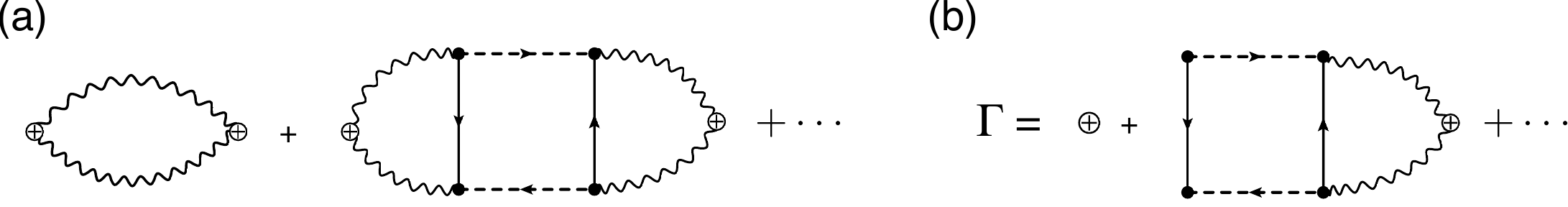}
    \caption{Diagrammatic representation of the first two leading terms for (a) the  current-current correlator and  (b) the current vertex for the $\xi$ fermion of our model. Right figure of (a) represents the bare current-current correlator of the $\xi$ fermion and the left figure shows its leading-order correction.}
    \label{fig:vertex-JJ}
\end{figure}
We estimate the significance of the vertex correction to the electrical conductivity (or equivalently the scattering rate) from the $\xi$ fermion (twice of the contribution of the conduction $c$ electron) in the Planckian strange metal state. We find that this contribution is negligible, consistent with the earlier studies in Refs. \cite{George-NatComm-SM,Khurana-PRL-vertex,PRB-vertex-ignore}. The leading-order contribution of vertex correction to the current-current correlator is diagrammatically depicted in Supplementary Figure (\ref{fig:vertex-JJ}). 
Without tackling the awkward diagrammatic calculation of the vertex correction $\Gamma_x$, here we use a simpler power-counting to evaluate how significant the contribution of $\Gamma_x$ is. The rule for the power-counting in the wide-band limit can be demonstrated as follows: 
\begin{align}
 \bar{g} = \frac{g}{\sqrt{\zeta}} \propto D, \quad \frac{1}{V}\sum_{\bm{k}} =\rho_0 \int d\varepsilon \propto D^0, \quad \frac{1}{\beta}\sum_{\omega_n}\propto D^0 ,\quad \mathcal{G}_f, \, \bar{\mathcal{G}}_\chi \propto D^{-1}, \quad \mathcal{G}_b \propto D^0.
\end{align}
The validity of the above set of rules for power-counting can be confirmed by examining $\Sigma_f^{\prime\prime} \sim \bar{g}^2 \mathcal{G}_\xi\mathcal{G}_b\propto D^2 
\times \frac{1}{D} = D$ which is consistent with our calculation where the dominant term for $\Sigma_f^{\prime\prime}$ is proportional to $\alpha \propto D$ as shown in Eq. (\ref{eq:sigma_f-img-scatteringrate}). 

When considering vertex correction, the dc-conductivity $\sigma_{\xi}$ from the $\xi$ fermion can be computed by [see, e.g., Ref.  \cite{flensberg}] 
\begin{align}
    \sigma_{\xi}&=2\times\frac{e^{2}}{2\pi}\frac{1}{V}\sum_{\bm{k}}\Gamma_{0}(\bm{k},\bm{k})\left[\bar{G}_{\xi}^{A}(\bm{k},0)\bar{G}_{\xi}^{R}(\bm{k},0)\Gamma^{RA}(\bm{k},\bm{k};0,0)-\bar{G}_{\xi}^{A}(\bm{k},0)\bar{G}_{\xi}^{R}(\bm{k},0)\Gamma^{AA}(\bm{k},\bm{k};0,0)\right],
    \label{eq:conductivity-JJ-correlator}
\end{align}
where $\bar{G}_\xi$ denotes the dressed Green function for the $\xi$ field after normalization, as described in Section II. Here, $R \,(A)$ represents ``retarded" (``advanced").

Powering counting for the leading-order vertex correction shown in Supplementary Figure (\ref{fig:vertex-JJ})  can be obtained as 
\begin{align}
    \Gamma  & \sim \bar{g}^4 \mathcal{G}_b^2 \, \mathcal{G}_f^2 \,  \bar{\mathcal{G}}_\xi^2 \nn
    & \propto D^4 \times \frac{1}{D^2} \times \frac{1}{D^2} = D^0.
\end{align}
Substituting the above result into Eq. (\ref{eq:conductivity-JJ-correlator}), the conductivity from the contribution of vertex correction of Supplementary Figure (\ref{fig:vertex-JJ}) can be estimated as
\begin{align}
    \sigma_\xi \sim \frac{ n e^2 }{m_\xi} \left( \bar{G}_\xi^2  \times \Gamma \right) 
     \propto \frac{ n e^2 }{m_\xi} \left( \frac{1}{D^2}  \times D^0\right) ,
\end{align}
where $m_\xi$ coming from the bare current vertex of $\xi$, $\Gamma_0 \propto \frac{1}{m_\xi}$, of Eq. (\ref{eq:conductivity-JJ-correlator}) denotes the effective mass for $\xi$ and $n$ represents the particle density of $\xi$ which is estimated via $\sum_{\bm{k}}/V \propto n$. This leads to the following power-law-in-bandwidth dependence for the conductivity from the $\xi$ fermion,
\begin{align}
      \sigma_\xi \sim \frac{ n e^2 }{m_\xi} \left( \frac{1}{D^2} \right),
\end{align}
indicating that the vertex correction of Supplementary Figure \ref{fig:vertex-JJ}(b) contributes a term proportional to $1/D^2$ to the transport time. 

Together with the contribution from the self-energy (lifetime) of the $\xi$ field, denoted as $\Sigma_\xi^{\prime\prime}$ in Eq. (\ref{eq:Im-Sigma-xi-final}), the conductivity in the wide-band limit can be approximated by
\begin{align}
    \sigma_\xi & \sim \frac{ n e^2 }{m_\xi} \left( \frac{1}{D} + \frac{1}{D^2} \right) = \frac{ n e^2 }{m_\xi} \frac{1}{D} \left[1 + O\left(\frac{1}{D}\right)\right] \nn
    &  \approx  \frac{ n e^2 }{m_\xi} \frac{1}{D}.
\end{align}
We thus conclude that, in the wide-band limit, the contribution of the vertex correction to the conductivity is vanishingly small compared to that of the self-energy and can be neglected, consistent with Refs. \cite{George-NatComm-SM,Khurana-PRL-vertex,PRB-vertex-ignore}.

\section{Universal scaling}
In this section, we will demonstrate the universal  scaling behavior of the electronic scattering  rate at $T=0$ and its generalization to the finite temperatures by referring to Refs. \cite{georges-multichannel-kondo-PRB,Georges-2021-PRR-seeback}. Notations below are mostly adopted from Refs. \cite{georges-multichannel-kondo-PRB,Georges-2021-PRR-seeback}.

We start from the equation below for the spectral representation of correlation function \cite{georges-multichannel-kondo-PRB},
\begin{align}
    G_{\psi}(\tau)=-\int_{-\infty}^{\infty}\frac{e^{-\tau \varepsilon}}{1+e^{-\beta\varepsilon}}A_{\psi}(\varepsilon)d\varepsilon,
\label{eq:seeback-spectral}
\end{align}
where $0\leq\tau\le\beta$ and $A_\psi (\varepsilon) = (-1/\pi) G^{\prime\prime}_\psi( \omega+i0^+)$ denotes the spectral function for the $\psi$ field. When a system has conformal invariance, any fermionic correlation function, such as the self-energy, in the imaginary-time domain can be expressed as \cite{Georges-2021-PRR-seeback}
\begin{align}
    \Sigma(\tau)\propto e{}^{\alpha(\tau/\beta-1/2)}\left[\frac{\pi/\beta}{\sin\left(\pi\tau/\beta\right)}\right]^{1+\nu},
    \label{eq:seeback-self-energy-1}
\end{align}
where $\alpha$ and $\nu$ are constants. Here, $\alpha$ is a measure of particle-hole asymmetry. Its spectral representation is given by
\begin{align}
     e^{\alpha(\tau/\beta-1/2)}\left[\frac{\pi}{\sin\left(\pi\tau/\beta\right)}\right]^{1+\nu}=C_{\alpha,\nu}\int_{-\infty}^{\infty}dx\frac{e^{-x\tau/\beta}}{1+e^{-x}}g_{\alpha,\nu}(x) 
     \label{eq:seeback-self-energy-2}
\end{align}
with
\begin{align}
    g_{\alpha,\nu}(x) & =\left|\Gamma\left(\frac{1+\nu}{2}+i\frac{x+\alpha}{2\pi}\right)\right|^2\frac{\cosh(x/2)}{\cosh(\alpha/2)\Gamma[(1+\nu)/2]^{2}}, \nn
     C_{\alpha,\nu} & =\frac{(2\pi)^{\nu}\cosh(\alpha/2)\Gamma\left[\frac{1+\nu}{2}\right]^{2}}{\pi\Gamma[1+\nu]}.
     \label{eq:seeback-g}
\end{align}
Following Eqs. (\ref{eq:seeback-spectral})-(\ref{eq:seeback-g}), we will generalize $\Sigma_\xi$ that we obtain in the zero-temperature limit in the previous section  to the finite-temperature one. We will further derive its $\omega/T$ scaling behavior once the finite-temperature generalization  is acquired. 

Applying Eq. (\ref{eq:seeback-spectral}), we first evaluate the (imaginary) time dependence of $ \Sigma_{\xi}$ which shows a linear-in-frequency dependence in the frequency domain, $\Sigma_\xi \propto |\omega|$. It can be shown that $\Sigma_{\xi}$ exhibits a $\tau^{-2}$ dependence:
\begin{align}
    \Sigma_{\xi}(\tau)	= & \frac{1}{\pi}\int_{-\infty}^{\infty}\frac{|\varepsilon|e^{-\tau\varepsilon}}{1+e^{-\beta\varepsilon}}d\varepsilon \nn 
	= & \frac{1}{\pi}\left[\int_{0}^{\infty}|\varepsilon|e^{-\tau\varepsilon}+\int_{-\infty}^{0}|\varepsilon|e^{(\beta-\tau)\varepsilon}\right]d\varepsilon \nn
	=& \frac{1}{\pi}\left[\frac{1}{\tau^{2}}\int_{0}^{\infty}|x|e^{-x}dx+\frac{1}{(\beta-\tau)^{2}}\int_{-\infty}^{0}|x|e^{x}dx\right] \nn
	= & \frac{1}{\pi \tau^{2}}.
\end{align}
It implies that this situation corresponds to the case of $\alpha = 0$ (particle-hole symmetric) and $\nu=1$ (Planckian) as discussed in Ref \cite{georges-multichannel-kondo-PRB,Georges-2021-PRR-seeback}. The second term of the above equation vanishes when $T \to 0$.

Once $\alpha=0$ and $\nu = 1$  are decided, we can  generalize $\Sigma_{\xi}$ to the finite-temperature region by conformal transformation governed by Eqs. (\ref{eq:seeback-spectral})-(\ref{eq:seeback-g}), leading to the following expression for $\Sigma_{\xi}(\omega,T)$,
\begin{align}
    \Sigma_{\xi}^{\prime\prime}(\omega,T)=\lambda_{0}\beta^{-1}g_{1,0}(x)=\frac{\lambda_{0}}{2}\omega\coth\left(\frac{\omega}{2T}\right)
    \label{eq:Sigma_xi-img-generalscaling}
\end{align}
with $\lambda_0$ being an unknown constant. $\lambda_0$ can be determined from its zero-temperature limit, where $\Sigma_{\xi}^{\prime\prime}(\omega,T=0) = -(2/\pi) |\omega| $. We thus find $\lambda_0 = -4/\pi$, leading to 
\begin{align}
     \Sigma_{\xi}^{\prime\prime}(\omega,T)=\lambda_{0}\beta^{-1}g_{1,0}(x)=-\frac{2}{\pi}\omega\coth\left(\frac{\omega}{2T}\right).
     \label{eq:Sigma_xi-img}
\end{align}
Using the relation of $\Sigma_c = (1/2)\Sigma_\xi$, the  scattering rate for the conduction $c$ electron can be obtained by $\hbar/\tau_c = -2\Sigma_c^{\prime\prime} (\omega)$ (here we restore $\hbar$ and $k_B$), $\frac{\hbar}{\tau_c} =\frac{2}{\pi}\hbar \omega\coth\left(\frac{\hbar \omega}{2k_B T}\right)$. It shows the following frequency-to-temperature scaling  behavior
\begin{align}
     \frac{\hbar/\tau_c}{k_B T} =\frac{2}{\pi}x\coth\left(\frac{x}{2}\right),
\end{align}
where $x\equiv \hbar \omega/k_B T$. The results of the above scaling relation for electronic scattering rate is plotted in Fig. 4(a) of the main text. In the high-frequency, low-temperature limit $x \gg 1$, the scattering rate divided by $k_B T$ shows a universal scaling behavior, $\frac{\hbar/\tau_c}{k_B T} \approx (2/\pi)x$. Conversely, in the DC-limit ($x \to 0$), the scattering rate manifests the Planckian scattering rate, revealing a universal feature that is insensitive to microscopic coupling constants:
$1/\tau_c \approx \alpha_P k_B T/\hbar$ with $\alpha_P  = 8/\pi \approx 2.55$.

\subsection{Resistivity scaling and the equivalence of frequency and magnetic field in scaling regime}

In this section, we observe that the scaling behavior of the resistivity (derivative) at the strange-metal region for cuprate superconductor Tl2201 \cite{hussey-incoherent-nature-2021} at zero frequency but finite applied magnetic field \added{shows the same quantum critical scaling form as we find within our theory at a finite frequency but zero magnetic field. This implies an equivalent role played by frequency and magnetic field near quantum criticality associated with the strange metal state. The temperature-dependent AC-resistivity in the absence of magnetic fields is obtained by the inverse of AC-conductivity as: $\rho(\omega,T)=1/\sigma(\omega, T) = 1/\tau(\omega,T) \times (m^\star/ne^2)$. \added{We assume $m^\star/n$ is independent of temperature, frequency and doping in the Planckian state though $m^\star$ and $n$, in general, depend on these variables.} Using scaling form for electron scattering rate Eq. (3) of the main text, we obtain the scaling of the derivative of AC-resistivity $d\rho/d(\hbar \omega)$.} The results are shown in Supplementary Figures \ref{fig:fig:2j} and \ref{fig:fig:2iSB}. The blue dashed lines in Supplementary Figures \ref{fig:fig:2j} and \ref{fig:fig:2iSB} are fitted to the scaling function for $d\rho(\hbar \omega/k_B T)/d (\hbar \omega)$ shown below \cite{George-NatComm-SM}: 
\begin{subequations}
	\begin{align}
		\frac{ne^2}{m^\star}\times \frac{\pi}{2}\frac{\dd \rho \left(\frac{\hbar\omega}{4 k_B T}\right)}{\dd (\hbar\omega)}&=\coth \left( \frac{\hbar \omega}{4 k_B T}\right) -\frac{\hbar \omega}{4 k_B T}\rm{csch}^2 \left(\frac{\hbar\omega}{4 k_B T}\right) \;, \\
		\frac{ne^2}{m^\star} \times \frac{\pi}{2}  \frac{\dd \rho(X)}{\dd (\hbar\omega)}&=\coth(X) -X\rm{csch}^2(X) 
  \label{eq:d_rho_SB_X_b}
	\end{align}\label{eq:d_rho_SB_X}
\end{subequations}
with $X= \hbar \omega /4k_B T$, while the green ones are fitted to the following equations of the marginal Fermi liquid,
\begin{subequations}
	\begin{align}
	\frac{ne^2}{m^\star} \times \frac{\pi}{2}\frac{\dd \rho\left(\frac{\hbar\omega}{4 k_B T}\right)}{\dd (\hbar\omega)}&=\frac{\frac{\hbar\omega}{4k_B T}}{\sqrt{1+(\frac{\hbar\omega}{4k_B T})^2}} \;, \\
	\frac{ne^2}{m^\star} \times \frac{\pi}{2}\frac{\dd \rho[X]}{\dd (\hbar\omega)}&=\frac{X}{\sqrt{1+X^2}}\label{eq:d_rho_MFL_X_b} \np
	\end{align}\label{eq:d_rho_MFL_X}
\end{subequations}

\added{In Ref. \cite{hussey-incoherent-nature-2021}, the magnetoresistivity of overdoped cuprates Bi2201 and Tl2201 is well described by the MFL form: $\rho(H,T)= F(T)+ \sqrt{(\alpha k_B T)^2+(\gamma \mu_B \mu_0 H)^2}$ where $\alpha$ and $\gamma$ are constants insensitive to field, temperature and doping. The authors in Ref. \cite{hussey-incoherent-nature-2021} discovered that the derivative of $\rho(H,T)$ with respect to the magnetic field (upon proper normalization) shows quantum critical $H/T$-scaling over a finite-range in doping, signature of a quantum critical strange metal ``phase". As shown in Supplementary  Figures \ref{fig:fig:2j} and \ref{fig:fig:2iSB}, the data in Ref. \cite{hussey-incoherent-nature-2021} for $(1/\gamma) d\rho(\beta \mu_B H/T)/d(\mu_B H)$ exhibits a MFL scaling form $X/\sqrt{1+X^2}$ with $X=\beta\mu_B H/T(\beta = \gamma \mu_B/\alpha k_B)$ over a wide range in doping for two distinct cuprates (Bi2201 and Tl2201)  \cite{hussey-incoherent-nature-2021}. Strikingly, it is clear from Supplementary  Figures \ref{fig:fig:2j} and \ref{fig:fig:2iSB} as well as Eq. (\ref{eq:d_rho_MFL_X_b}) that this scaling function for  $(1/\gamma)d\rho(\beta \mu_B H/T)/d(\mu_B H)$ found in Ref. \cite{hussey-incoherent-nature-2021} is nicely reproduced  within our theory for the AC-resistivity (in the absence of magnetic fields) for $(\pi/2) d\rho(X)/d(\hbar \omega)$. This result indicates that in the scaling regime $\hbar \omega /4k_B T$ plays the role of $\beta \mu_B H/ T$, and prefactor $(m^\star/ne^2)\times (\pi/2)$ here plays the role of $\gamma$.}

\begin{figure}[h]
	\includegraphics[width=0.8 \textwidth]{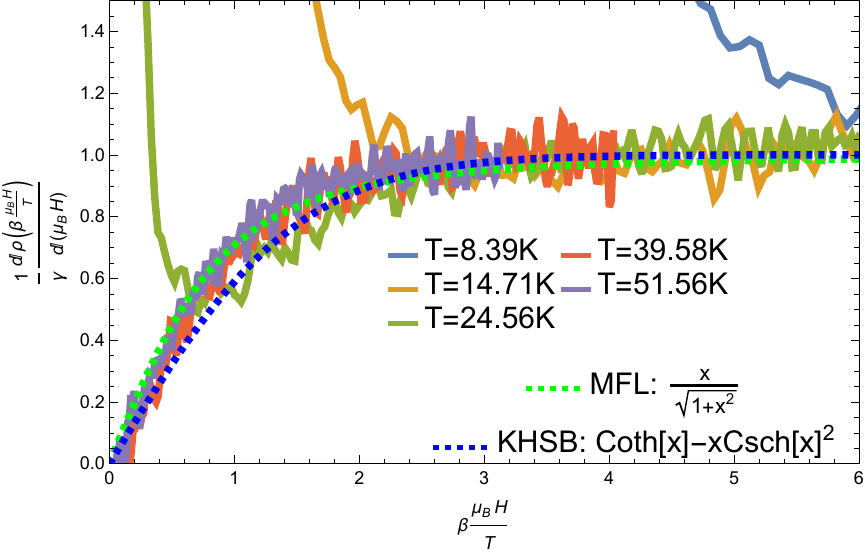}
	\caption{Derivative of the resistivity data for Tl2201 with $T_c=23~\rm{K}$ divided by $\gamma\mu_B$ plotted in function of $\frac{\beta\mu_0 H}{T}$. Here, $\beta \simeq 4.66$ and $\gamma \simeq 0.0049$ are fitting parameters. We also show the behavior of the Marginal Fermi-Liquid (MFL) form given in Eq. (\ref{eq:d_rho_MFL_X}) as well as the analytic form via our Kondo-Heisenberg approach to the slave-boson  \textit{t-J} model (KHSB)  [Eq. (\ref{eq:d_rho_SB_X_b})]. We note that both curves nicely describe the experimental data.  The resistivity data shown here is reproduced from Ref. \cite{hussey-incoherent-nature-2021}.}
	\label{fig:fig:2j}
\end{figure}

\begin{figure}[h]
 	\includegraphics[width=0.8 \textwidth]{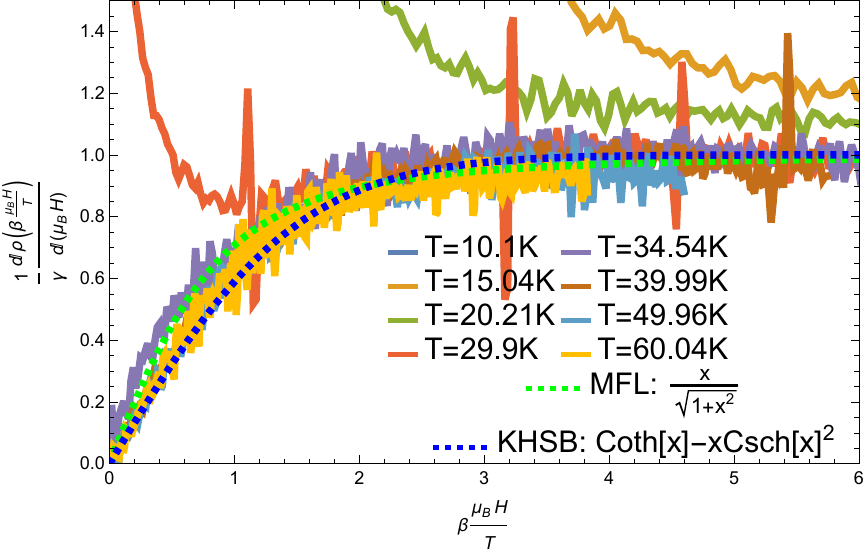}
 	\caption{Derivative of the resistivity data divided by $\gamma\mu_B$ plotted in function of $\frac{\beta\mu_0 H}{T}$. For the sample  Tl2201 with $T_c=26.5~\rm{K}$ \cite{hussey-incoherent-nature-2021}, for  $\beta \simeq 6.6$ and $\gamma \simeq 0.00562$. We note that the KHSB model [Eq. (\ref{eq:d_rho_SB_X_b})] nicely describes the experimental data. }
 	\label{fig:fig:2iSB}
 \end{figure}

 \subsection{Mass enhancement}
According to Ref. \cite{George-NatComm-SM} and its supplementary material,  the mass enhancement is given by
\begin{equation}
	\frac{m^{\ast}(\omega, T)}{m}-1\simeq -\frac{2}{\hbar\omega}\left[\Sigma^{\prime}\left(\frac{\hbar\omega}{2}\right)-\Sigma^{\prime}(0,T)\right]\np
\end{equation}
This implies the knowledge of the real part of the electrons' self energy, which can be usually be obtained by using Kramers-Kroning relations
\begin{align}
	\Sigma^{\prime}_{c}(\omega, T)&=\frac{1}{\pi}\mathcal{P}\int \frac{ \Sigma''_{c}(\omega_1)}{\omega_1-\omega} d \omega_1\label{eq:S10_KK}.
\end{align}
 Unfortunately, within our KHSB model, the imaginary part of self energy obtained from Eq. (\ref{eq:Sigma_xi-img}) and  \mbox{$\Sigma^{\prime}_{c}(\omega, T)=\frac{1}{2}\Sigma^{\prime}_{\xi}(\omega, T)$} yielding 
\begin{equation}
 \Sigma^{\prime\prime}_{c}(\omega, T)=-\frac{\omega}{\pi}\coth \left( \frac{\omega}{4 k_B T}\right) \label{eq:S2_def1}
\end{equation}
 is not holomorphic. As s result, an analytic derivation of the real part of the self energy is not possible. As explained in Ref. \cite{George-NatComm-SM}, this is overcome by considering a similar yet different scaling function $f(x)=\norm{x}+2\exp{\frac{\norm{x}}{2}}$ instead of  $f(x)=\frac{x}{2}\coth(\frac{x}{2})$ with $x=\frac{\hbar\omega}{k_BT}$ allows to evaluate the real part of self energy. The resulting real part of self energy  leads to a scaling of the mass enhancement $m^*/m-1 \propto g(x)$ with $g(x)$ taking the following form [see Eq. (S16) in Ref. \cite{George-NatComm-SM}]: 
 \begin{equation}
	g(x)=2g\left( 1-\gamma \ln(\frac{x}{4})+\frac{2}{x}\left[\exp{\frac{x}{4}}\textrm{Ei}(-\frac{x}{4})-\exp{-\frac{x}{4}}\textrm{Ei}(\frac{x}{4})\right]\right)\;,\label{eq:g_x}
\end{equation}
where $\gamma=0.577$ is Euler's constant, $g$ is a constant fitting parameter, and $\textrm{Ei}(x)$ is the exponential integral function.

 Now, we compare our scaling function in the imaginary part of self energy with that shown in Ref. \cite{George-NatComm-SM} and discuss the correspondence between  the fine-tuned prefactor $g$ of the scaling function in Ref. \cite{George-NatComm-SM} and the universal prefactor in the scaling function of our approach. Following the notations of  Ref. \cite{George-NatComm-SM}, the imaginary part of self energy takes the form 
 \begin{equation}
 	\Sigma^{\prime\prime}_{c}(\omega, T)=-\pi g k_BT f\left(\frac{\hbar\omega}{k_BT}\right) \; .\label{eq:S2_def2-2}
 \end{equation}
 A direct comparison of Eq. (\ref{eq:S2_def1}) and  Eq. (\ref{eq:S2_def2-2}) gives 
 \begin{equation}
 g=\frac{2}{\pi^2}\simeq0.203\;,
 \end{equation}
within our KHSB model, while within the work presented in Ref. \cite{George-NatComm-SM} $g$ is a fitting parameter evaluated to $g=0.23$.  

In Fig. 4b of the main text, we plotted the  mass enhancement scaling function $g(x)$ on top of the data set reproduced from Ref. \cite{George-NatComm-SM}. The dashed red line corresponds to the fit of Ref. \cite{George-NatComm-SM} with $g=0.23$, and the solid black line corresponds to the prediction for $g(x)$ within the KHSB model with $ g=\frac{2}{\pi^2}$.

\section{The spectral weight}
In this section, we present our results on single-electron spectral weight via RG renormalized perturbative approach to our Kondo-Heisenberg formulated slave-boson \textit{t-J} model. We further compare our results with the recent ARPES measurements on the overdoped cuprates in Ref. \cite{zxshen-sicience-incoherent-cuprate}. Excellent agreement between our theoretical predictions and the experimental observations is achieved.

\subsection{Imaginary part of the self energy at zero temperature}
We assume the dispersion of the electrons to be given by a tight-binding model with nearest $t$ and second--nearest $t'$ hopping parameters on the square lattice, same as that used in Ref. \cite{zxshen-sicience-incoherent-cuprate}:
\be
\varepsilon(\bm{k})=-2t(\cos(k_x)+\cos(k_y))-4t'\cos(k_x)\cos(k_y)\;. \label{eq:epsilon_def}
\ee
The zero-temperature imaginary part of the self energy for the $\xi$ field is given by Eq. (\ref{eq:Im-Sigma-xi-final}), using $\Sigma_c^{\prime\prime}(\omega) = \Sigma_\xi^{\prime\prime}/2$, we obtain
\be
\Sigma^{\prime\prime}_c(\omega)=-\frac{1}{\pi}\norm{\omega}+\Gamma\label{eq:S20_def2}\; .
\ee
The real part of the self energy is given by the Kramers-Kroning relations
\begin{align}
\Sigma^{\prime}_c(\omega) =\frac{1}{\pi}\mathcal{P}\int \frac{ \Sigma''_c(\omega_1)}{\omega_1-\omega}\dd \omega_1
=\frac{2}{\pi^2}\omega\ln \left(\frac{\norm{\omega}}{D}\right)\; ,\label{eq:S10_def}
\end{align}
where $D\gg\omega$ is a large cutoff.

\subsection{Imaginary part of the self energy at finite temperatures}
At finite temperature, the imaginary part of the self energy is given by Ref. \cite{George-NatComm-SM} [cf. Eqs. (\ref{eq:S2_def1}) and (\ref{eq:S2_def2-2})]
\be
\Sigma^{\pp}_c(\omega,T)=-\pi g k_B T f\left(\frac{x}{2}\right)+\Gamma,
\label{eq:S2_def}
\ee
where $f(\frac{x}{2})\equiv \frac{x}{2}\coth(\frac{x}{4})$ and $x=\frac{\hbar \omega}{k_B T}$. In our case of the slave-boson approach to the \textit{t-J} model, the parameter $g$ corresponds to $g=\frac{2}{\pi^2}$, leading to
\be
\Sigma^{\prime\prime}_c(\omega, T)=-\frac{\omega}{\pi}\coth\left(\frac{\omega}{4 k_B T}\right)+\Gamma, 
\label{eq:S2_def2}
\ee
which is equivalent to Eq. (\ref{eq:S2_def1}) supplemented by a constant term $\Gamma$.

The real part is given by the  Kramers–Kronig  relations 
\be
\Sigma^{\prime}_c(\omega,T)=\frac{1}{\pi}\mathcal{P}\int \frac{ \Sigma^{\prime\prime}_c(\omega_1,T)}{\omega_1-\omega}\dd \omega_1\; .\label{eq:S1_KK}
\ee
Because $f(x)$ is not a holomorphic function, the exact analytical integration in Eq. (\ref{eq:S1_KK}) is not possible. However, one can evaluate it numerically. Meanwhile, we note that $\Sigma''_c(\omega)$ in Eq. (\ref{eq:S2_def2}) is well approximated by marginal Fermi liquid-like (MFL) form
\be
\Sigma_{MFL}^{\prime\prime}(\omega,T)=-g\frac{\pi}{2}\sqrt{\omega^2+(\pi  k_B T)^2}+\Gamma\; ,\label{eq:S2_MFL1}
\ee
but with different prefactors:
\be
\Sigma_{c}^{\prime\prime}(\omega,T)\approx -\frac{1}{\pi}\sqrt{\omega^2+(\alpha  k_B T)^2}+\Gamma\; ,\label{eq:S2_MFL}
\ee
where the prefactor $g$ in the MFL form becomes a universal constant in our case: $g=2/\pi^2$ and $\alpha = 4$, as will be explained below.

One can  now use the  Kramers–Kronig relations with the approximate expression in the marginal Fermi-liquid form to find an approximate form of the real part of the self-energy
\begin{align}
	\Sigma_{c}^{\prime}(\omega,T)&=\frac{1}{\pi}\mathcal{P}\int \frac{ \Sigma_{c}^{\prime\prime}(\omega_1,T)}{\omega_1-\omega}\dd \omega_1 \nn
	&=\frac{2}{\pi^2}\omega\ln \left(\frac{\sqrt{\omega^2+(\alpha  k_B T)^2}}{D} \right)\; , \label{eq:S1_MFL}
\end{align}
where again $D\gg\omega$ is a large energy cutoff.

From the marginal Fermi--liquid approximations, Eq. (\ref{eq:S2_MFL}) and Eq. (\ref{eq:S1_MFL}), we note that in the limit $T \to 0$, we recover the zero temperature expressions in Eqs. (\ref{eq:S20_def2}) and (\ref{eq:S10_def}) of our KHSB approach , and that at finite temperature the limit $\omega \to 0$ of the marginal Fermi--liquid imaginary part of the self-energy [Eq. (\ref{eq:S2_MFL})] recovers the slave-boson expression [Eq. (\ref{eq:S2_def2})]  for $\alpha=4$:
\begin{align}
	\lim_{T \to 0}	\Sigma_{c}^{\prime\prime}(\omega,T)&= -\frac{\norm{\omega}}{\pi}=\lim_{T \to 0}\Sigma^{\prime\prime}_c(\omega, T)=\Sigma''_c(\omega)\; ,\label{eq:LimSigmaIm}\\
	\lim_{T \to 0}	\Sigma_{c,MFL}^{\prime}(\omega,T)& =\frac{2}{\pi^2}\omega\ln \left(\frac{\norm{\omega}}{D}\right)=\Sigma^{\prime}_c(\omega)\; , \\
	\lim_{\omega  \to 0}\Sigma_{c,MFL}^{\prime\prime}(\omega,T)&=	-\frac{\alpha  k_B T}{\pi}=\lim_{\omega  \to 0}\Sigma^{\prime\prime}_c(\omega, T)=	-\frac{4 k_B T}{\pi}\; ,
\end{align}
hence $\alpha=4$. This is represented in Supplementary Figure \ref{fig:S2MFLvsS2SB} where we compare the imaginary part of the self-energy for the marginal Fermi-liquid and our KHSB model .

\begin{figure*}[h]
	\includegraphics[width=\linewidth]{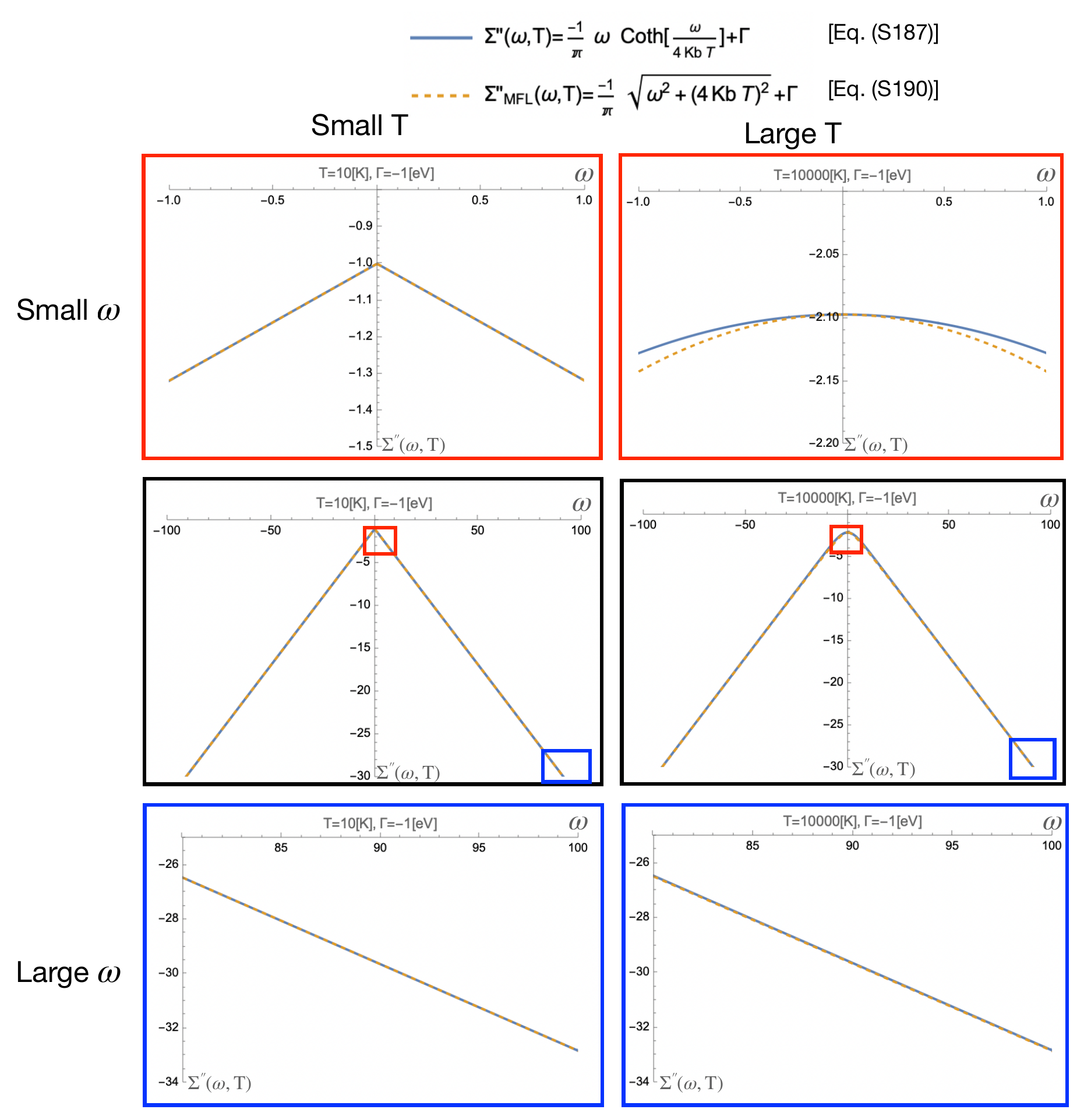}
	\caption{\added{Comparison of the imaginary part of the self-energy between Eq. (\ref{eq:S2_def2}) and the approximated marginal Fermi-liquid-like form in Eq. (\ref{eq:S2_MFL}). Within the physically relevant range in $\omega$ and $T$, the marginal Fermi-liquid-like form Eq. (\ref{eq:S2_MFL}) is a good approximation to Eq. (\ref{eq:S2_def2}).}}
	\label{fig:S2MFLvsS2SB}
\end{figure*}

In Supplementary Figure \ref{fig:S1MFLvsS1SB}, we also see that the numerical evaluation of Eq. (\ref{eq:S1_KK}) is almost identical to the  marginal Fermi-liquid form in Eq. (\ref{eq:S1_MFL}). The numerical integration of Eq. (\ref{eq:S1_KK}) is performed between $-D$ and $D$. We indeed note that  for a sufficiently large cutoff $D\geq1000$ eV, and a not too high temperature $T<10^{5}~\rm{K}$ (i.e. $k_B T\simeq 1~\rm{eV}$) the numerically evaluated  Eq. (\ref{eq:S1_KK}) and the  marginal Fermi-liquid form in Eq. (\ref{eq:S1_MFL}) lie very close to each other. 

Indeed, from inspection of the  Kramers–Kronig relation of Eq. (\ref{eq:S1_KK}), it is clear that, if the integration boundary $D$ is too close to the energy $\omega$ of interest, the pole lies very close to the integration boundary leading to fallacious results. We also note that for relatively ``low" temperature $k_B T\ll \hbar\omega$, i.e in the limit $\frac{ \hbar \omega}{k_B T}\to \infty$ or in the limit $T\to 0$, the value of the imaginary part of self-energy approaches to $-\frac{1}{\pi}\norm{\omega}$ [see Eq. (\ref{eq:LimSigmaIm})] for both marginal Fermi-liquid-like form  [Eq. (\ref{eq:S2_MFL})] and the form via our KHSB model.  We expect this agreement to also appear for the real part of the self energy. 
In the last column of Supplementary Figure \ref{fig:S1MFLvsS1SB}, we observe that, when the condition $k_B T \ll \hbar\omega$ is no longer satisfied, a discrepancy begins to emerge in the real part of the self-energy between the marginal Fermi liquid-like form and the form derived via our KHSB approach.

The main difference between the analytical form of our self-energy [Eq. (\ref{eq:S2_def2})] and the approximated form by the marginal Fermi liquid-like expression [Eq.  (\ref{eq:S2_MFL})] is that, at finite temperature, the real part of the self-energy is well defined at $\omega=0$ in the approximated marginal Fermi-liquid-like form shown in Eq. (\ref{eq:S1_MFL}), but it is not well-defined via the analytic form in Eq. (\ref{eq:S1_KK}).  Indeed, we note from Eq. (\ref{eq:S1_KK}) that, for $\omega=0$, the integrand becomes simply $\coth(\frac{\hbar\omega_1}{4 k_B T})$, which has a discontinuity in $\omega_1=0$ lying inside the integration interval $\left[-D, D\right]$.

\begin{figure*}[h]
	\includegraphics[width=\linewidth]{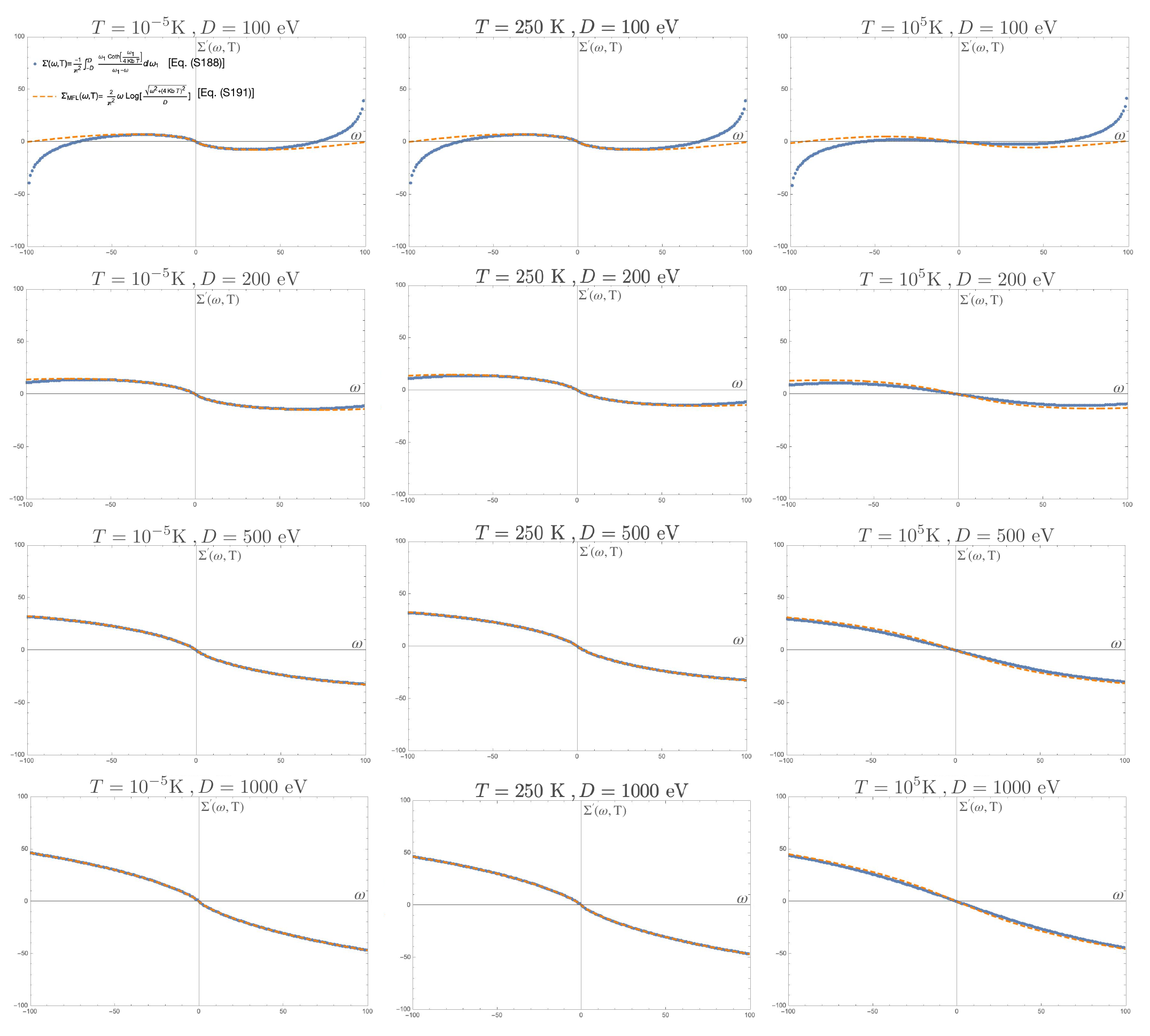}
	\caption{Comparison of the real part of the self-energy for the marginal Fermi-liquid [Eq. (\ref{eq:S1_MFL})] (dashed orange line) and Eq. (\ref{eq:S1_KK}) (blue dots) obtained by numerical integration of imaginary part according to the Kramers-Kronig relations for different value of the cutoff parameter $D=100 \, \rm{eV}, \,  200\, \rm{eV}, \, 500\, \rm{eV},$ and $ 1000 \, \rm{eV}$  and different temperatures $T=10^{-5} \rm{K}, 250 \rm{K},$ and $ 10^{5} \rm{K}$. We note that for a large enough cutoff $D\geq1000$ eV, \added{the numerically obtained results by Eq. (\ref{eq:S1_KK}) is well-fitted by the marginal Fermi-liquid-like form in Eq. (\ref{eq:S1_MFL}).}}
	\label{fig:S1MFLvsS1SB}
\end{figure*}

\subsection{Single-particle spectral function}
The spectral function of the single-particle's Green's function is given by
\be
A(\bm{k},\omega)=\frac{-1}{\pi}\rm{Im}\left[\frac{1}{\omega-(\varepsilon(\bm{k})-\mu)-\Sigma(\omega,T)}\right]\; .
\label{eq:A_def}
\ee
Our aim is to compare spectral function $A(\bm{k},\omega)$ via our KHSB approach to the ARPES data measured in Ref. \cite{zxshen-sicience-incoherent-cuprate} for Bi2212 $\rm{[(Bi, Pb)}_2 \rm{Sr}_2 \rm{CaCu}_2 \rm{O}_{8+\delta}]$ in the strange metal phase with hole doping $p=0.196$ at $T=250$K. This is done in the strange metal phase of this compound (doping $p=0.196$) at 250 K. This material features bilayer Cu-O planes and shows  two parabolic bands (bonding and anti-bonding bands) in the  heavily overdoped case which corresponds to the strange metal phase. Therefore, the total spectral function is the  sum of the two spectral functions associated with each band
\begin{align}
A(\bm{k},\omega)=\sum_{i=1}^{2}\frac{-C_i}{\pi}\rm{Im}\left[ \frac{1}{\omega-\varepsilon_i(\bm{k})+\mu_i-\Sigma_{c,i}(\omega,T)}\right] ,
\label{eq:A2_def}
\end{align}
\added{where $\rm{Im} \left[ \Sigma_{c,i} \right]$ is given by Eq. (\ref{eq:S2_def2}) and it is well-approximated by the marginal Fermi-liquid-like form  Eq. (\ref{eq:S2_MFL}) (see the comparison between these two forms in Supplementary Figures \ref{fig:S2MFLvsS2SB} and \ref{fig:S1MFLvsS1SB}),} and $C_i$ are the normalization weights of the experimental data. We also assume the two dispersion bands to be given by Eq. (\ref{eq:epsilon_def}) with different parameter for each bands
\be
\varepsilon_i(\bm{k})=-2t_i \left[\cos(k_x)+\cos(k_y)\right]-4t_i'\cos(k_x)\cos(k_y)\;, 
\label{eq:epsilon_def_i}
\ee
\added{with $t_i $ and $t_i^\prime$ as defined in Eq. (\ref{eq:epsilon_def}). In the following, the self-energy $\Sigma(\omega,T)$ is treated by the marginal Fermi-liquid approximation, since it is a faithful representation of the scaling form in Eqs. (\ref{eq:S2_MFL}) and (\ref{eq:S1_MFL}) [with $(x/2)\coth\left(x/4\right)$], and it is a smooth and well-defined function over the entire regions of $\omega$ and $T$, convenient for our calculations:  } 
\begin{subequations}
\begin{align}
	\Sigma_i(\omega,T)&=	\Sigma_{i,c}^{\prime}(\omega,T)+i\Sigma_{i,c}^{\prime\prime}(\omega,T)\\
	&=\frac{2}{\pi^2}\omega\ln \left(\frac{\sqrt{\omega^2+(\alpha  k_B T)^2}}{D_i} \right) 
  -i\frac{1}{\pi}\sqrt{\omega^2+(\alpha  k_B T)^2}+i\Gamma_i\; , \label{eq:Stot_MFL}
\end{align}\label{eq:Sigmafit}
\end{subequations}
with $\alpha=4$.

We fit our results for the spectral weight [Eqs. (\ref{eq:A2_def})-(\ref{eq:Sigmafit})] to the experimental data from Ref. \cite{zxshen-sicience-incoherent-cuprate} in the following two cases. The first one is the energy density curve along $\omega$ at the anti-nodal point $(k_x=\frac{-\pi}{a_0}, k_y\approx0)$, and the second one is the momentum density curve along $k_y$ at zero-energy $\omega\approx0$ and $k_x=\frac{-\pi}{a_0}$. These are shown in Figs. 5(a) and 5(b) of the main text, respectively, and correspond to the blue lines which are associated to a doping value of $\delta=0.196$. The fitting of our model is done by simultaneously fitting both curves together. Our model consist of a total of 12 parameters that need to be found in order to reproduce the experimental data. Namely the parameters are
\be
\begin{matrix}
\mu_1, & \Gamma_1& D_1,&t_1, &t_1',&C_1,\\
\mu_2, &\Gamma_2,& D_2,&t_2, &t_2',&C_2\;,
\end{matrix}\label{eq:parameters}
\ee
where the subscripts refers to the two bands.

\added{We first set $T=250$ K, same as that taken in experiment in Ref. \cite{zxshen-sicience-incoherent-cuprate}. The energies $\omega$ are measured from the chemical potential $\mu = \mu_1 =\mu_2$. Here, we set $\omega \to \omega-\mu$ since experimentally the energies are measured from the chemical potential $\mu_1 , \, \mu_2$.} We also a large bandwidth cutoff for the two bands, $D_1=D_2=100$ eV, since for the energy region $\omega\in [-0.25\, \rm{eV}, 0.05\, \rm{eV}]$ that we are interested in, these are sufficiently big enough cutoff for the marginal Fermi-liquid version of the real part of the self-energy in Eq. (\ref{eq:S1_MFL}) to reliably reproduce the features of Eq. (\ref{eq:S1_KK}) from our KHSB approach, as depicted in Supplementary Figure \ref{fig:S1MFLvsS1SB}. We find that the set of parameters that reproduces the best the experimental data (blue lines in Fig. 5 of the main text) are given by 
\be
\begin{matrix}
	 & \Gamma_1=-0.1~\rm{eV},& D_1=1000\, \rm{eV},&t_1=2.2~\rm{eV}, &t_1'=-0.005~\rm{eV},&C_1 =5.8,\\
	 &\Gamma_2=-0.2~\rm{eV},& D_2=1000\, \rm{eV},&t_2=6.5~\rm{eV}, &t_2'=-0.07\, \rm{eV},&C_2=8\;.
\end{matrix}\label{eq:parameters_fit}
\ee

 The predictions from our approach plotted for the parameter set given in Eq. (\ref{eq:parameters_fit}) are depicted in Figs. 5(a) and 5(b) of the main text (green lines) on top the experimental data (blue lines) for easy comparison. In Fig. 5(a) of the main text, we show the energy density curve of the spectral function  applied to Bi2212, predicted by our theoretical approach  [Eqs. (\ref{eq:A2_def}) - (\ref{eq:Sigmafit})] and plotted for the parameter set given in Eq. (\ref{eq:parameters_fit}). In Fig. 5(b) of the main text, we plotted  the momentum density curve, and in Fig. 5(c), we show the single-particle spectral function for the same parameters as used in  Fig. 5(a) of the main text.

\twocolumngrid
\bibliographystyle{apsrev4-2}
\bibliography{SBtJ.bib}

\end{document}